\documentclass[aps,prr,reprint,twocolumn,superscriptaddress,floatfix,nofootinbib,longbibliography]{revtex4-1}
\usepackage{amsmath,amssymb,color,comment,physics,mathtools}
\usepackage[makeroom]{cancel}
\usepackage[caption=false]{subfig}
\usepackage{mathrsfs}
\usepackage{graphicx}
\usepackage{subfig}
\usepackage[countmax]{subfloat}
\usepackage[english]{babel}
\usepackage[bookmarks=true,colorlinks,linkcolor=OrangeRed,urlcolor=NavyBlue,citecolor=RoyalBlue]{hyperref}
\usepackage[dvipsnames]{xcolor}
\usepackage{ulem} 
\usepackage{braket}
\usepackage{dsfont}

\newcommand{\Fig}[1]{Fig.~\ref{#1}}

\definecolor{mygreen}{rgb}{0,0.5,0}
\definecolor{myblue}{rgb}{0,0,0.75}
\definecolor{mymagenta}{cmyk}{0,1,0,0.12}
\definecolor{mygray}{rgb}{0.5,0.5,0.5}

\def\d{\mathrm d}

\definecolor{mypink1}{rgb}{0.858, 0.188, 0.478}
\definecolor{mypurple}{rgb}{0.49,0.18,0.56}
\definecolor{mygold}{rgb}{0.93,0.69,0.13}
\definecolor{mygreen}{rgb}{0,0.5,0}
\definecolor{myred}{rgb}{1,0,0}
\definecolor{myblue}{rgb}{0,0,0.75}
\definecolor{mymagenta}{cmyk}{0,1,0,0.12}
\definecolor{mygray}{rgb}{0.5,0.5,0.5}

\newcommand{\ignore}[1]{}
\usepackage{geometry}

\begin{document}

\title{Robustness of gauge-invariant dynamics against defects in ultracold-atom gauge theories}
\author{Jad C.~Halimeh}
\affiliation{INO-CNR BEC Center and Department of Physics, University of Trento, Via Sommarive 14, I-38123 Trento, Italy}

\author{Robert Ott}
\affiliation{Institute for Theoretical Physics, Ruprecht-Karls-Universit\"{a}t Heidelberg, Philosophenweg 16, 69120 Heidelberg, Germany}

\author{Ian P.~McCulloch}
\affiliation{School of Mathematics and Physics, The University of Queensland, St.~Lucia, QLD 4072, Australia}

\author{Bing Yang}
\affiliation{Physikalisches Institut, Ruprecht-Karls-Universit\"{a}t Heidelberg, Im Neuenheimer Feld 226, 69120 Heidelberg, Germany}
\affiliation{Institut f\"ur Experimentalphysik, Universit\"at Innsbruck, Technikerstra{\ss}e 25, 6020 Innsbruck, Austria}

\author{Philipp Hauke}
\affiliation{INO-CNR BEC Center and Department of Physics, University of Trento, Via Sommarive 14, I-38123 Trento, Italy}

\begin{abstract}
Recent years have seen strong progress in quantum simulation of gauge-theory dynamics using ultracold-atom experiments. A principal challenge in these efforts is the certification of gauge invariance, which has recently been realized in [B.~Yang \textit{et al.}, \href{https://arxiv.org/abs/2003.08945}{arXiv:2003.08945}]. One major but poorly investigated experimental source of gauge-invariance violation is an imperfect preparation of the initial state. Using the time-dependent density-matrix renormalization group, we analyze the robustness of gauge-invariant dynamics against potential preparation defects in the above ultracold-atom implementation of a $\mathrm{U}(1)$ gauge theory. We find defects related to an erroneous initialization of matter fields to be innocuous, as the associated gauge-invariance violation remains strongly localized throughout the time evolution. A defect due to faulty initialization of the gauge field leads to a mild proliferation of the associated violation. Furthermore, we characterize the influence of immobile and mobile defects by monitoring the spread of entanglement entropy. Overall, our results indicate that the aforementioned experimental realization exhibits a high level of fidelity in the gauge invariance of its dynamics at all evolution times. Our work provides strong evidence that ultracold-atom setups can serve as an extremely reliable framework for the quantum simulation of gauge-theory dynamics.
\end{abstract}
\date{\today}
\maketitle

\section{Introduction}
The simulation of gauge theories \cite{Weinberg_book,Cheng_book} in quantum hardware promises insights into subatomic phenomena that are forbiddingly difficult to access by classical computers \cite{Gattringer2009,Wiese_review,Zohar_review,Dalmonte_review,Pasquans_review,Alexeev_review}.
This potential has incited a strong experimental effort to engineer the dynamics of gauge theories in low-energy devices based on trapped ions, superconducting qubits, and ultracold atoms, following mainly two approaches.
The first experimental approach exploits the target gauge symmetry to eliminate either the matter \cite{Bernien2017,Surace2019,Zohar2019} or gauge fields \cite{Martinez2016,Muschik2017,Kokail2019} at the cost that mechanisms breaking gauge invariance cannot be tested and that local errors in the physical hardware may represent highly nonlocal errors in the target model \cite{Muschik2017}. Similar considerations hold for implementations in quantum computers that remove all unphysical, gauge-violating states from the Hilbert space \cite{Klco2018}.
A second direction of experiments generates local gauge invariance as a constraint on microscopic interactions between different degrees of freedom representing matter and gauge fields \cite{Dai:2017,Goerg2019,Schweizer2019,Mil2019,Yang2020}, providing the possibility to observe the gauge symmetry.
In the absence of unrealistic fine tuning, however, such experiments will always show some degree of gauge violation.
While some results do exist on error terms that break gauge invariance during the time evolution \cite{Banerjee2012,Hauke2013,Stannigel2014,Kuehn2014,Kuno2015,Yang2016,Kuno2017,Negretti2017,Barros2019,Halimeh2020a,Halimeh2020b,Halimeh2020c,Halimeh2020e}, defects in the preparation of the initial state are sorely unexplored. A major concern about such defects is that they may proliferate and compromise gauge invariance throughout the entire system, in analogy, e.g., to the destruction of topological order in the toric code by propagation of defects \cite{Stark2011,Hastings2011} or to the domain-wall melting in the presence of hole and spin-flip defects in the two-species Bose--Hubbard realization of XXZ chains \cite{Halimeh2014}. Indeed, it has been experimentally shown in the latter realization that a single spin impurity can give rise to significantly different dynamics, depending on how accurately the realization maps onto the target model \cite{Fukuhara2012}. Consequently, such defects can also be used as a probe of how faithful ultracold-atom implementations of target models are. This motivates us to carry out a rigorous analysis of the effect of defects in such implementations on gauge-theory dynamics.

\begin{figure*}[!ht]
	\centering
	\includegraphics[width=\textwidth]{{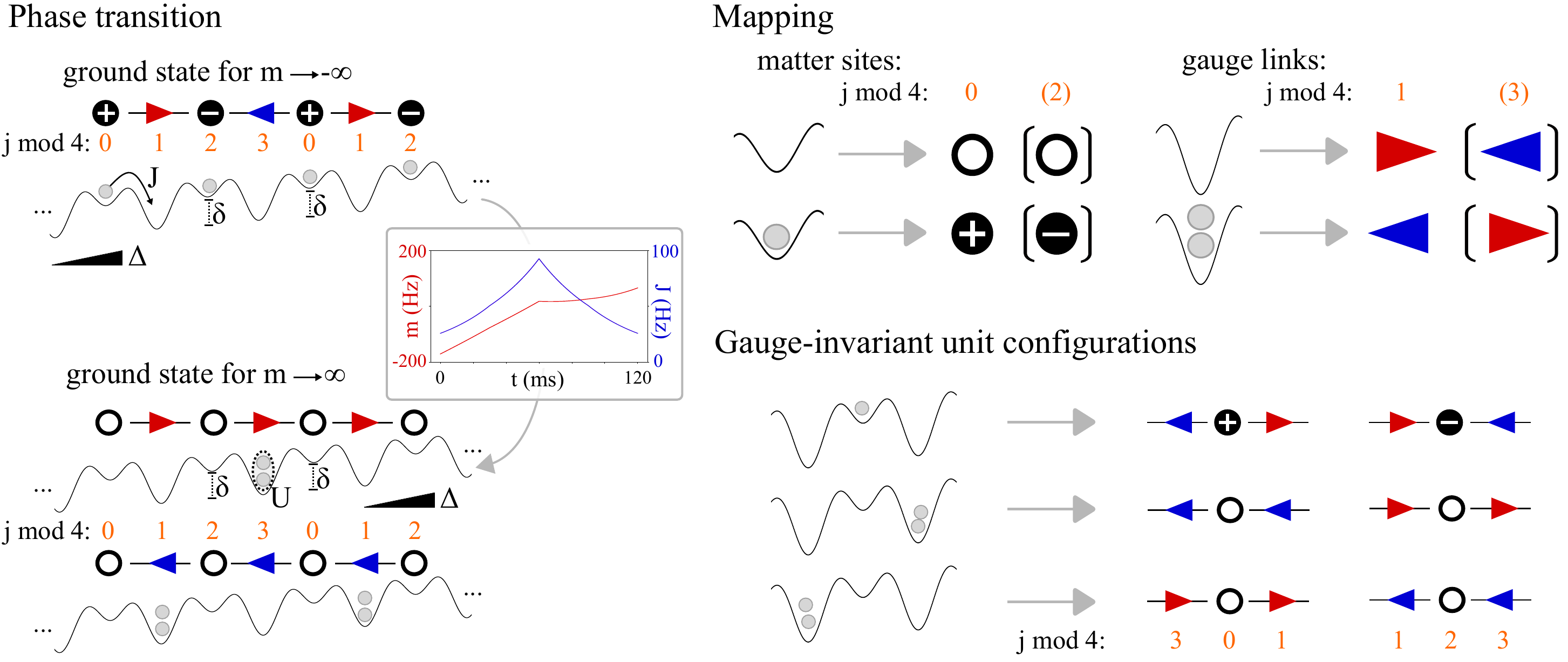}}
	\caption{(Color online). Mapping a U(1) gauge theory to a Bose--Hubbard model and ramp protocol. Left: The ramp protocol in the experiment of Ref.~\cite{Yang2020} sweeps across Coleman's phase transition, which separates a nondegenerate charge- and parity-symmetric phase, where charges proliferate, from a doubly degenerate phase where both charge and parity (C/P) symmetries are broken, and the electric field passes through the system unhindered.
		The sketches each display the states in the gauge theory (top) and the corresponding configuration in the optical superlattice including the Hubbard parameters $J$, $U$, as well as the superlattice staggering $\delta$ and a tilt $\Delta$ (bottom). The mapping of sites of the optical lattice, labelled by $j=1,\dots,L$, to degrees of freedom of the gauge theory has a periodicity of four sites.
		Right: Mapping between the $\mathrm{U}(1)$ gauge theory and the bosonic system realized experimentally. In our indexing, odd sites of the bosonic superlattice are deep sites that represent the gauge fields after a staggered rotation. Thus, when the gauge site hosts a doublon, this indicates an electric field pointing to the right (left) when the site index $j\bmod4=1$ ($3$). The absence of doublons indicates an electric field with the opposite orientation. Even sites of the bosonic superlattice are shallow and represent matter fields. When a matter site hosts a boson this indicates the presence of an ``electron'' (``positron'') on that site in an alternating fashion, i.e., when its index $j\bmod4=2$ ($4$). Zero bosons indicates the absence of matter. Bottom: The configurations of the Bose--Hubbard quantum simulator that satisfy Gauss's law.}
	\label{fig:mapping}
\end{figure*}

\begin{figure*}[!ht]
	\centering
	\includegraphics[width=\textwidth]{{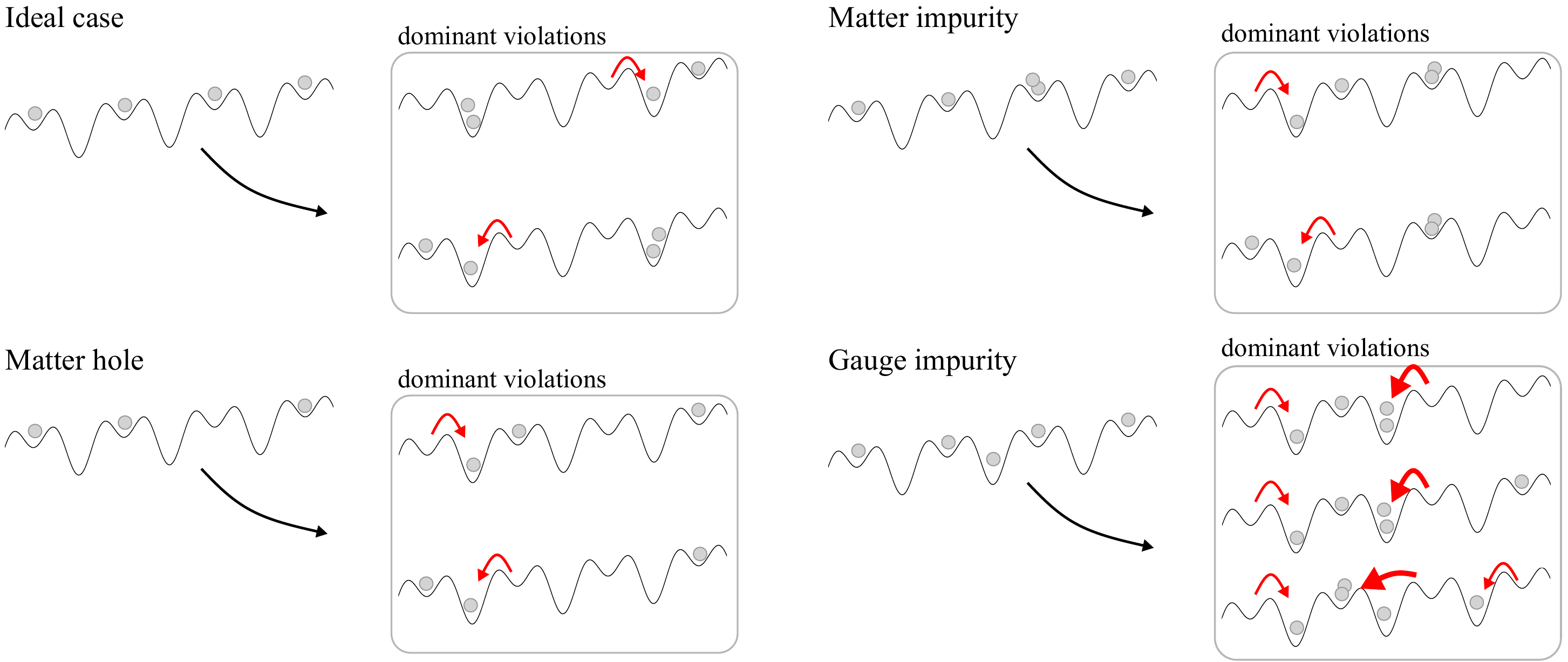}}
	\caption{(Color online). The various initial states considered in this work, along with the associated most probable gauge-invariance-violating product states in the final wave function after the ramp.
	The influence of matter-hole and matter-impurity defects on the subsequent dynamics-induced gauge violation is extremely small (by the end of ramp it remains comparable to that of the clean case;~cf.~Fig.~\ref{fig:FigTotalError} and Sec.~\ref{sec:TotalViolation}).
	The effect of gauge impurities, on the other hand, is less benign, because such defects facilitate further gauge-violating processes that are otherwise energetically too costly. See Appendix~\ref{sec:hist} for a histogram analysis of the product states most highly populated by the final wave function 
	for the four initial states shown above.
	}
	\label{fig:violations}
\end{figure*}

Here, we thoroughly investigate the effect of initial state-preparation errors on the gauge-invariant dynamics of a $\mathrm{U}(1)$ gauge theory as has recently been realized in the ultracold-atom experiment of Ref.~\cite{Yang2020}.
There, a lattice gauge theory \cite{Rothe_book} describing fermionic charged matter and electric fields in the quantum link model formalism \cite{Wiese_review,Chandrasekharan1997} has been realized by a mapping onto a Bose--Hubbard system (Fig.~\ref{fig:mapping}, top right).
The purpose of the present work is to provide an in-depth study of defects in the initial atom configuration that violate the $\mathrm{U}(1)$ gauge symmetry, as encoded by Gauss's law (Fig.~\ref{fig:mapping}, bottom right).
We consider three main defects that carry nonnegligible probabilities to occur experimentally;~see Fig.~\ref{fig:violations} and associated discussion in Sec.~\ref{sec:setup}. The first two are matter-field defects in which a given matter site of the initial state has either no bosons [matter-hole (MH) defect] or two bosons [matter-impurity (MI) defect], instead of the single boson configuration at each matter site that is in accordance with Gauss's law, as shown in Fig.~\ref{fig:violations}. The third considered defect that violates Gauss's law consists of one boson residing on a gauge-field site, defined as gauge-impurity (GI) defect, whereas gauge sites are supposed to be initially empty.
Using the time-dependent density-matrix renormalization group, we analyze the influence of these defects on the $\mathrm{U}(1)$ gauge symmetry, numerically simulating the ramp of Ref.~\cite{Yang2020} through a symmetry-breaking quantum phase transition (Fig.~\ref{fig:mapping}, left).
Importantly, the matter defects remain localized throughout the entire time evolution while the gauge-impurity defect shows a small spreading. This finding also enables us to propose a simple but effective procedure to remove the most detrimental defects (gauge impurities) and thus to improve the reliability and scalability of upcoming gauge-theory quantum simulator experiments.

The paper is organized as follows. In Sec.~\ref{sec:model}, we describe the $\mathrm{U}(1)$ quantum link model (QLM), its mapping onto a staggered Bose--Hubbard model (BHM), and the ramp through Coleman's phase transition \cite{Coleman1976}, as has recently been experimentally realized in an optical superlattice \cite{Yang2020}. In Sec.~\ref{sec:setup}, we discuss the experimental system employed in Ref.~\cite{Yang2020} and how the defects considered in this work can arise in the initial-state preparation. In Sec.~\ref{sec:results}, we present our main numerical results obtained using the time-dependent density-matrix renormalization group. We then provide in Sec.~\ref{sec:future} proposals for future experiments based on our results, and conclude in Sec.~\ref{sec:conclusion}. We also include a detailed derivation of the mapping of the $\mathrm{U}(1)$ QLM onto the BHM in Appendix~\ref{sec:mapping}, as well as a histogram analysis of the dominant processes in the final wave function at the end of the ramp in Appendix~\ref{sec:hist}.

\section{Model and ramp protocol}\label{sec:model}
The target model we wish to investigate is a $\mathrm{U}(1)$ lattice gauge theory in one spatial dimension described by the Hamiltonian \cite{Hauke2013,Yang2016} 
\begin{align}
\label{eq:H-QLM}
H_\text{QLM}=\sum_\ell\left[-\frac{\mathrm{i}\tilde{t}}{2}\left(\psi_\ell S^+_{\ell,\ell+1}\psi_{\ell+1}-\text{H.c.}\right)+m\psi_\ell^\dagger\psi_\ell\right].
\end{align}
(See also Appendix~\ref{sec:mapping}.) Here, the matter fields on sites $\ell$ are represented by the fermionic creation and annihilation operators $\psi_\ell^{\dagger}$ and $\psi_\ell$, respectively. We employ the Hamiltonian formulation of lattice gauge theories by Kogut and Susskind with staggered fermions \cite{Kogut1975} followed by a particle--hole transformation on every second matter site \cite{Hauke2013,Yang2016}. Within the QLM framework \cite{Wiese_review,Chandrasekharan1997}. the gauge fields on the links $(\ell,\ell+1)$ are denoted by spin-$1/2$ ladder operators, while the electric field $E_{\ell,\ell+1}=(-1)^{\ell+1}S^z_{\ell,\ell+1}$ has been coarse-grained to two eigenvalues, represented by the blue ($-\frac{1}{2}$) and red ($+\frac{1}{2}$) arrows in Fig.~\ref{fig:mapping}.

We encode the gauge invariance of the Hamiltonian \eqref{eq:H-QLM} under local spatial gauge transformations via the so-called Gauss's law operators
\begin{equation}
	G_\ell = (-1)^{\ell+1}\big( S^{z}_{\ell-1,\ell} + S^{z}_{\ell,\ell+1} +  \psi^{\dagger}_\ell \psi_\ell\big).
\end{equation}
Ideally, the Hilbert space of our model is constrained to a gauge-invariant subspace by requiring
\begin{equation}
	G_\ell \ket{\psi_\text{i}} = 0,
\end{equation}
for the initial state $\ket{\psi_\text{i}}$ of the system.
Gauge invariance of the dynamics ($[G_{\ell},H]=0$) then retains this condition throughout the evolution.
There are only three gauge-invariant configurations on a local constraint, defined by a matter site $\ell$ with its two gauge links $(\ell-1,\ell)$ and $(\ell,\ell+1)$. These are shown in the bottom right of Fig.~\ref{fig:mapping}.
A violation of gauge symmetry may emerge as a consequence of gauge-noninvariant dynamics, i.e., when $[G_{\ell},H]\neq0$, or of an ill-prepared initial state, i.e., $G_\ell \ket{\psi_\text{i}} \neq 0$ for at least one $\ell$.

The experiment quantum-simulates this gauge theory by mapping it onto an effective Hamiltonian derived from a staggered tilted Bose--Hubbard model (BHM), given by
\begin{align}\nonumber
H_\text{BHM}=&\,\sum_j\bigg\{-J\big(b_j^\dagger b_{j+1}+\text{H.c.}\big)+\frac{U}{2}n_j\big(n_j-1\big)\\\label{eq:BHM}
&+\frac{1}{2}\big[(-1)^j\delta+2j\Delta\big]n_j\bigg\},
\end{align}
where the staggering strength is represented by $\delta$, $J$ denotes the strength of the direct tunneling, and $U\sim 2\delta\gg J$ is the on-site interaction strength. Furthermore, we use a linear tilt $\Delta$ to suppress unwanted hopping processes.

The gauge symmetry of this system emerges as an effective description for a near-degenerate subspace of the Hilbert space (not the ground subspace of the Bose--Hubbard model). The system is initialized in this subspace through the choice of the initial configuration shown in the top left of \Fig{fig:mapping}, which ideally fulfils Gauss's law everywhere. The choice of parameters of the Bose--Hubbard model ensures the energetic separation of `gauge-invariant' configurations (see \Fig{fig:mapping}, bottom right) from all other, unwanted ones.
The gauge-invariant interaction between a pair of oppositely charged fermions and a gauge field between them is realized by a resonant second-order process with strength $\sim J^2/U$. Specifically, this corresponds to either formation or destruction of doublons on the deep `gauge sites' of the superlattice involving the two single bosons on the neighboring (shallow) `matter sites'. The process is made resonant by tuning the interaction energy $U$ of the doublon to the sum of two separate single boson energies from the staggering ($U\approx2\delta$). As detailed in Appendix~\ref{sec:mapping}, at leading order in degenerate perturbation theory we obtain the effective Hamiltonian
\begin{equation}
\label{eq:effective-Hamiltonian}
	H_{\text{eff}} = \sum_{j_\text{m}} \left\{m b_{j_\text{m}}^{\dagger}b_{j_\text{m}} + \frac{\tilde{t}}{\sqrt{8}} \Big[b_{j_\text{m}-1} \big(b^{\dagger}_{j_\text{m}}\big)^2 b_{j_\text{m}+1} +\text{H.c.} \Big]\right\},
\end{equation}
with $j_\text{m}$ an even site index of the optical lattice. 
This Hamiltonian acts on an energy manifold where occupations are constrained to $\left\{\ket{0},\ket{1}\right\}$ on even sites and to $\left\{\ket{0},\ket{2}\right\}$ on odd sites of the optical lattice. 
A Jordan--Wigner transformation maps the ``hard-core'' bosons on even sites to fermionic matter operators, while we associate the number states on the odd sites with two spin-$1/2$ eigenstates, which we reinterpret as the quantum links. 
With every odd (even) site on the bosonic superlattice representing a gauge (matter) site, and the additional alternating interpretations of gauge fields, as well as matter fields as charges and anti-charges, we obtain a mapping of ultracold-atom degrees of freedom onto the gauge theory with period four, as indicated in \Fig{fig:mapping}. 
The above mapping allows to identify the effective couplings $m = \delta -U/2$ and $\tilde{t}=8\sqrt{2}J^2/U$ with the fermion mass and the gauge-invariant interaction strength as introduced in Eq.~\eqref{eq:H-QLM}. 

Initially, we prepare the gauge-invariant state shown in \Fig{fig:mapping} (left) by initializing the Bose--Hubbard system for lattice parameters $U\gg \delta \gg J$. In the QLM, this corresponds to the limit $m\rightarrow -\infty$, where the lowest-energy eigenstate has matter sites filled alternatingly with either positively or negatively charged fermions. During the time evolution, we ramp the lattice parameters to generate time-dependent profiles for mass and interaction strength as shown in \Fig{fig:mapping} (left, inset).
We cross a quantum critical point (QCP) \cite{Coleman1976,Pichler2016} and end the evolution deep in the new phase, anticipating the limit $m \rightarrow \infty$. Here, fermion pairs are too costly and the $\mathrm{Z}_2$-symmetric ground state is characterized by a homogeneous electric-field configuration in a superposition of pointing left and right.
In our coherent evolution, the system firmly follows the instantaneous ground state until close to the QCP, where it evolves out of equilibrium. Distinct signatures of the new phase are then recovered towards the end of the ramp, such as a drastically reduced fermion number and doublon correlations indicating domains of homogeneous electric fields \cite{Yang2020}.

In our microscopic derivation of Eq.~\eqref{eq:effective-Hamiltonian}, we have assumed $U\gg J$ and discarded correction terms of order $\order{J^3/U^2}$, which will appear in any realistic implementation of this model. In addition, one expects the direct tunneling process not to be fully suppressed. This leads to single occupations $\ket{1}$ on gauge sites, which have no counterpart in the QLM. Moreover, we have introduced the linear tilt $\Delta$ to suppress second-order hoppings of bosons on matter sites to empty neighboring matter sites. Since parametrically $\Delta \sim m \sim \tilde{t}$ in some parts of the evolution, this condition is not always strictly fulfilled. However, it has been experimentally demonstrated that all these gauge violations are well controlled throughout the time evolution \cite{Yang2020}. Conversely, starting from defective initial states, with one or several misplaced bosons, the state resides in another `sector' of the Hilbert space where our identification with an effective gauge theory description may break down locally. In this regard, we find qualitative differences for matter defects and the gauge impurity in the strict perturbation-theory analysis leading to Eq.~\eqref{eq:effective-Hamiltonian}. There, we assume that a sufficiently strong linear tilt suppresses second-order tunnelings associated with a transport of the matter defect (MH or MI). Hence, we can expect these cases to be dominated by the same gauge-violating processes that occur in the clean case, as illustrated by the small red arrows in \Fig{fig:violations} (top left, top right, and bottom left). For the gauge impurity, however, one encounters additional gauge-noninvariant configurations, which resonantly couple to the initial state due to the presence of the GI defect (small and bold arrows in \Fig{fig:violations}, bottom right). We provide in Appendix~\ref{sec:hist} a quantitative histogram analysis of the dominant gauge-invariant and noninvariant processes in the final wave function at the end of the ramp for initial states with either no defect or a single (MH, MI, or GI) defect. In the following, we introduce the experimental platform of Ref.~\cite{Yang2020} together with its main error sources leading to the defects.

\section{Experimental initialization procedure and main defects}\label{sec:setup}
The experimental sequence in Ref.~\cite{Yang2020} begins with a two-dimensional unity-filled Mott insulator of ultracold bosonic $^{87}$Rb atoms.
A recently developed staggered-immersion cooling technique in optical lattices has enabled the average filling factor on $10^4$ lattice sites to be $0.992(1)$ \cite{Yang:2019}.
To initialize the state for quantum-simulating a lattice gauge theory, the experiment employs a site-selective addressing technique to remove the atoms on all of the gauge sites \cite{Yang:2017}.
Since the cold ensemble holds extremely low thermal entropy, all atoms occupy the ground band of the optical lattice.
To good approximation, the initial state has then unity filling on matter sites while there are no atoms on gauge sites, which maps to the target QLM ground state at $m\to-\infty$ where the system is filled with negative and positive charges alternatingly on matter sites, with an overscreened electric field alternating its direction at each link;~cf.~Fig.~\ref{fig:mapping}, top left.

In the initialization stage, imperfections in the state manipulation may imprint defects onto the initial state.
In particular, defects in the Mott insulator may appear due to an atom splitting after the staggered-immersion cooling \cite{Yang:2019}.
The basic concept of this cooling method is to transfer the entropy into contacting superfluid reservoirs, which can absorb almost all of the thermal entropy and are later removed from the system.
In this way, the experiment obtains a stripe Mott insulator with doublons on gauge sites and vacuum on matter sites, giving an occupation of $\ket{202020\ldots}$ along the one-dimensional chain.
The probability of a doublon is as close to unity as $0.999(1)$, which is derived from the temperature of the sample.
From the stripe state, the doublons are split into two sites to fill the vacancies of matter sites with single atoms, realizing the transformation $\ket{202020\ldots} \rightarrow \ket{111111\ldots}$.
This splitting operation has an efficiency of $0.993(1)$, which leads to a final probability for unity filling on each site of $0.992(1)$. Errors in this operation result in two kinds of defects with almost equal probability, one is a doublon $\ket{2}$ and the other is a vacuum state $\ket{0}$.
Since the gauge and matter sites are symmetric in the splitting operation, we can conclude that both defects have $0.4(1)\%$ probability per site.

The second main error source in the state initialization stems from the site-dependent cleansing.
The experiment uses a spin-dependent optical superlattice specifically developed for selective addressing of the gauge or matter sites \cite{Yang:2017}.
A microwave pulse drives a rapid adiabatic passage to transfer the atoms on gauge sites into another hyperfine state, with an efficiency of $99.5(3)\%$ per site.
These are then blown out from the lattice confinement by a resonant laser light.
Imperfections in the state transfer and subsequent atom removal leave gauge sites with a residual single-atom occupation of $0.5(3)\%$ per site.

These two stages of state manipulations cause independent defects on both gauge and matter sites.
Locally, the probability for a defect on gauge sites (state $\ket{1}$) is $0.5(3)\%$. On a matter site, the probability for each of the states $\ket{2}$ and $\ket{0}$ is $0.4(1)\%$. Accumulated over the entire chain with an experimental region of $71$ lattice sites \cite{Yang2020} ($36$ matter and $35$ gauge sites), the probabilities are: $13(4)\%$ for a single matter-hole or matter-impurity defect, $15(10)\%$ for a single gauge-impurity defect, and $4(2)\%$ for two defects of any kind.
As such, the probability of finding one of the MH, MI, or GI defects in the initial state is nonnegligible. Therefore, their effect on the subsequent gauge-invariant dynamics is a relevant aspect to investigate, which is what we do numerically in the following.

\section{Results and discussion}\label{sec:results}
\begin{figure*}[!ht]
	\centering
	\hspace{-.01 cm}
	\includegraphics[width=.48\textwidth]{{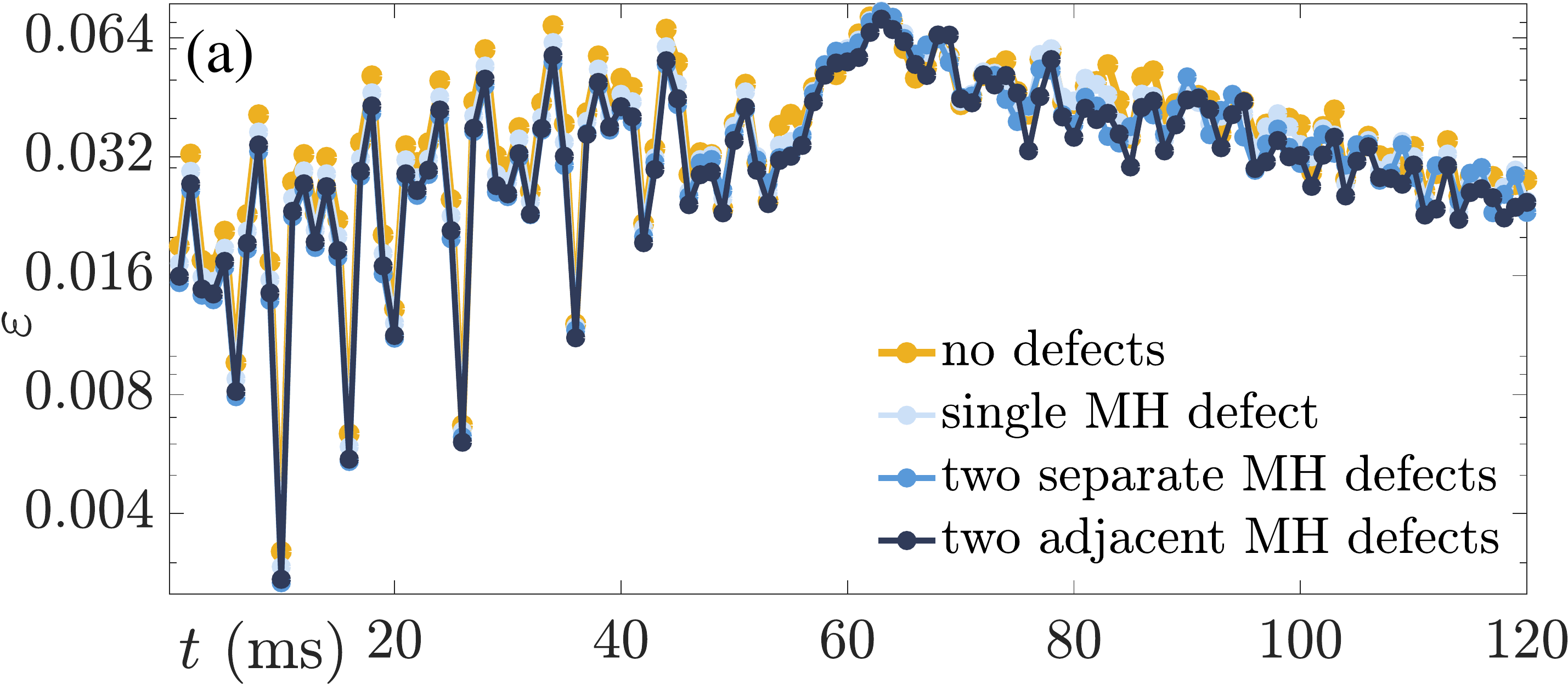}}\quad
	\includegraphics[width=.48\textwidth]{{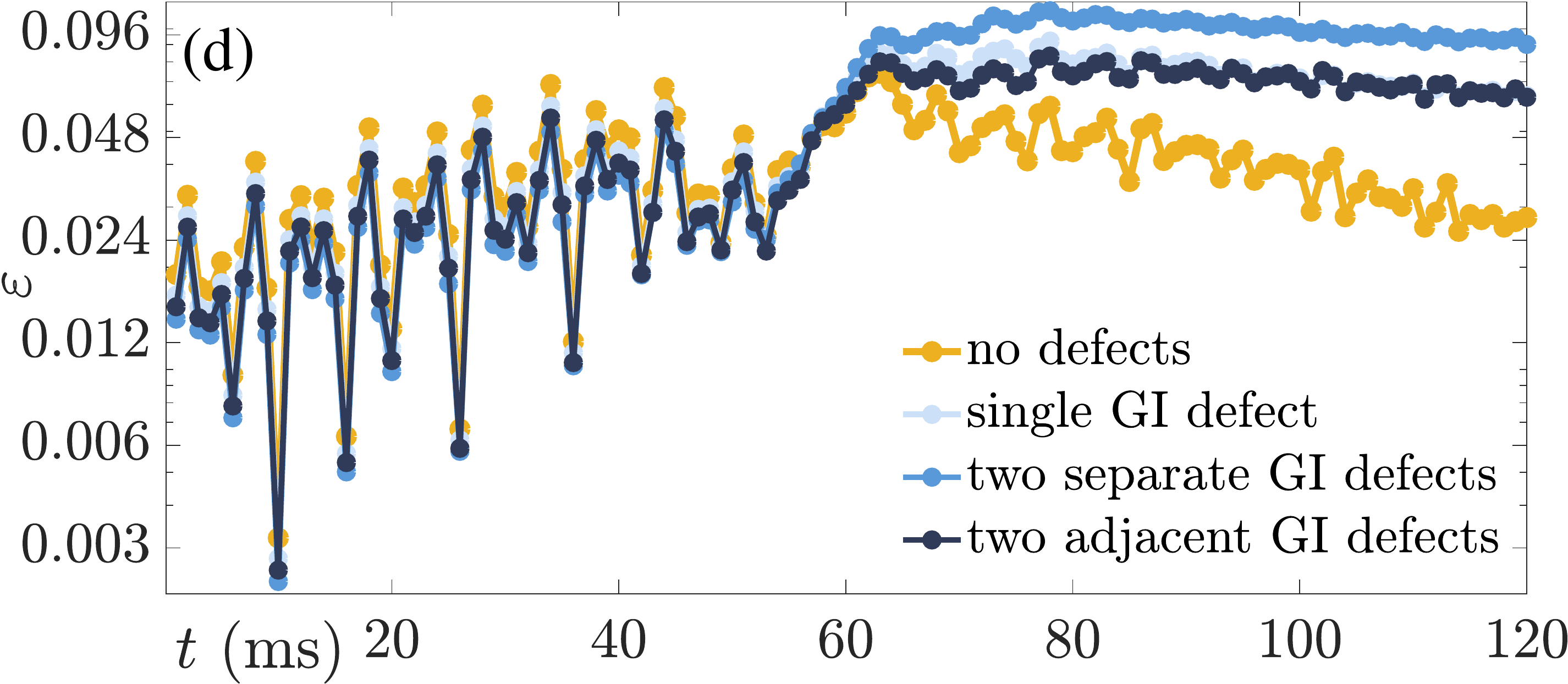}}\quad\\
	\hspace{-.01 cm}
	\includegraphics[width=.48\textwidth]{{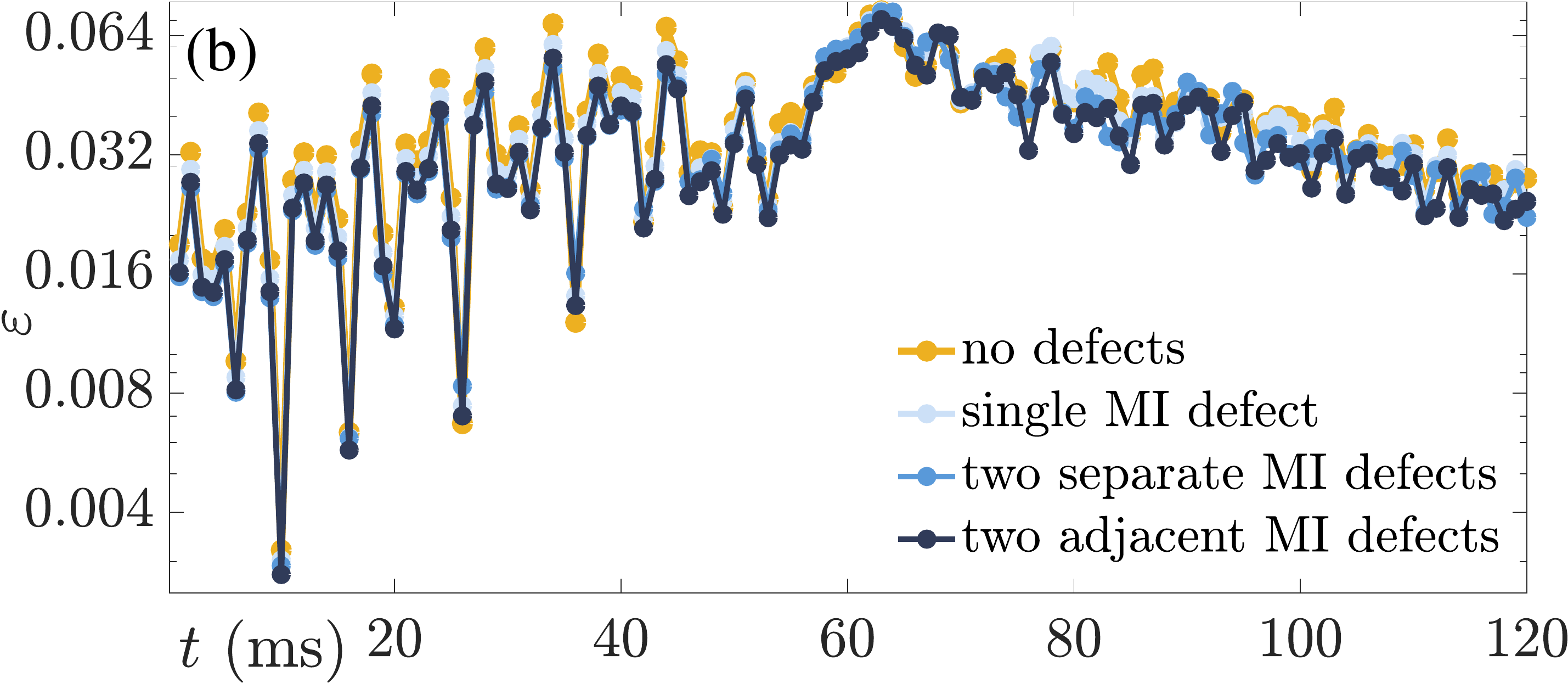}}\quad
	\includegraphics[width=.48\textwidth]{{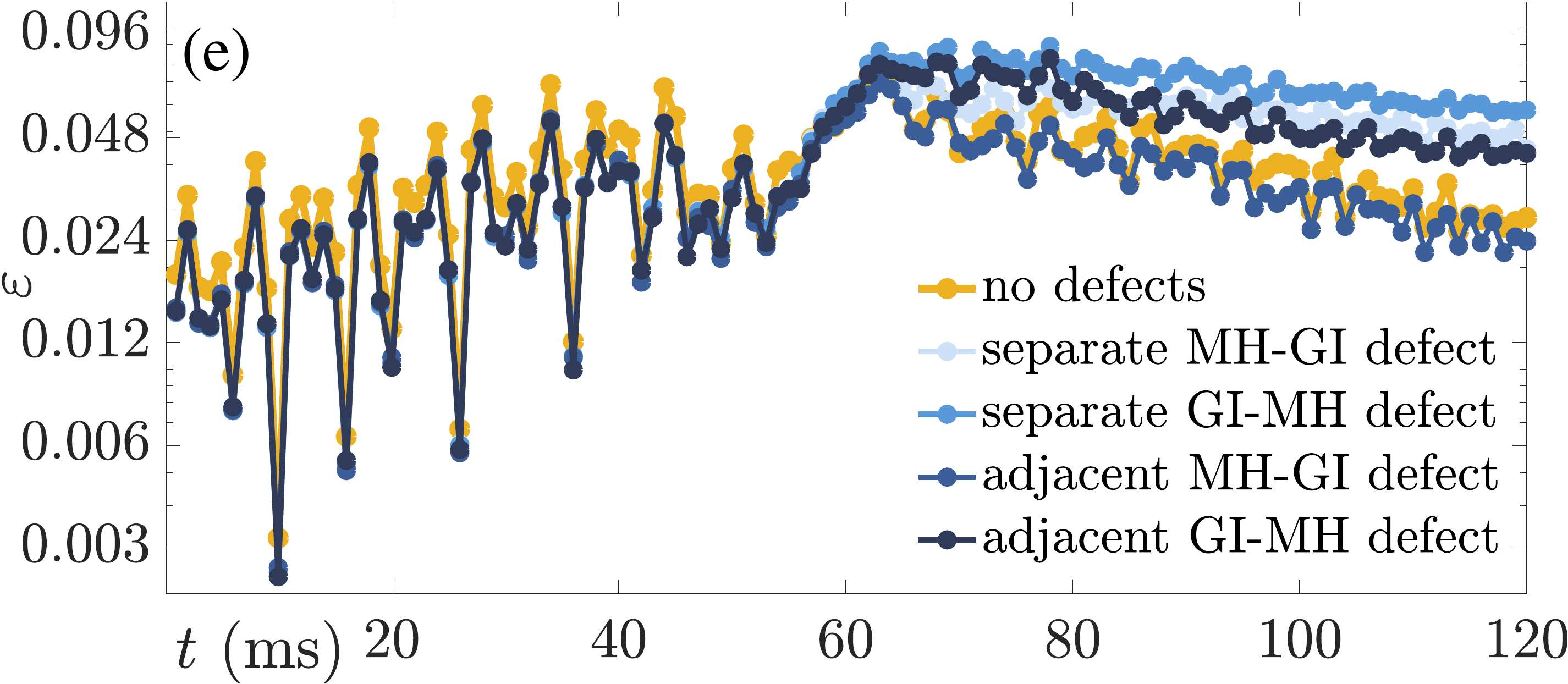}}\quad\\
	\hspace{-.01 cm}
	\includegraphics[width=.48\textwidth]{{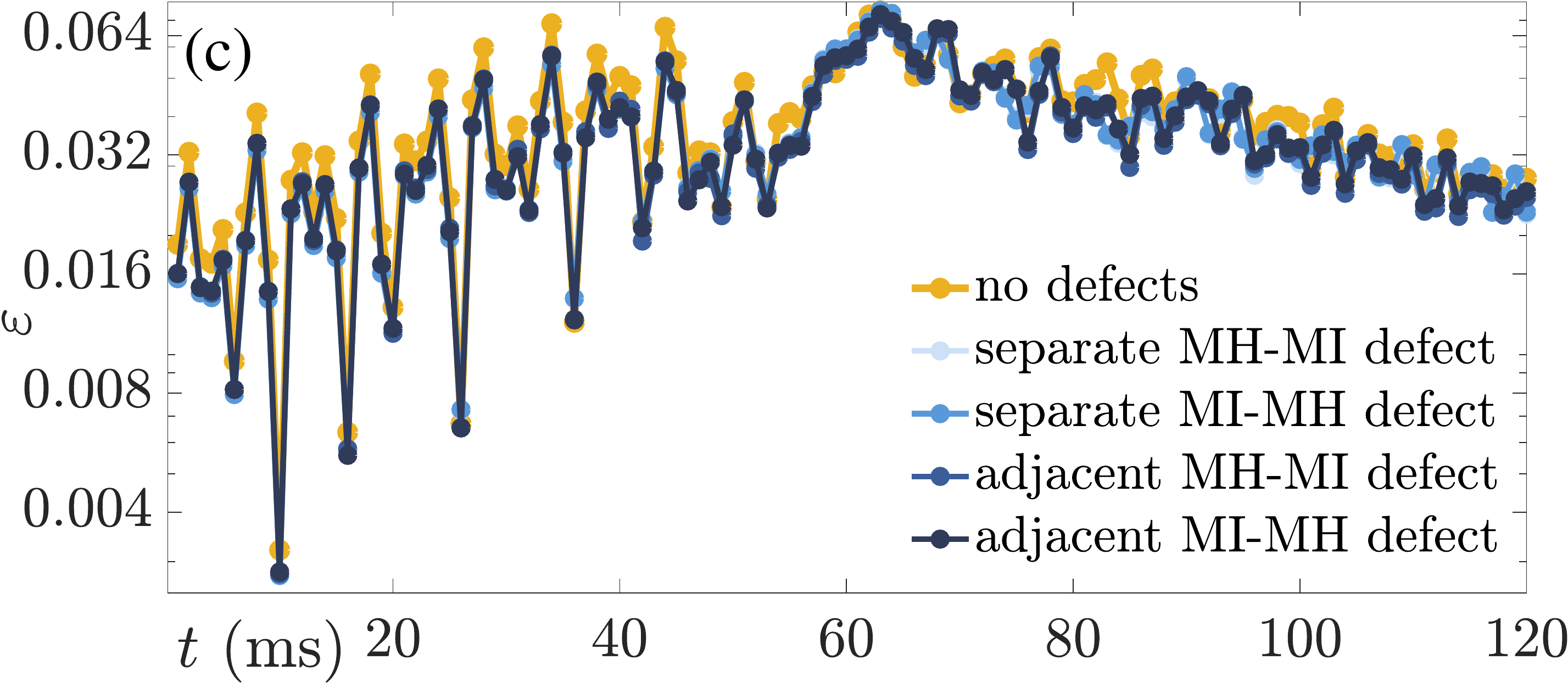}}\quad
	\includegraphics[width=.48\textwidth]{{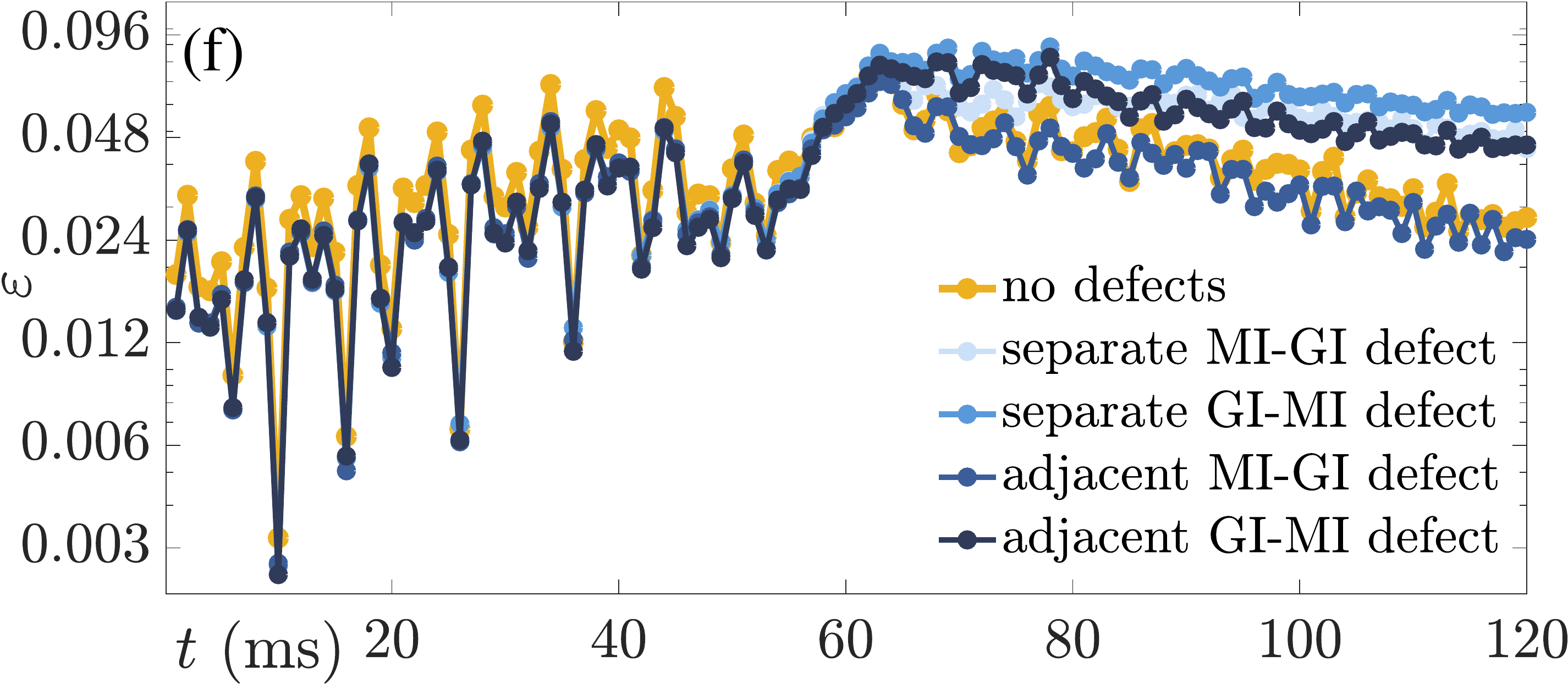}}\quad
	\caption{(Color online). The effect of various defects in the initial-state preparation on dynamics of the total dynamics-induced gauge-invariance violation given in Eq.~\eqref{eq:TotErr}. In the left column, we consider only matter-field defects, which include configurations of (a) matter holes (MH), (b) matter impurities (MI), or (c) various combinations of both. The right column displays results in the presence of (d) gauge-impurity (GI) defects as well as for the case when GI coexist with (e) MH or (f) MI defects.
	Matter-field defects induce no significant additional gauge violations during the dynamics. Although gauge impurities can increase the gauge violation, the effect remains bounded (does not continue increasing after crossing the critical point) for the simulated scenario.
	Interestingly, when a matter-field defect is placed closely to the left of a gauge impurity, this can block the rise of gauge violation due to the latter.
	(Note that the occurrence of double defects in the experiment of Ref.~\cite{Yang2020} is quite rare.) 	
	}
\label{fig:FigTotalError}
\end{figure*}

\begin{figure}[!ht]
	\centering
	\hspace{-.01 cm}
	\includegraphics[width=.48\textwidth]{{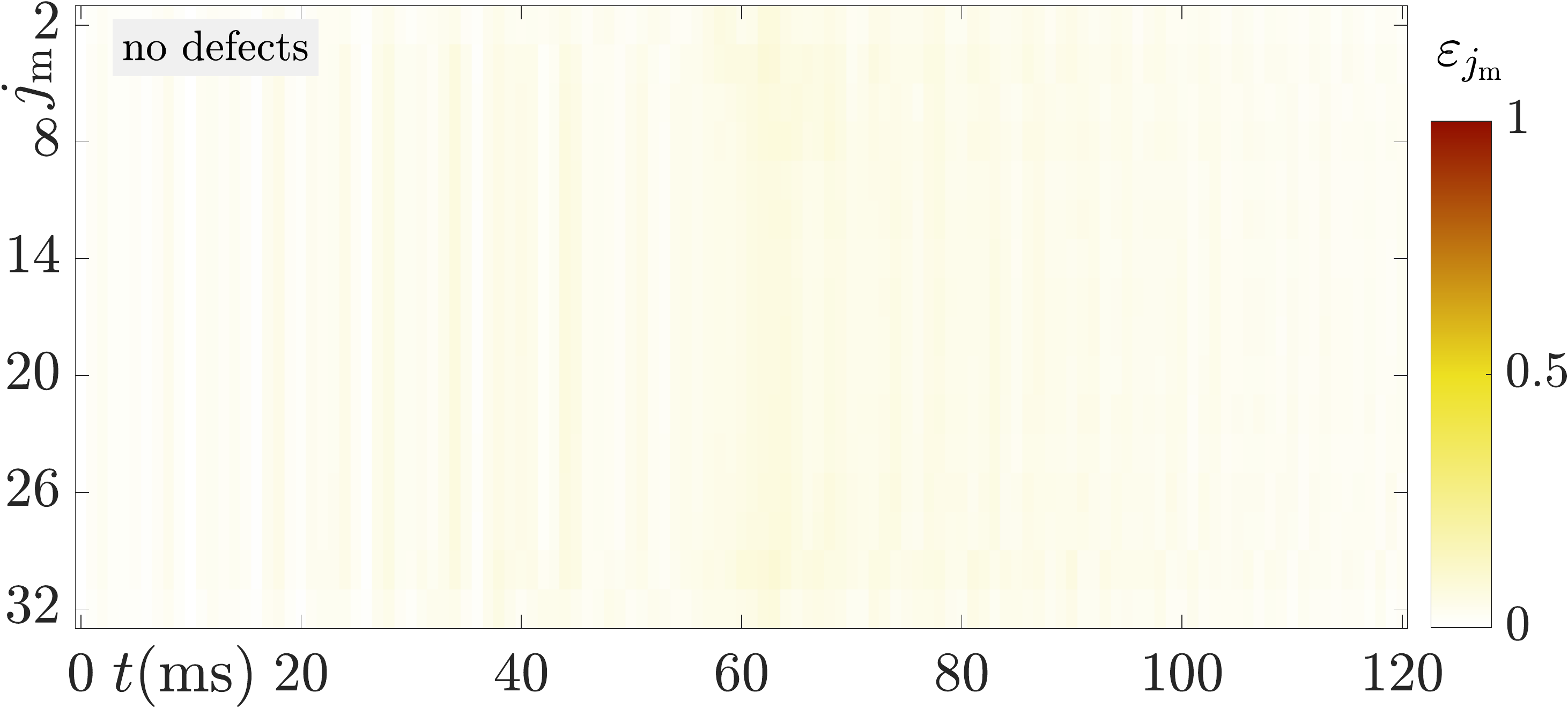}}\\
	\includegraphics[width=.48\textwidth]{{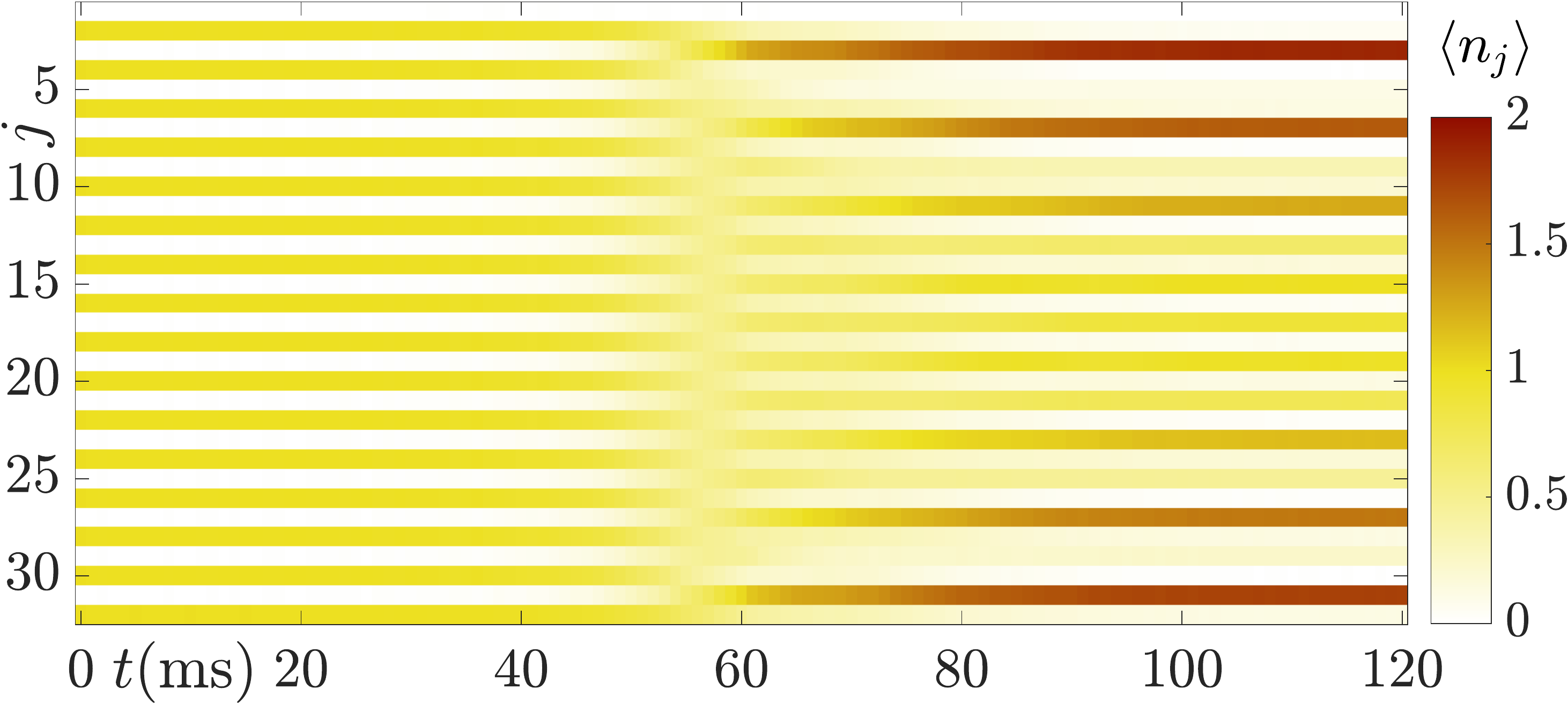}}\\
	\includegraphics[width=.48\textwidth]{{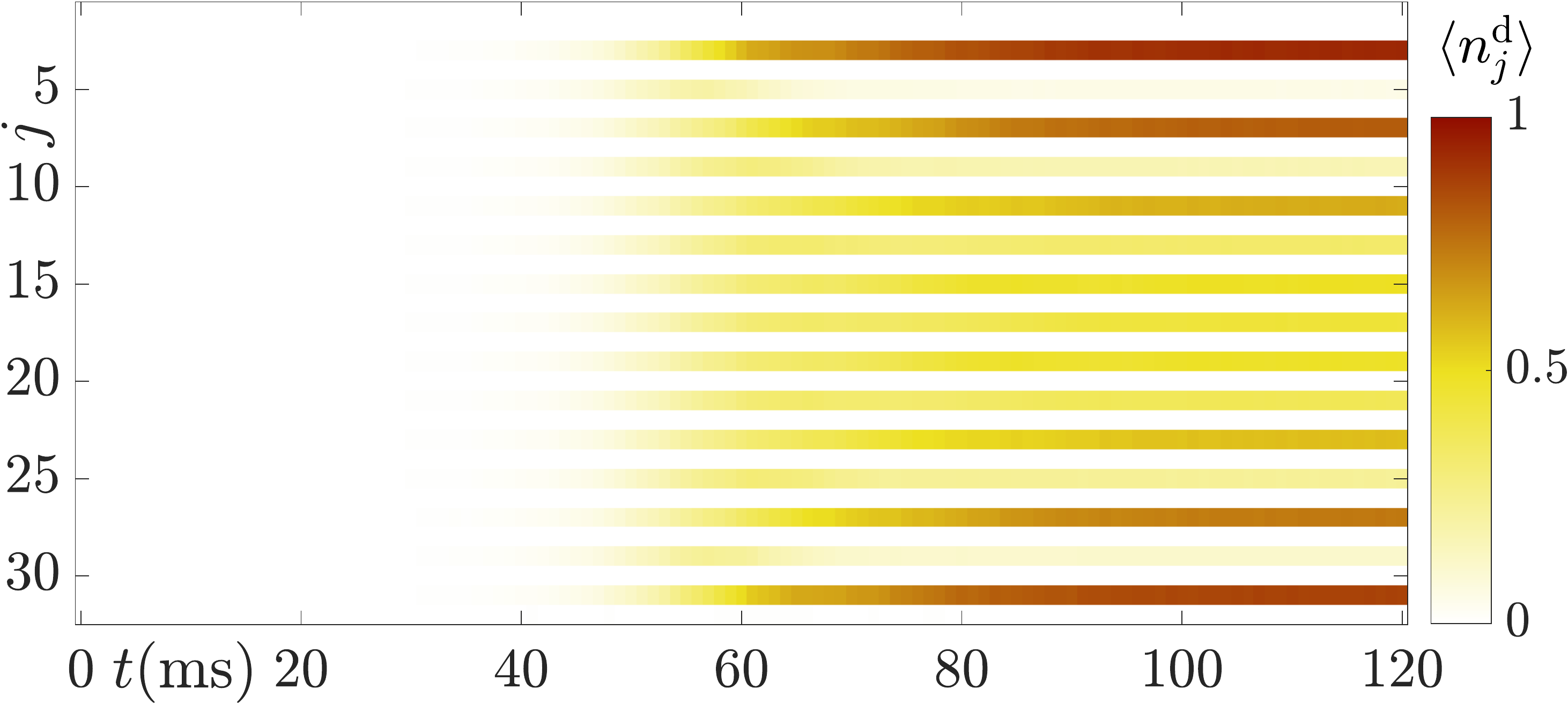}}
	\caption{(Color online). Site-resolved dynamics in the absence of defects. Top: The gauge-invariance violation is very reliable at all local constraints (defined at even sites, which represent matter fields), reaching a maximum as the ramp crosses the critical point at around $t\approx 60$ ms. See associated video \cite{HalimehChannel}. Middle: The particle number starts with a single boson on every even site. After the critical point, the particle number tends to diminish on even sites and increase on every other odd site ($j=3,7,11,\ldots$) especially at the edges. In the middle of the chain, there seems to be a competition between this configuration and its counterpart $j=1,5,9,\ldots$, since odd sites around $j=L/2$ all have nonvanishing particle numbers. This behavior is indicative of the spontaneous breaking of the $\mathrm{Z}_2$ symmetry characterizing Coleman's phase transition.
	Bottom: This picture is confirmed by the doublon occupation, which approaches unity at every other odd site near the edges, where the configuration of doublons is that of a right-pointing electric field in the QLM picture, but spreads more evenly over all odd sites in the middle of the chain, indicating a GHZ superposition state between the right- and left-pointing electric-field configurations (see Fig.~\ref{fig:mapping}, bottom left).}
	\label{fig:FigClean}
\end{figure}

\begin{figure*}[!ht]
	\centering
	\hspace{-.01 cm}
	\includegraphics[width=.48\textwidth]{{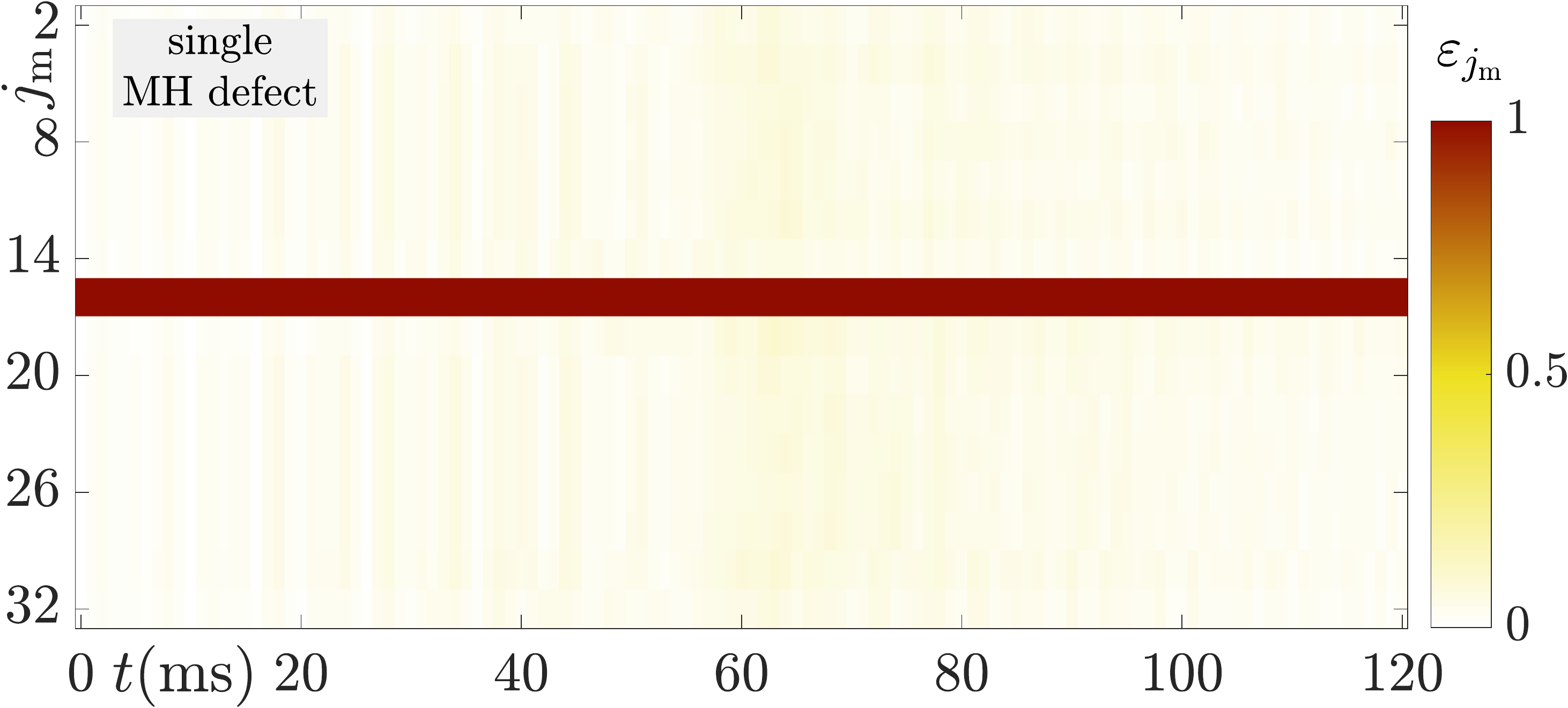}}\quad
	\includegraphics[width=.48\textwidth]{{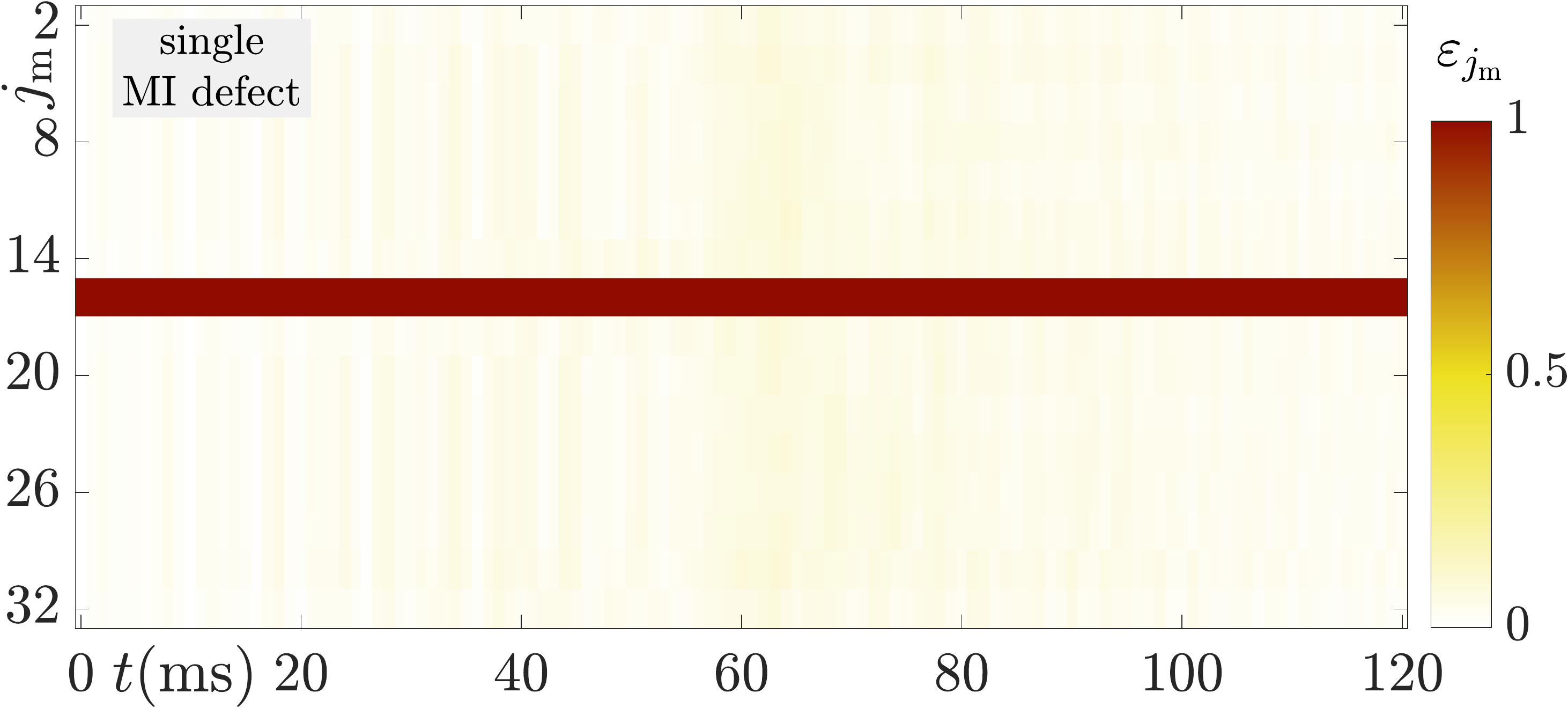}}\quad\\
	\hspace{-.01 cm}
	\includegraphics[width=.48\textwidth]{{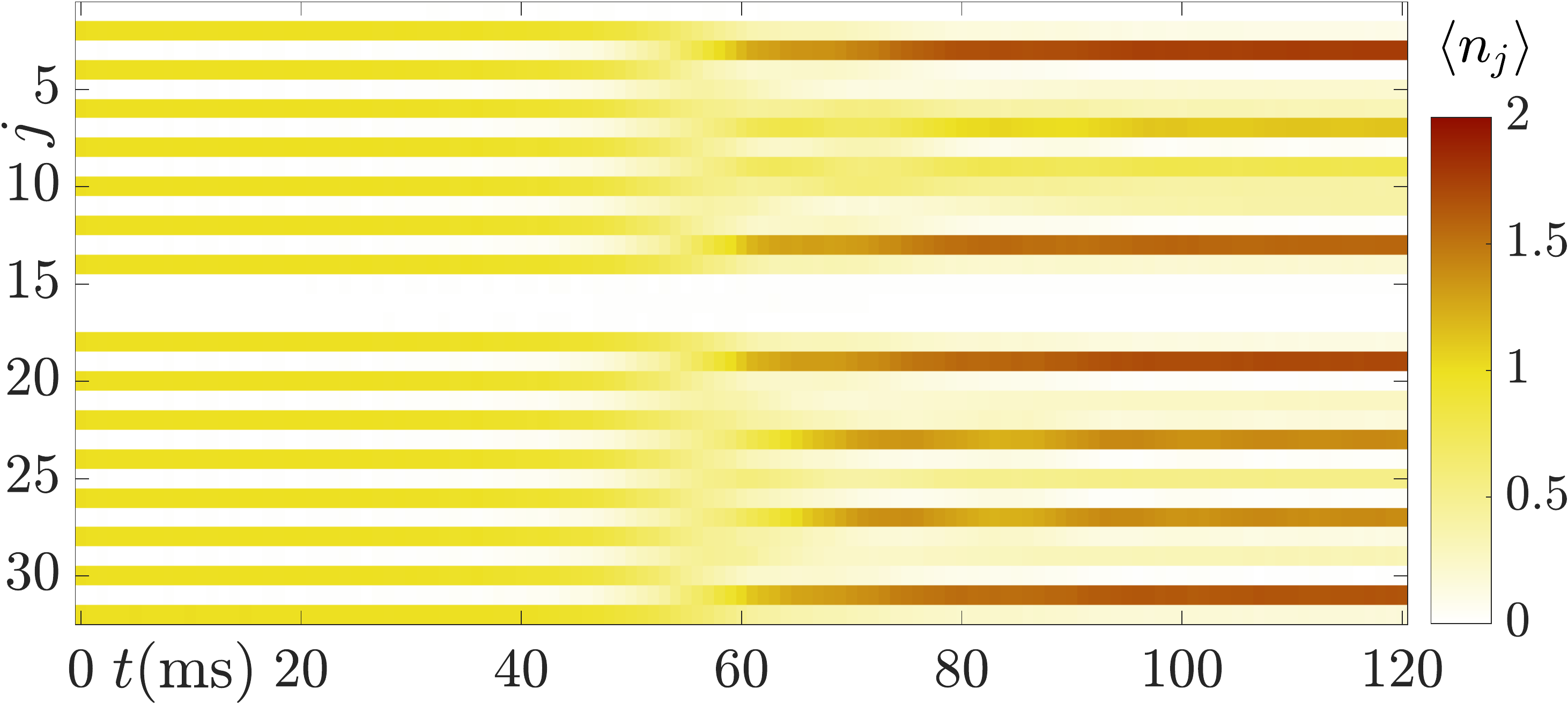}}\quad
	\includegraphics[width=.48\textwidth]{{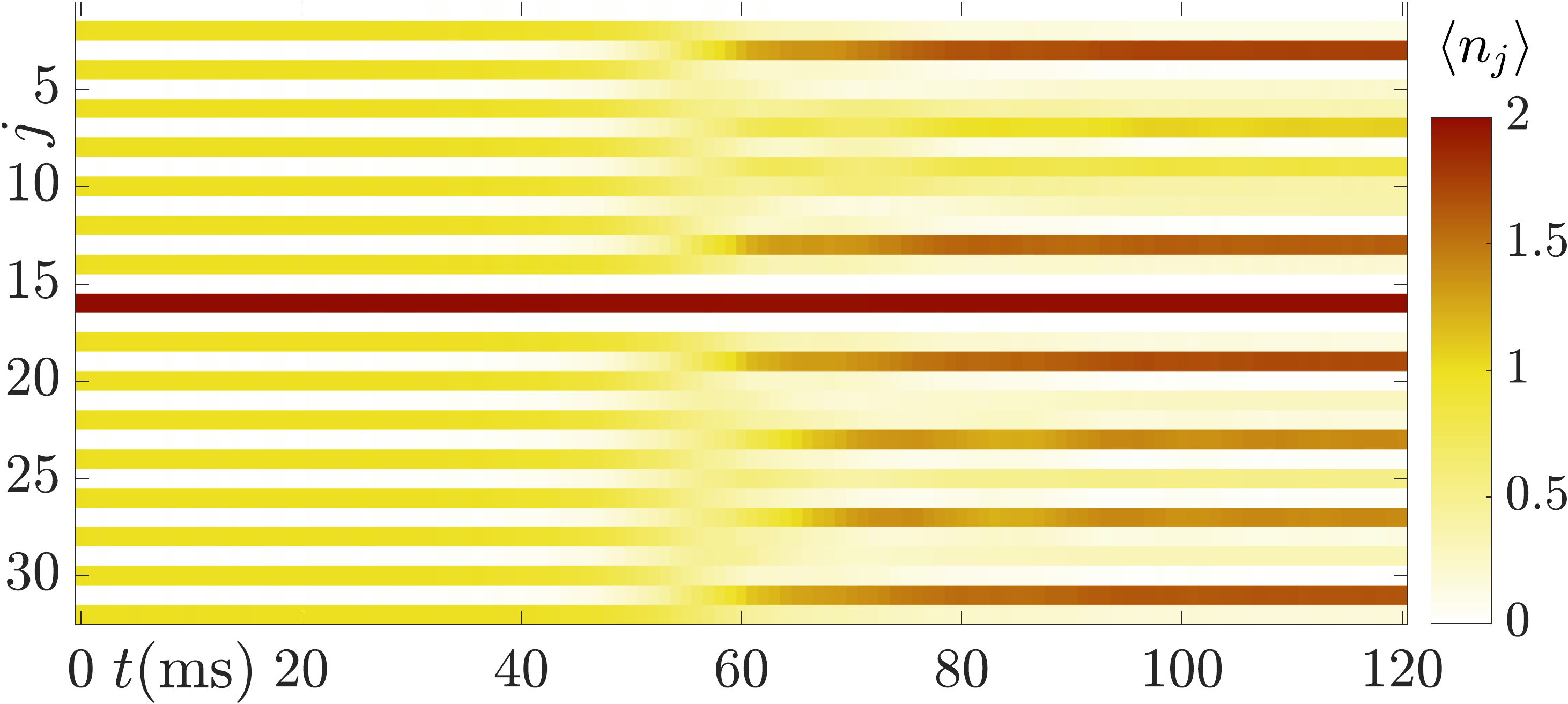}}\quad\\
	\hspace{-.01 cm}
	\includegraphics[width=.48\textwidth]{{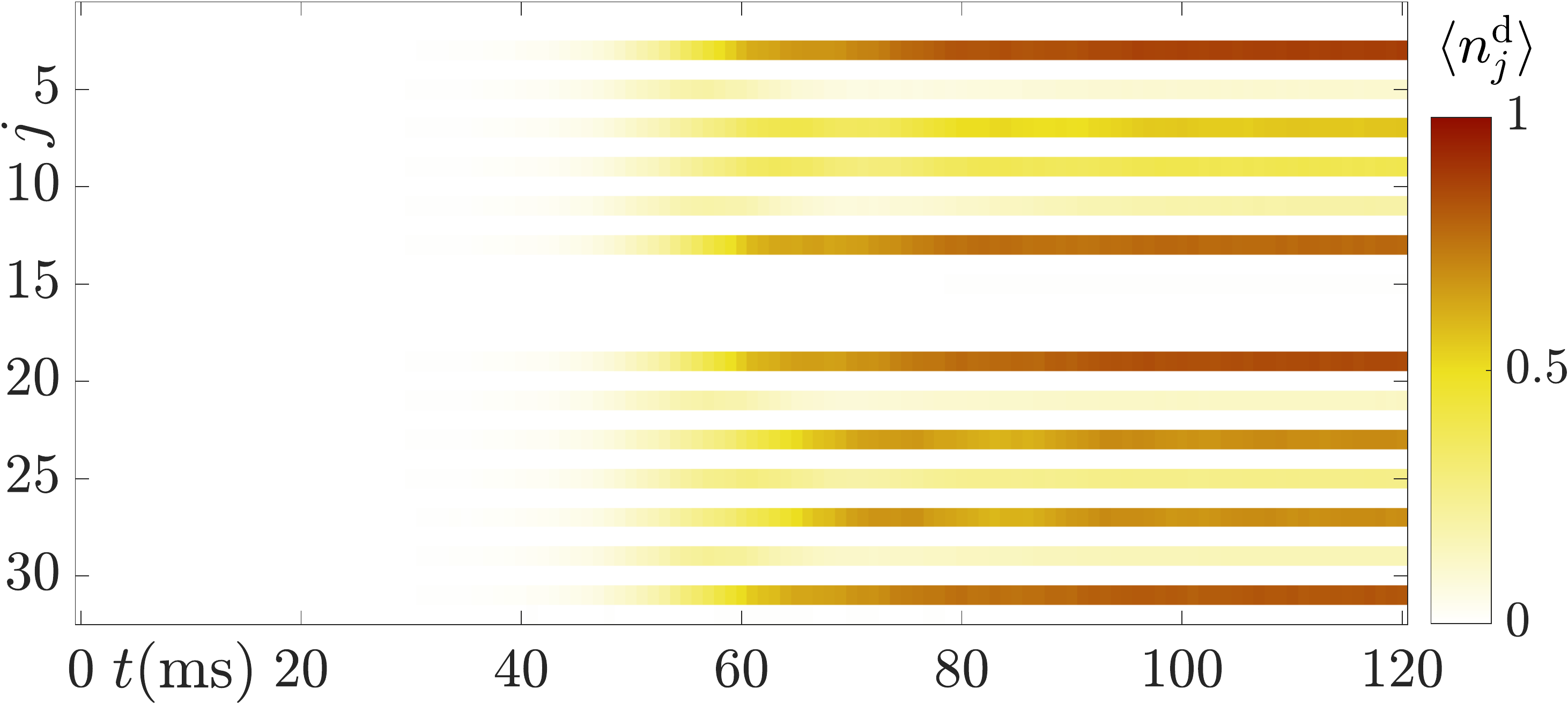}}\quad
	\includegraphics[width=.48\textwidth]{{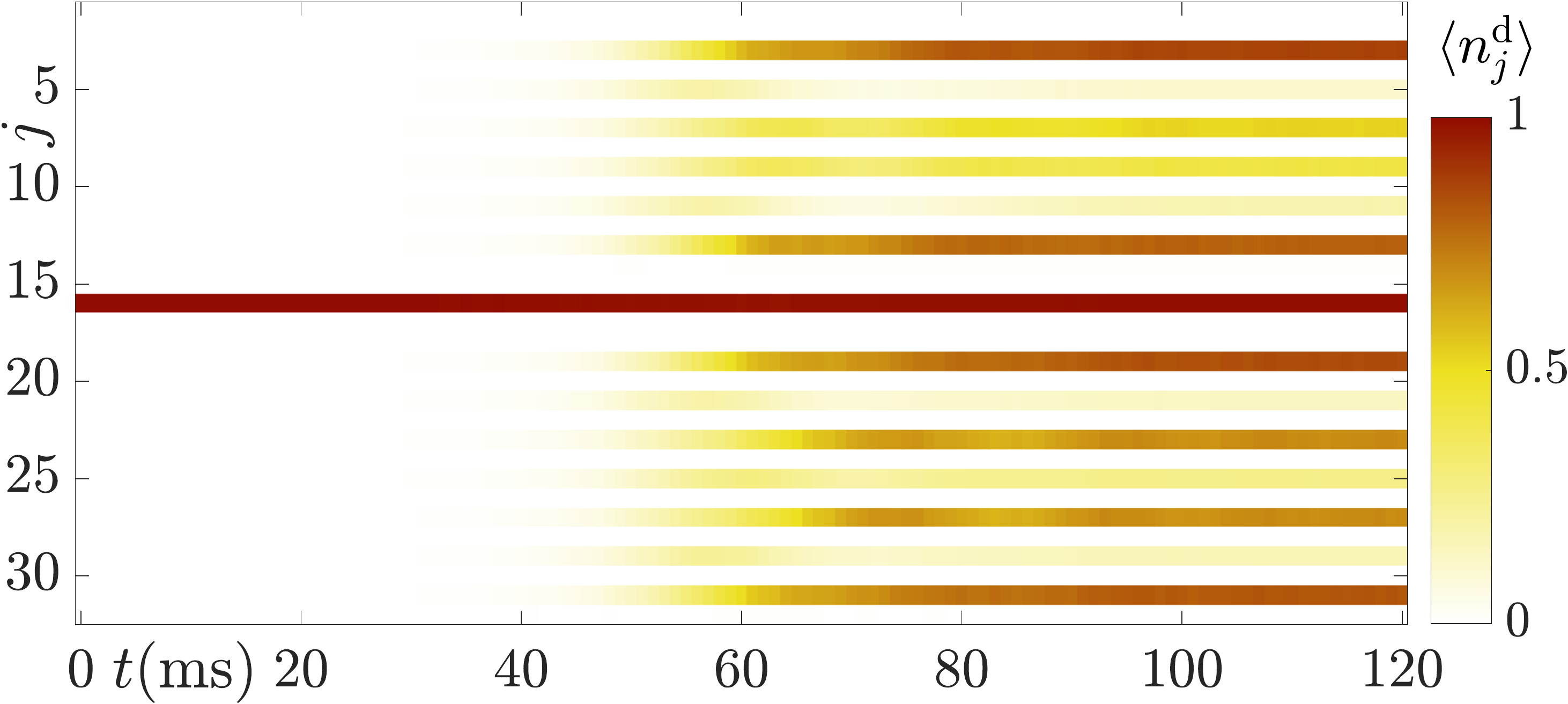}}\quad
	\caption{(Color online). Site-resolved dynamics in the presence of a single matter-hole (left column) or matter-impurity (right column) defect at site $j=16$. Both of the MH and MI defects give rise to a very localized gauge-invariance violation, which effectively shows no proliferation over all evolution times. See associated video \cite{HalimehChannel}.
	As can be seen in the particle-number dynamics, the defective atom occupation remains strongly localized, whereby it acts like a hard edge. At late times, this effective edge leads the system to settle into the right-pointing electric-field configuration in the middle of the chain whereas in the clean case it is a superposition of that configuration and its left-pointing electric-field counterpart. This picture is reflected in the doublon-number dynamics. (In the upper half of the chain, the right-pointing electric-field configuration is not clear since that side of the chain has only seven bosons.)}
	\label{fig:FigMatterDefects}
\end{figure*}

\begin{figure}[!ht]
	\centering
	\includegraphics[width=.48\textwidth]{{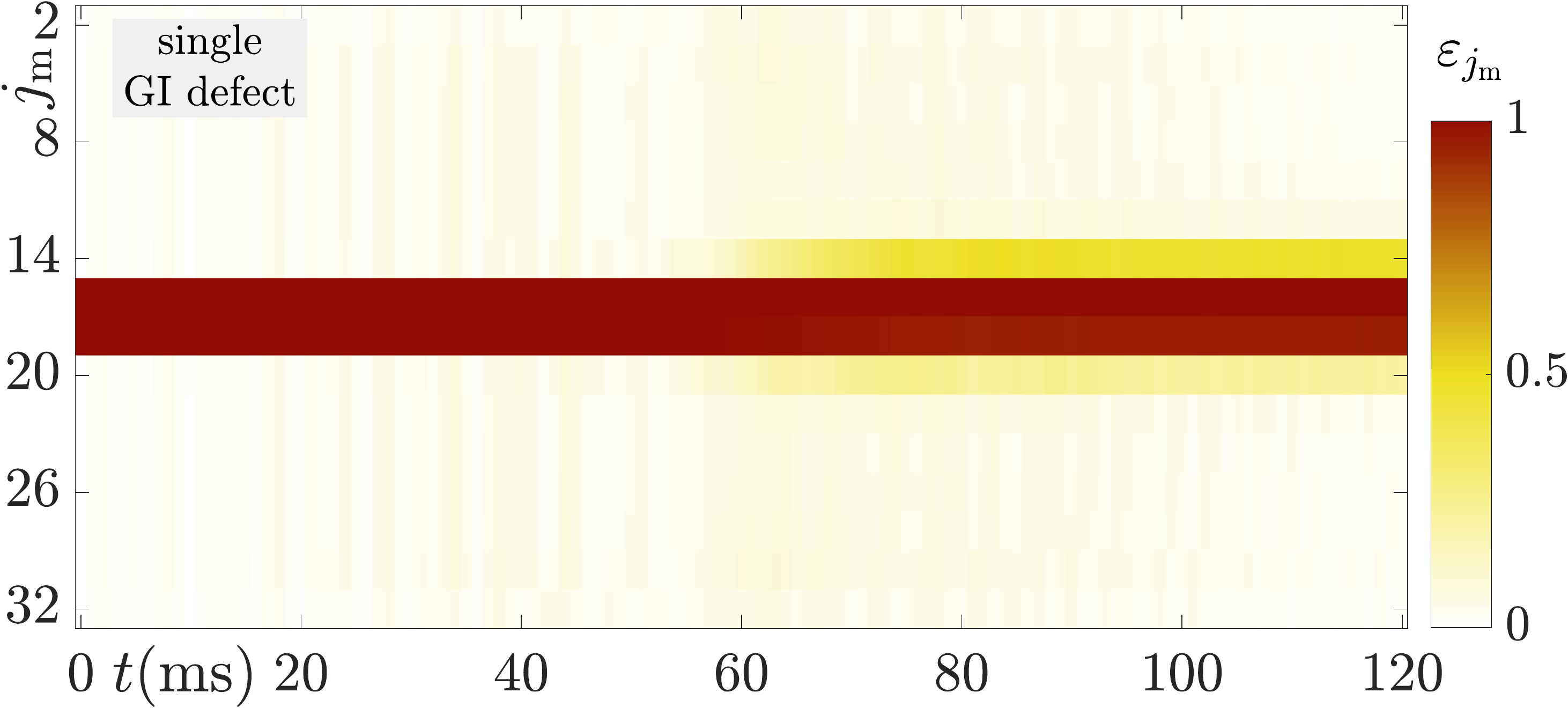}}\\
	\includegraphics[width=.48\textwidth]{{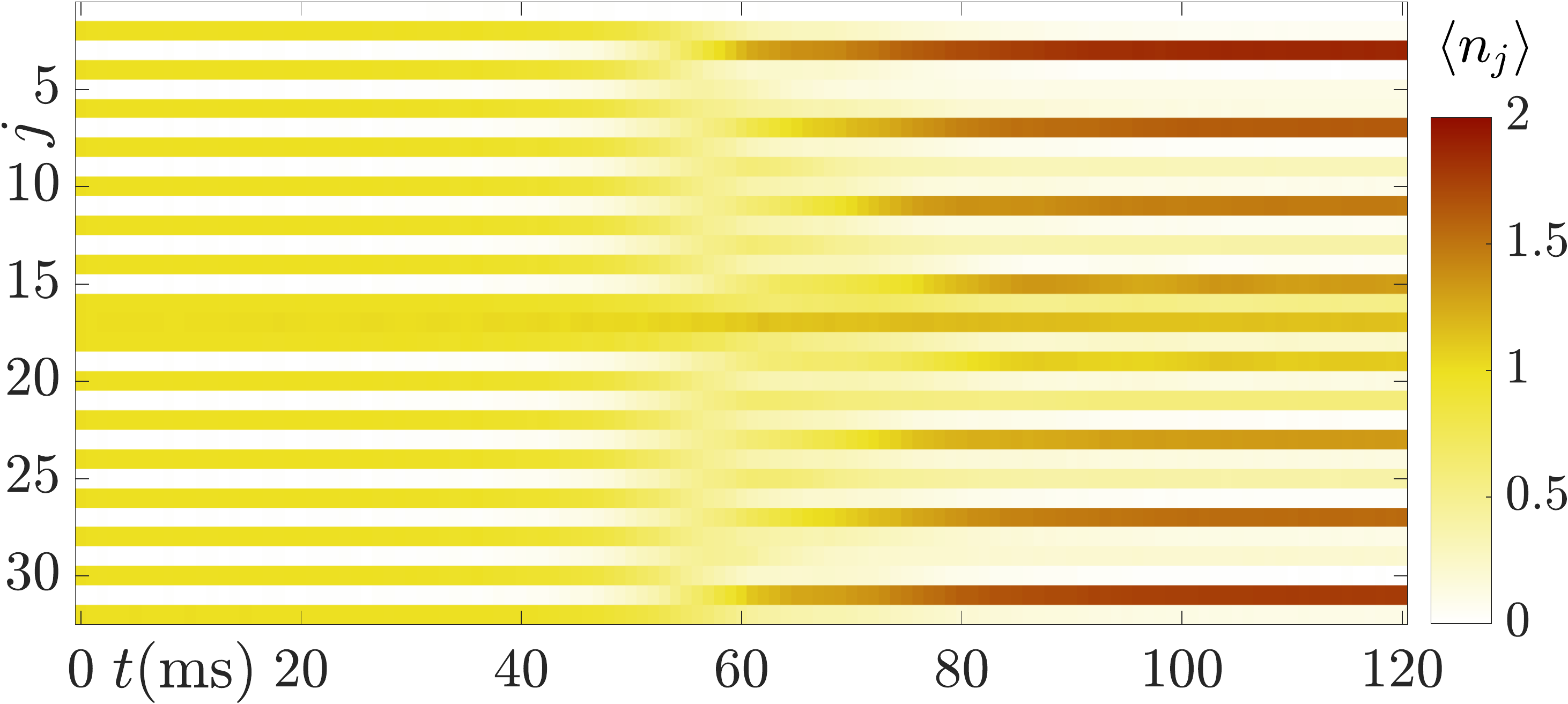}}\\
	\includegraphics[width=.48\textwidth]{{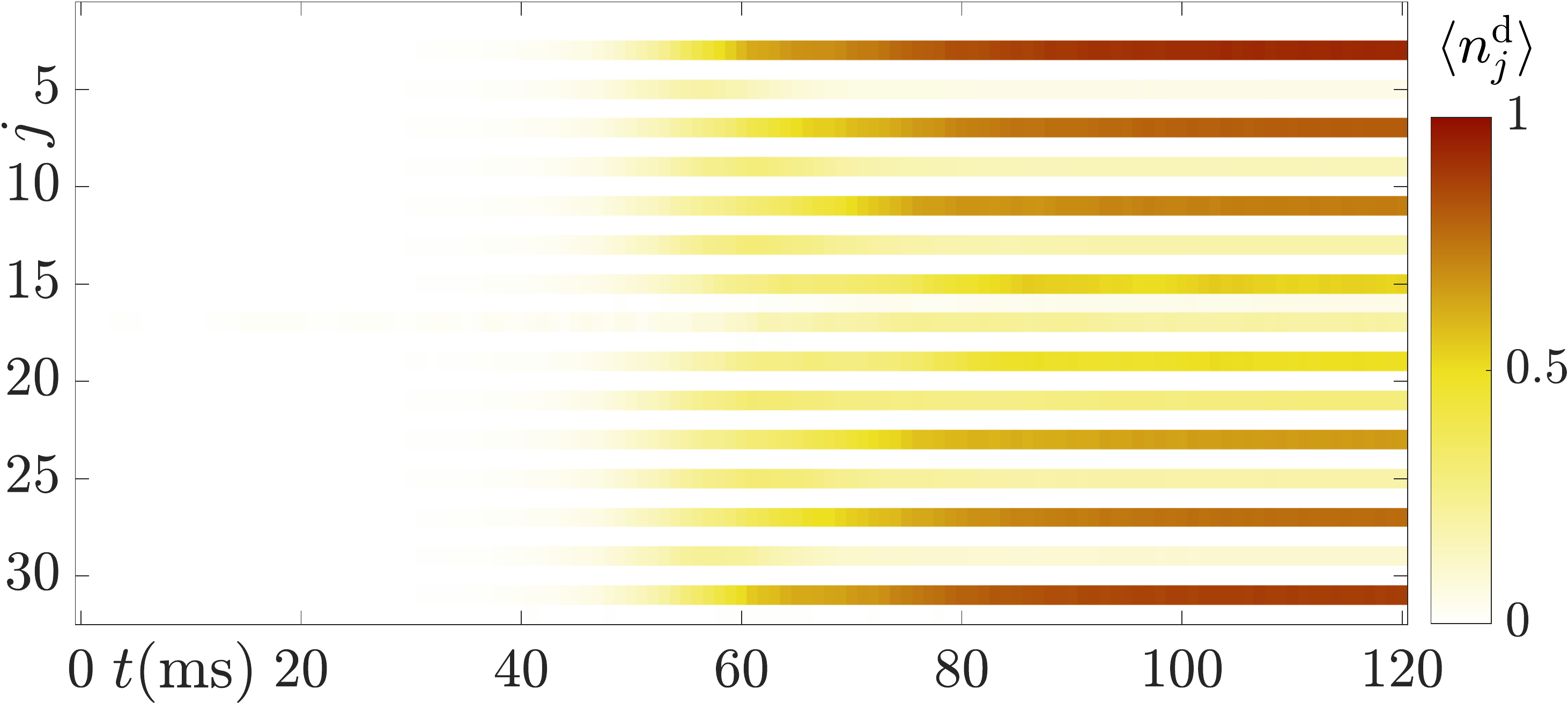}}
	\caption{(Color online). Site-resolved dynamics in the presence of a single gauge-impurity defect on site $j=17$, which breaks Gauss's law at the local constraints defined on the three-site unit cells centered at sites $j=16$ and $17$. Unlike the case of matter-field defects in Fig.~\ref{fig:FigMatterDefects}, the violation here slightly spreads after crossing the critical point (See associated video \cite{HalimehChannel}). Nevertheless, the GI defect also serves in part as an edge that tends to configure the center of the chain in the right-pointing electric-field state shown in the bottom left of Fig.~\ref{fig:mapping}, which becomes more prominent here in the late-time dynamics of the particle and doublon occupations than in the clean case of Fig.~\ref{fig:FigClean}.}
	\label{fig:FigGI}
\end{figure}

\begin{figure*}[!ht]
	\centering
	\hspace{-.01 cm}
	\includegraphics[width=.48\textwidth]{{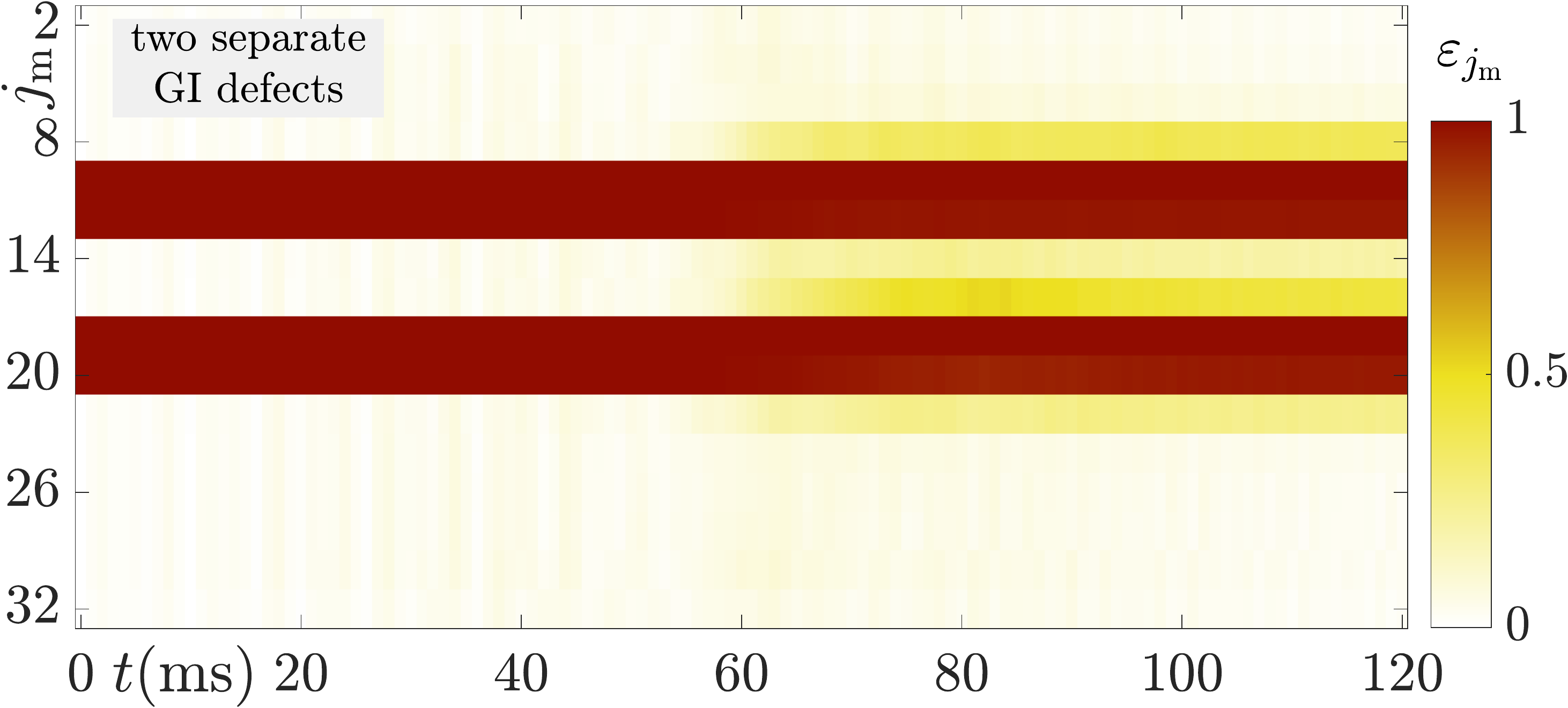}}\quad
	\includegraphics[width=.48\textwidth]{{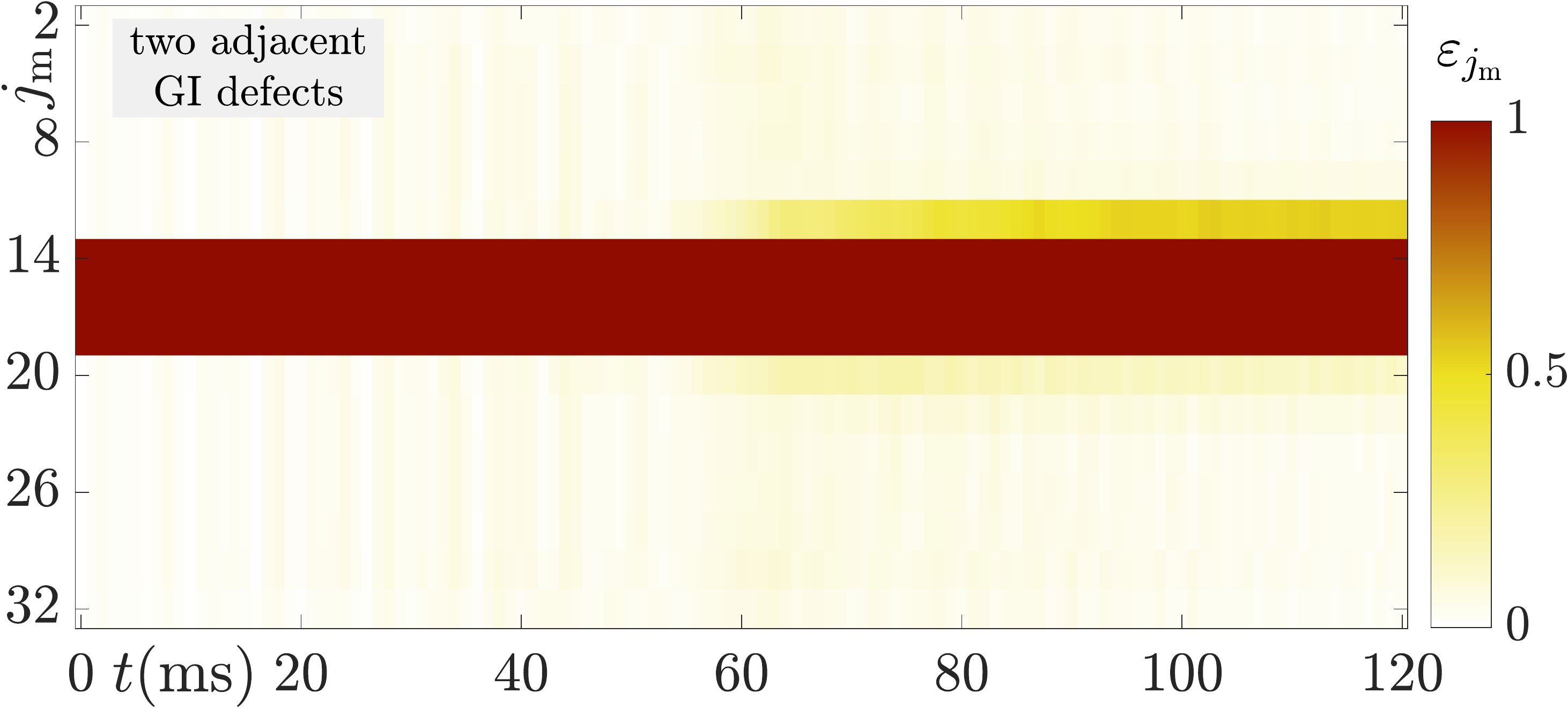}}\quad\\
	\hspace{-.01 cm}
	\includegraphics[width=.48\textwidth]{{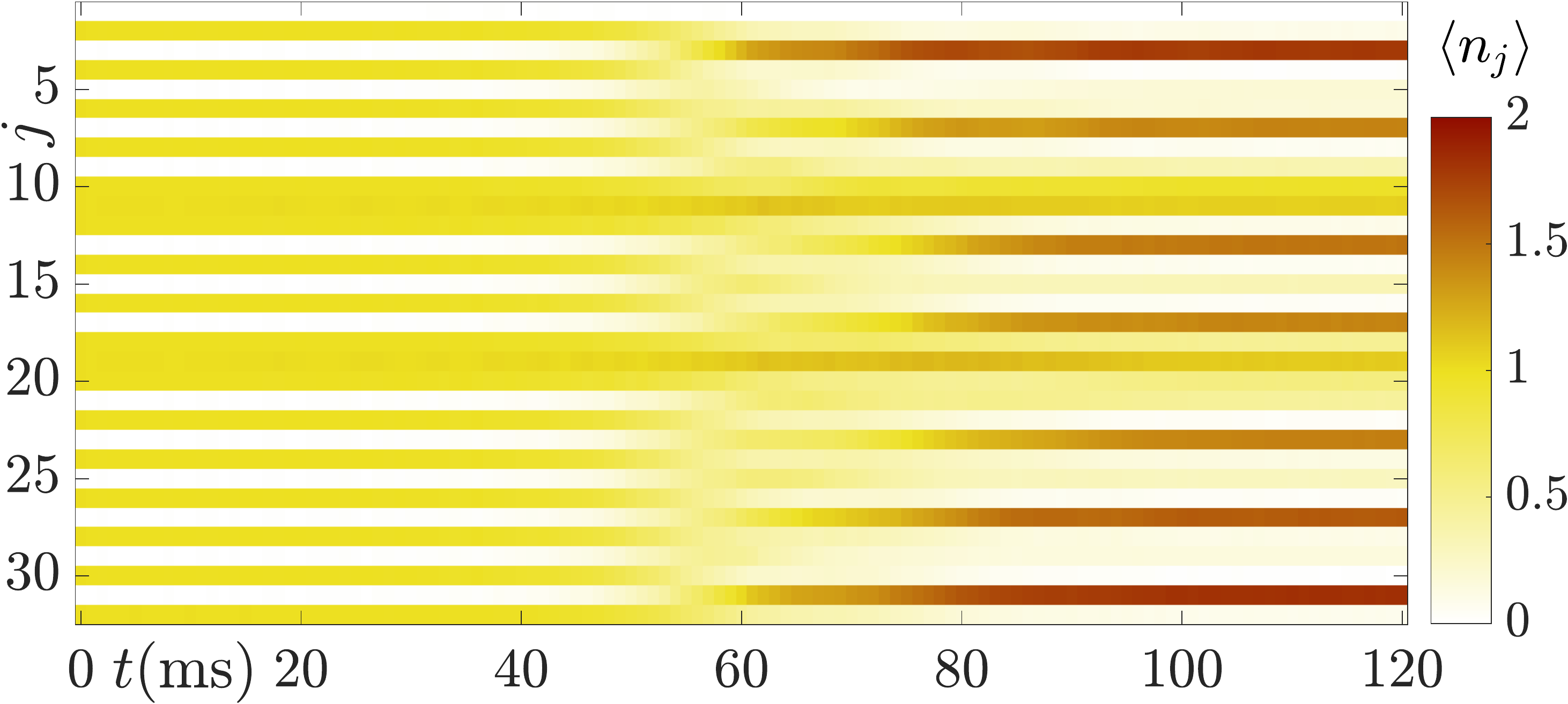}}\quad
	\includegraphics[width=.48\textwidth]{{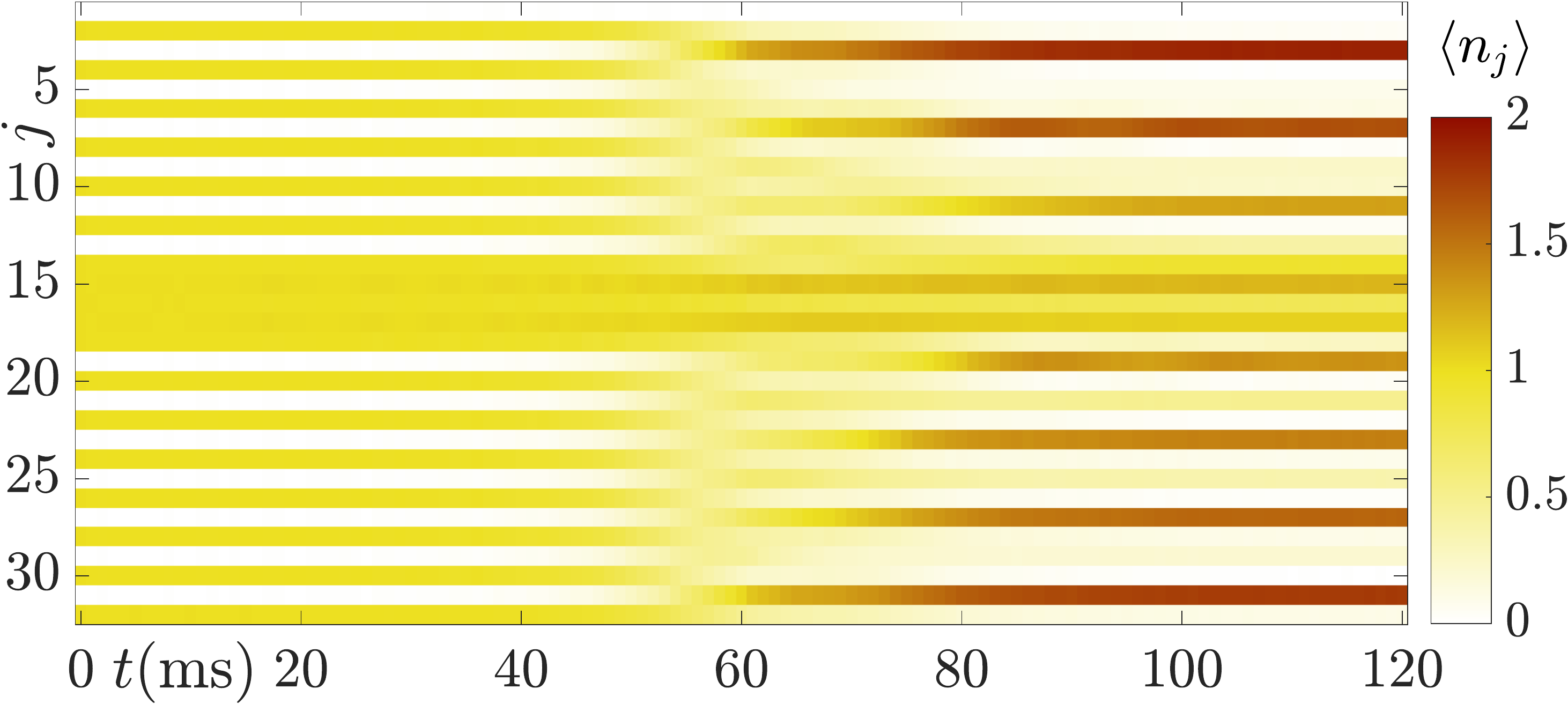}}\quad\\
	\hspace{-.01 cm}
	\includegraphics[width=.48\textwidth]{{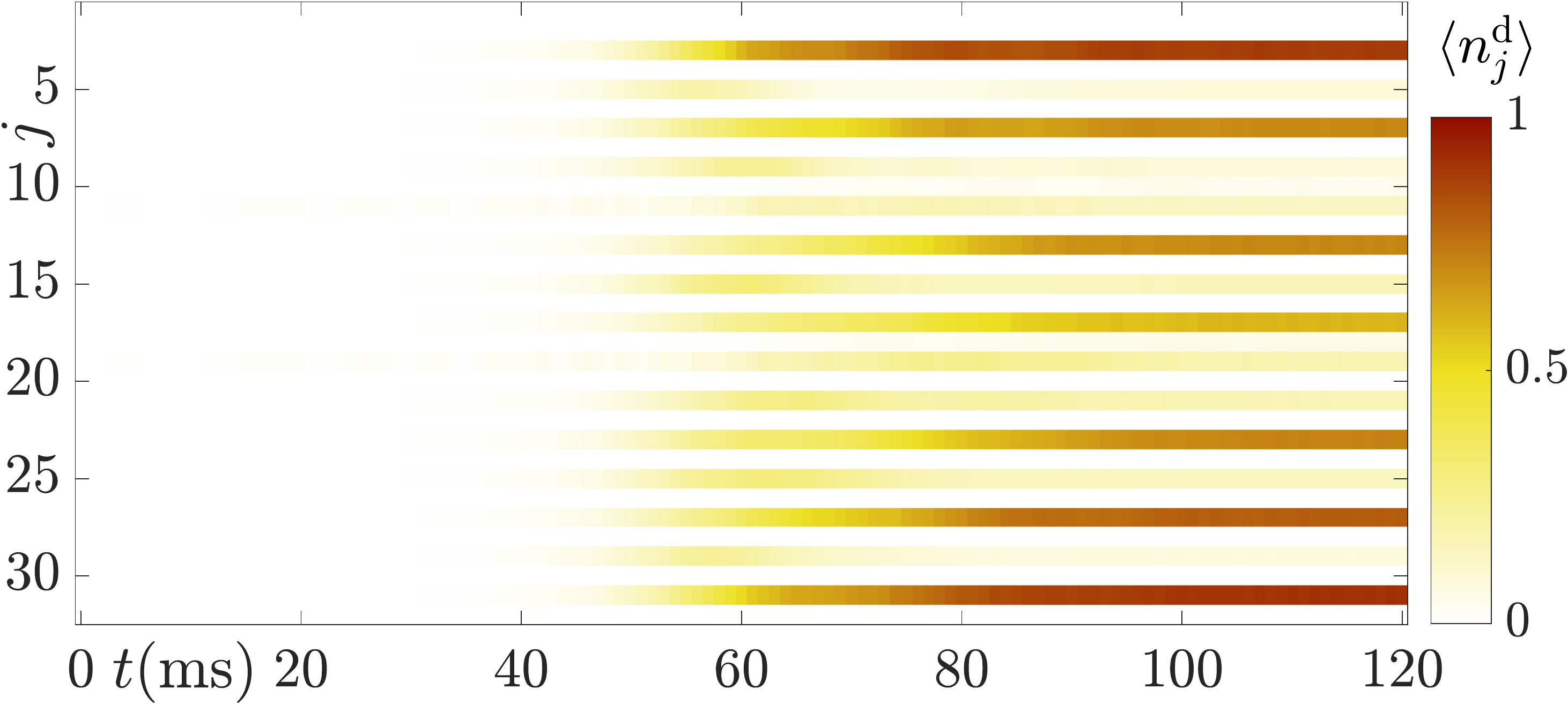}}\quad
	\includegraphics[width=.48\textwidth]{{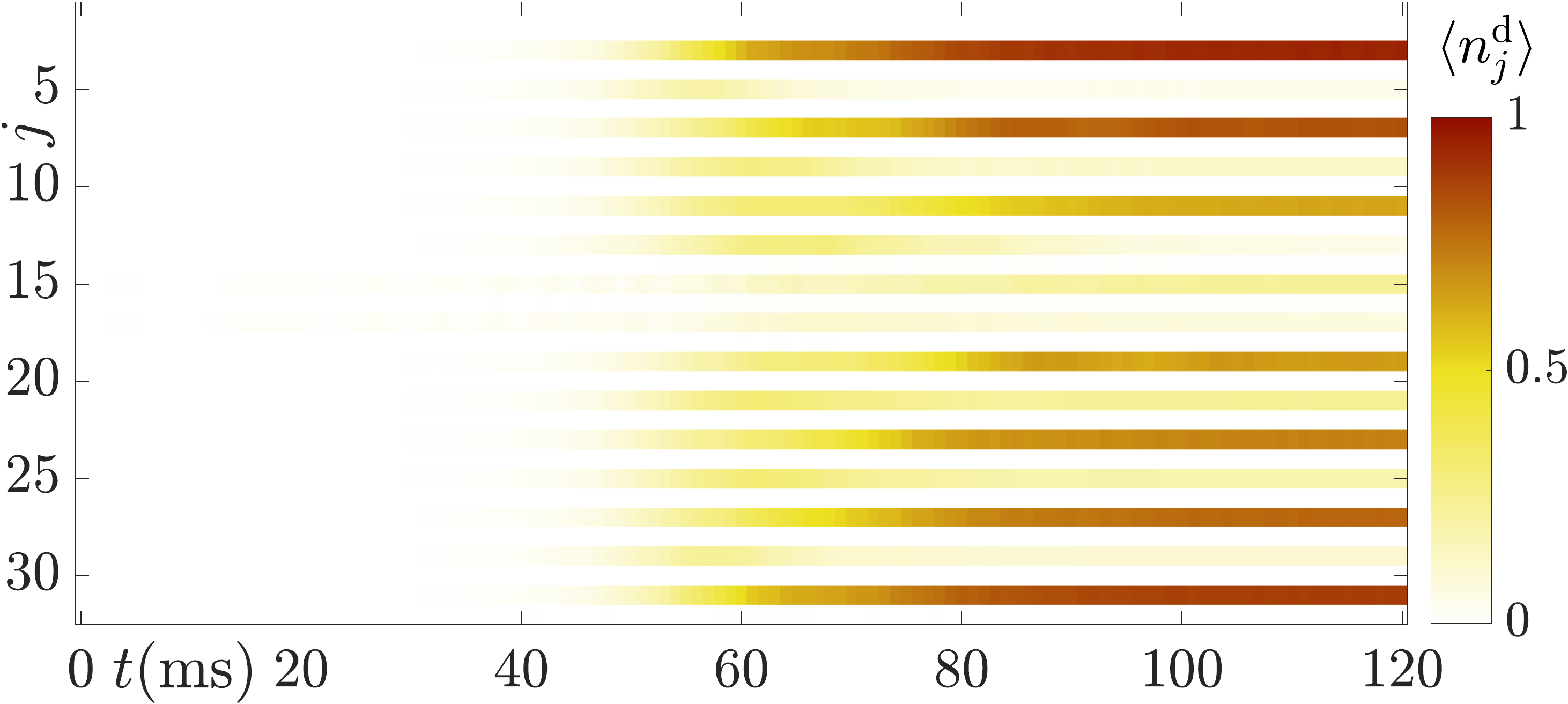}}\quad
	\caption{(Color online). Site-resolved dynamics in the presence of two separate (left column) gauge-impurity defects on sites $j=11,19$ and two adjacent (right column) gauge-impurity defects on sites $j=15,17$. When apart, the GI defects at late times lead to an overall gauge violation larger than that in the case of a single GI defect. In contrast, when the two defects are adjacent, the overall violation at large times is qualitatively similar to that of the single GI defect; cf.~Figs.~\ref{fig:FigTotalError}(d) and~\ref{fig:FigGI}, and see associated videos \cite{HalimehChannel}. The dynamics of the particle and doublon occupations share qualitatively similar behavior between the cases of adjacent GI defects and a single defect. In the case of separate GI defects, there is a subtle difference due to the specific location of the leftmost GI defect (see text).}
	\label{fig:FigMultipleGI}
\end{figure*}

\begin{figure*}[!ht]
	\centering
	\hspace{-.01 cm}
	\includegraphics[width=.48\textwidth]{{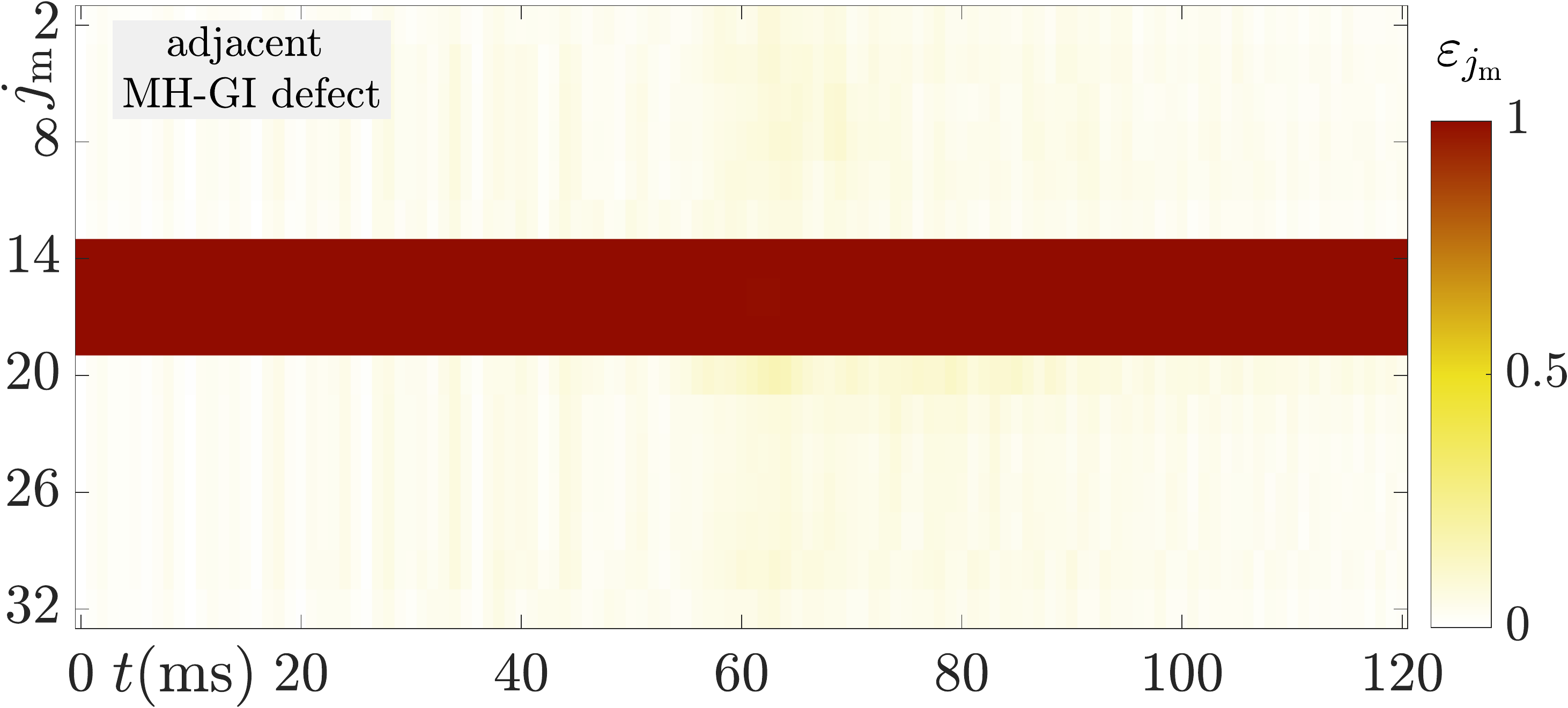}}\quad
	\includegraphics[width=.48\textwidth]{{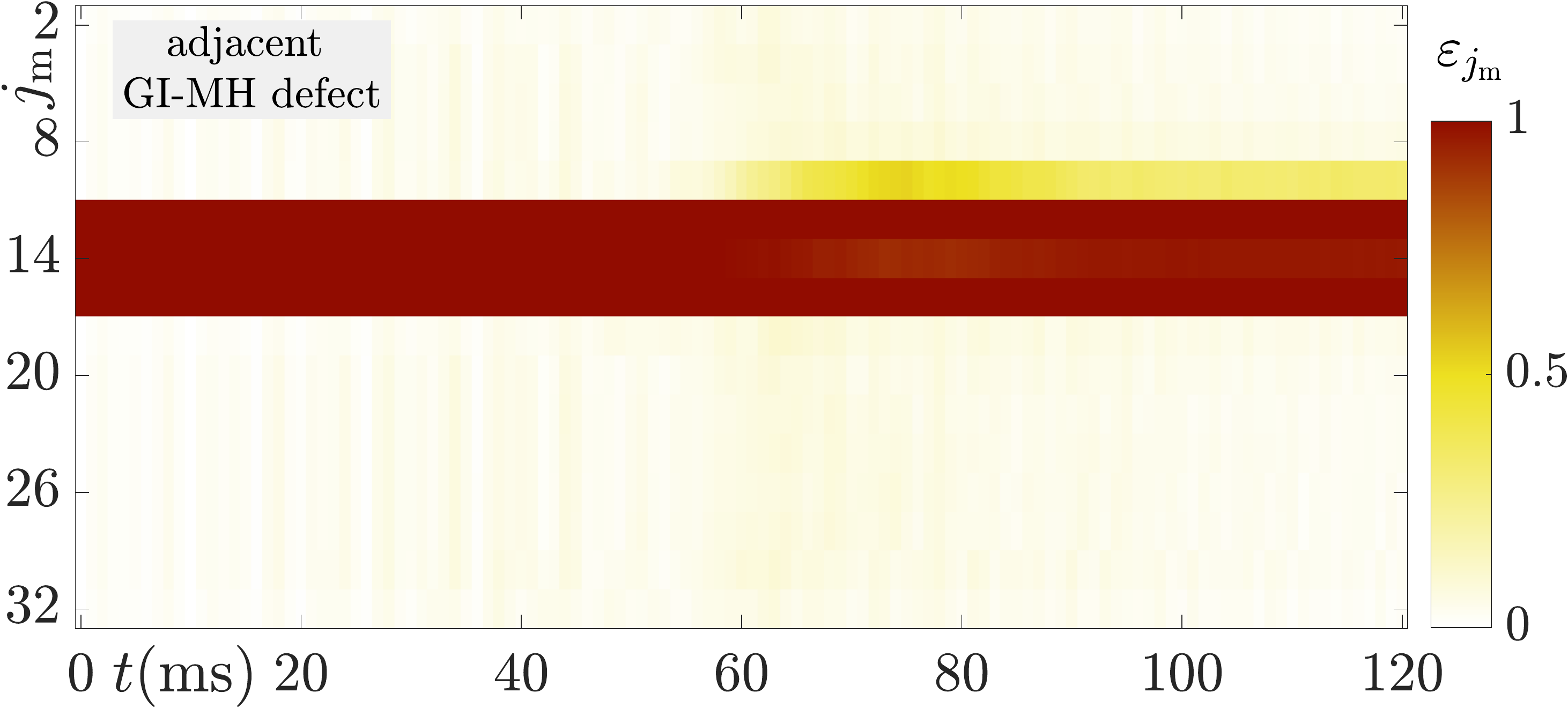}}\quad\\
	\hspace{-.01 cm}
	\includegraphics[width=.48\textwidth]{{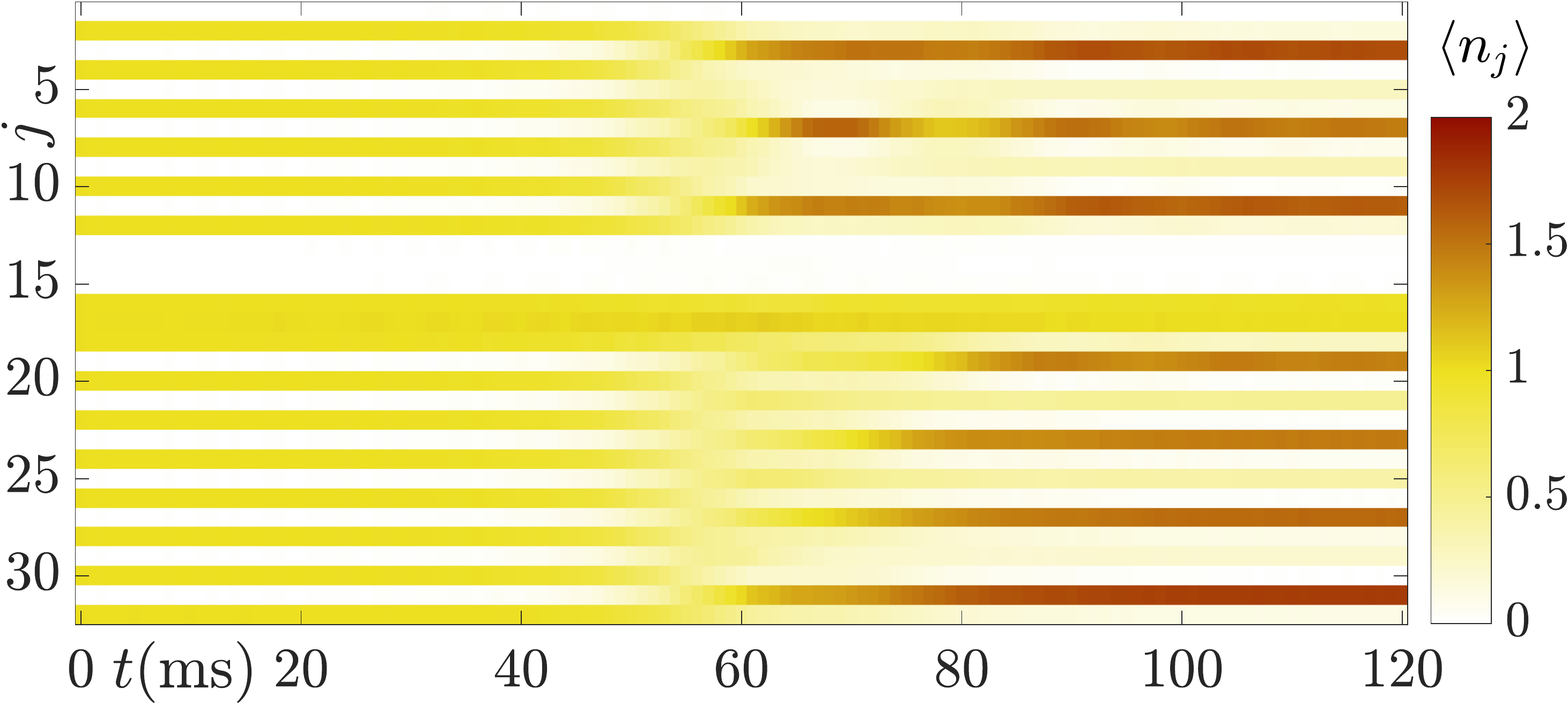}}\quad
	\includegraphics[width=.48\textwidth]{{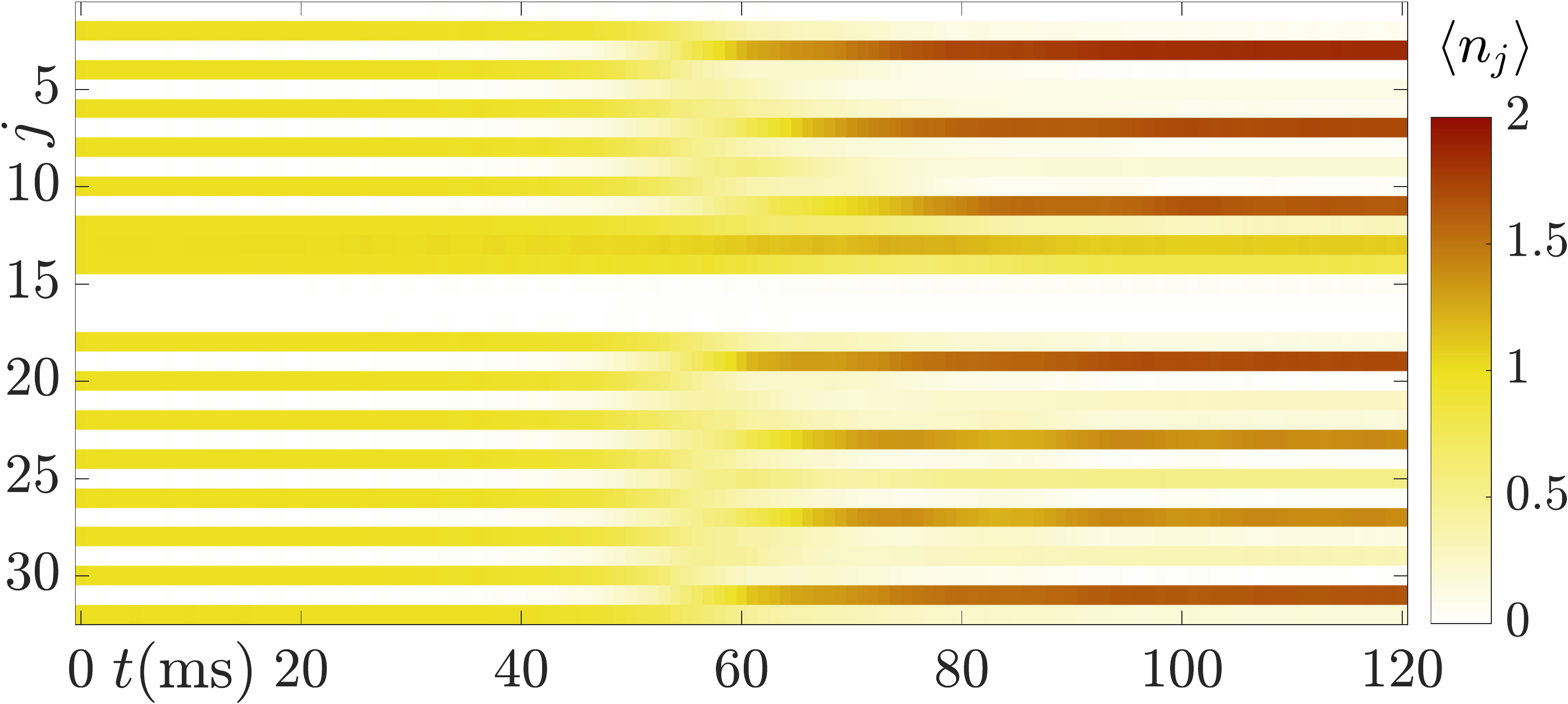}}\quad\\
	\hspace{-.01 cm}
	\includegraphics[width=.48\textwidth]{{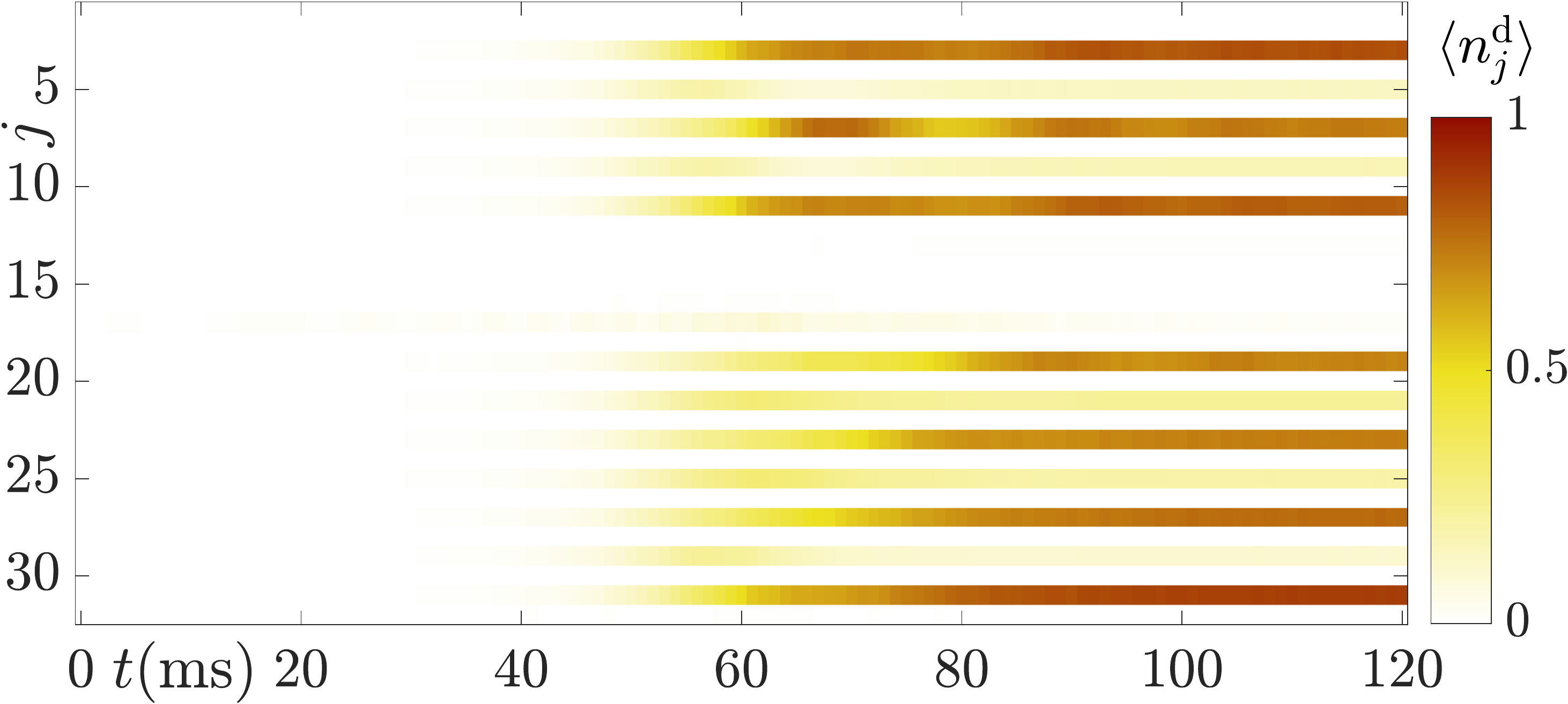}}\quad
	\includegraphics[width=.48\textwidth]{{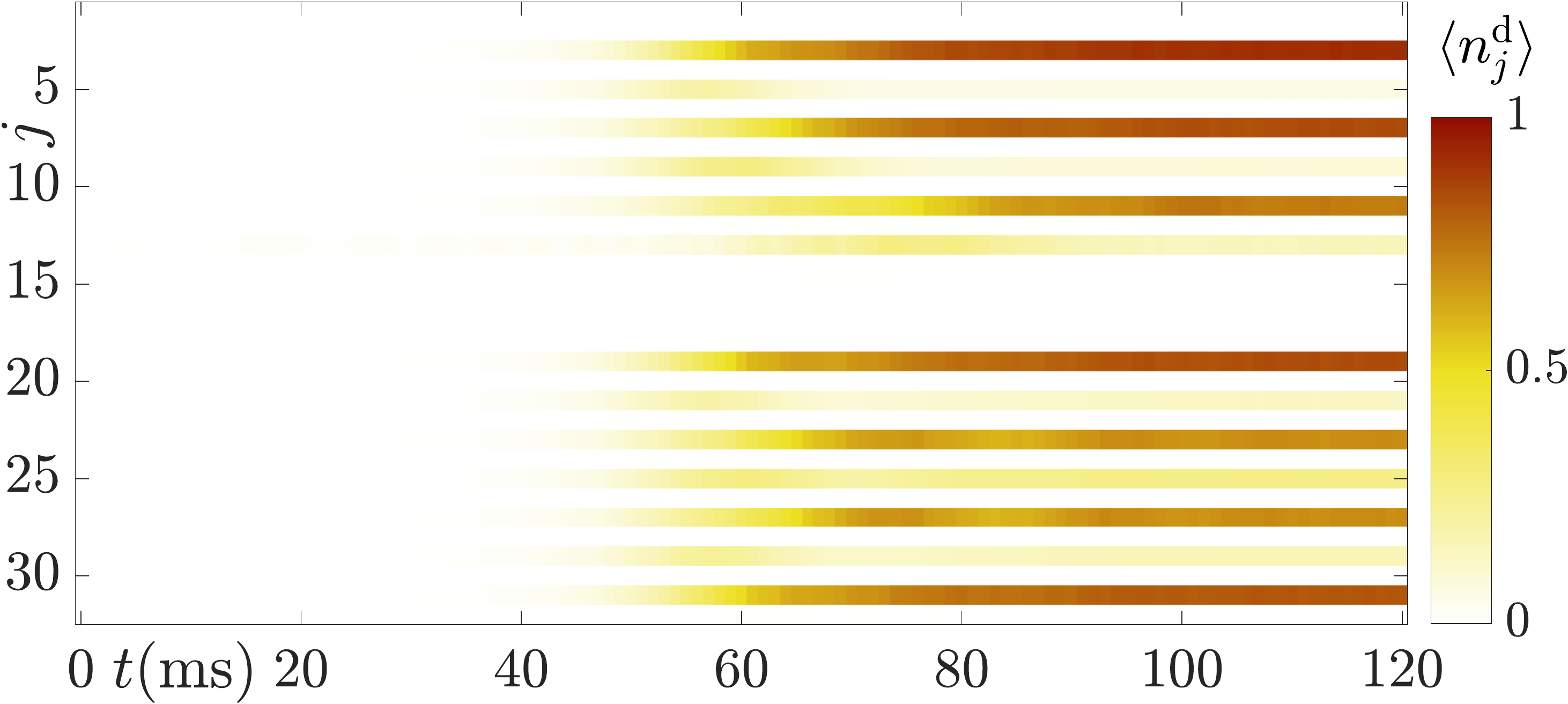}}\quad
	\caption{(Color online). A composite MH-GI defect on sites $j=14,17$ (left column) and GI-MH defect on sites $j=13,16$ (right column). In both cases the arrangement amounts to breaking Gauss's law at two adjacent local constraints. Due to the tilted lattice, we see that the gauge violation due to the GI defect is more suppressed when the MH defect is to its left. This result confirms the behavior of the spatially averaged total gauge violation displayed in Fig.~\ref{fig:FigTotalError}(e). For a different visualization of this site-resolved gauge-invariance dynamics, see the provided ``time-slice'' videos \cite{HalimehChannel}. In both cases, we see interesting dynamics in the particle and doublon occupations. The composite defect forms an edge with an even number of bosons in each subsystem at either side, leading to a clear right-pointing electric-field configuration at late times in both subsystems. The same qualitative behavior persists if the MH defect is replaced with an MI defect in the above calculations.}
	\label{fig:FigMHGI}
\end{figure*}

\begin{figure*}[!ht]
	\centering
	\hspace{-.01 cm}
	\includegraphics[width=.48\textwidth]{{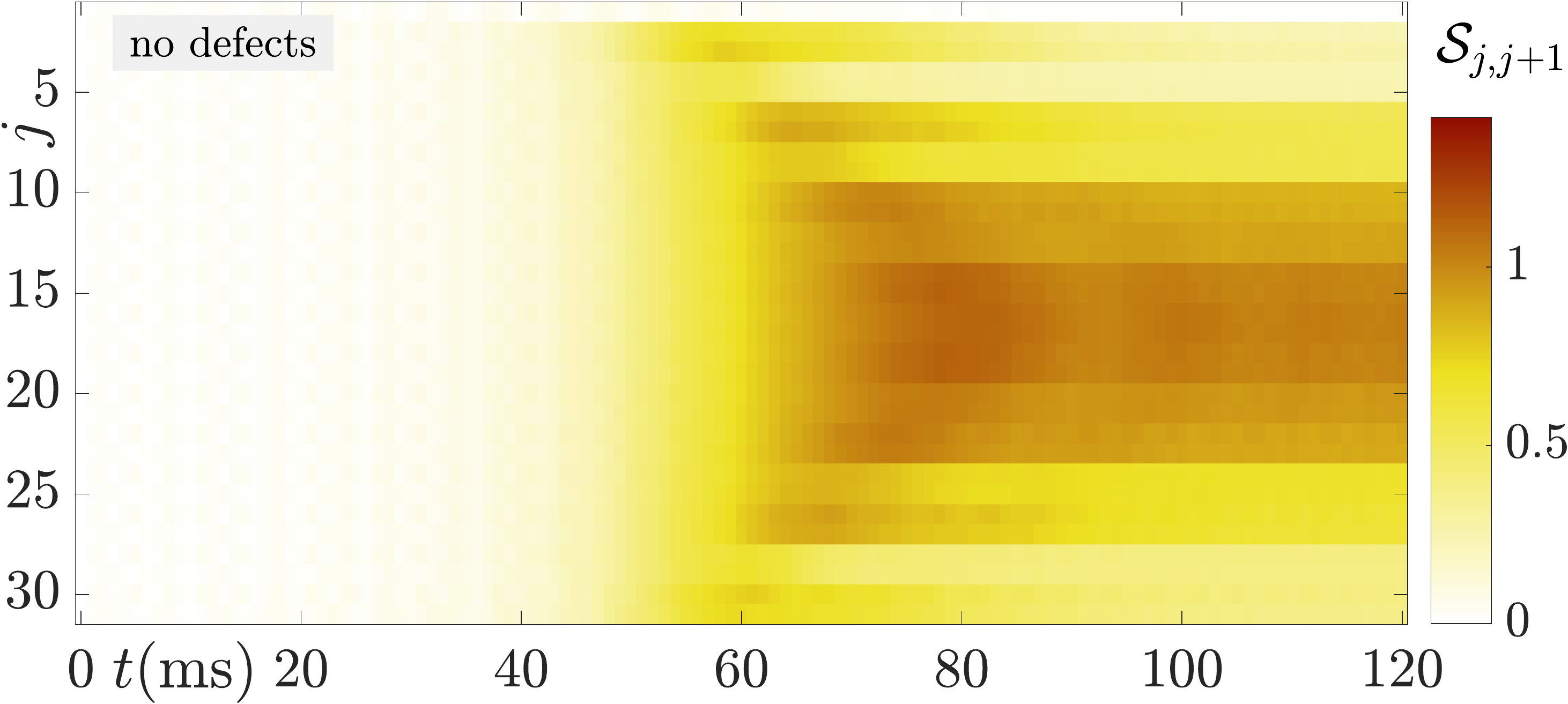}}\quad
	\includegraphics[width=.48\textwidth]{{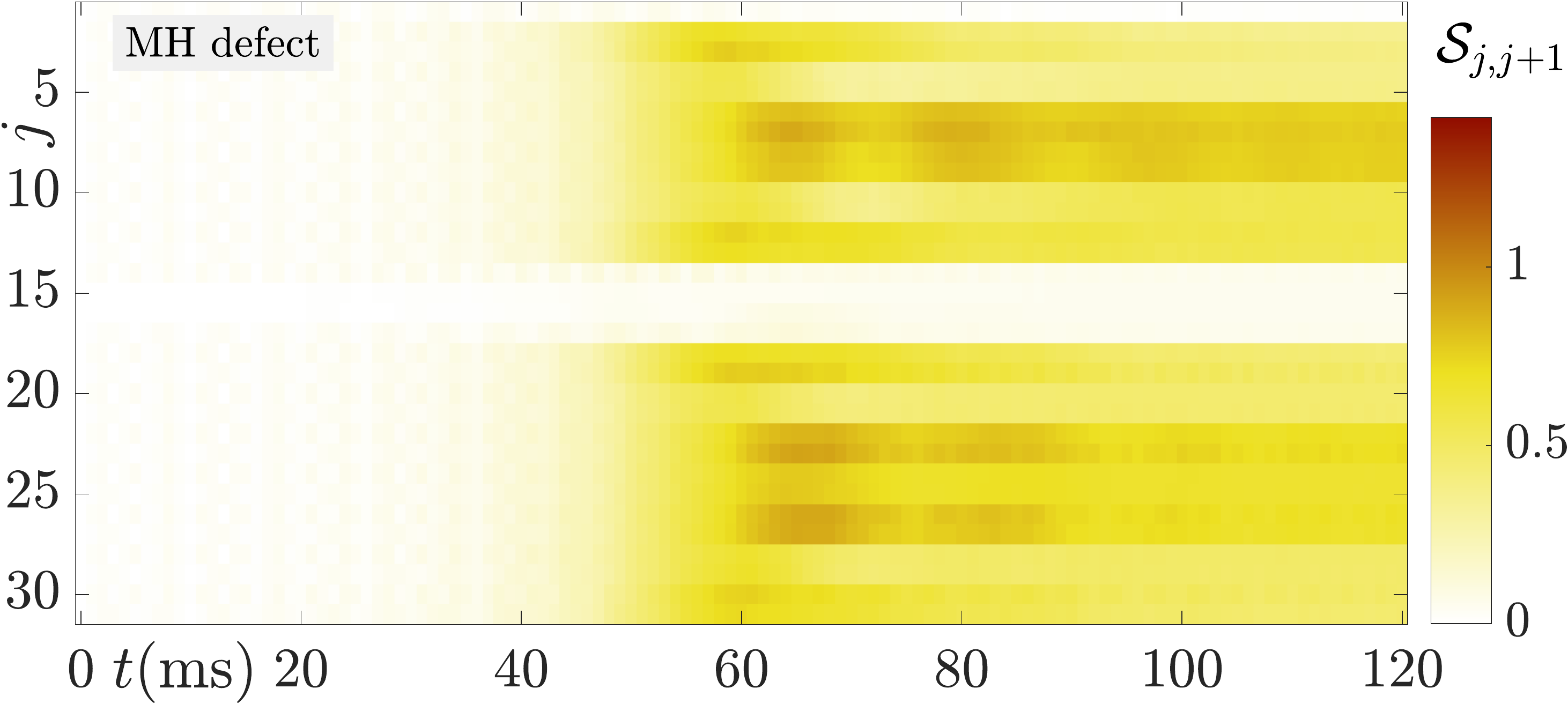}}\quad\\
	\hspace{-.01 cm}
	\includegraphics[width=.48\textwidth]{{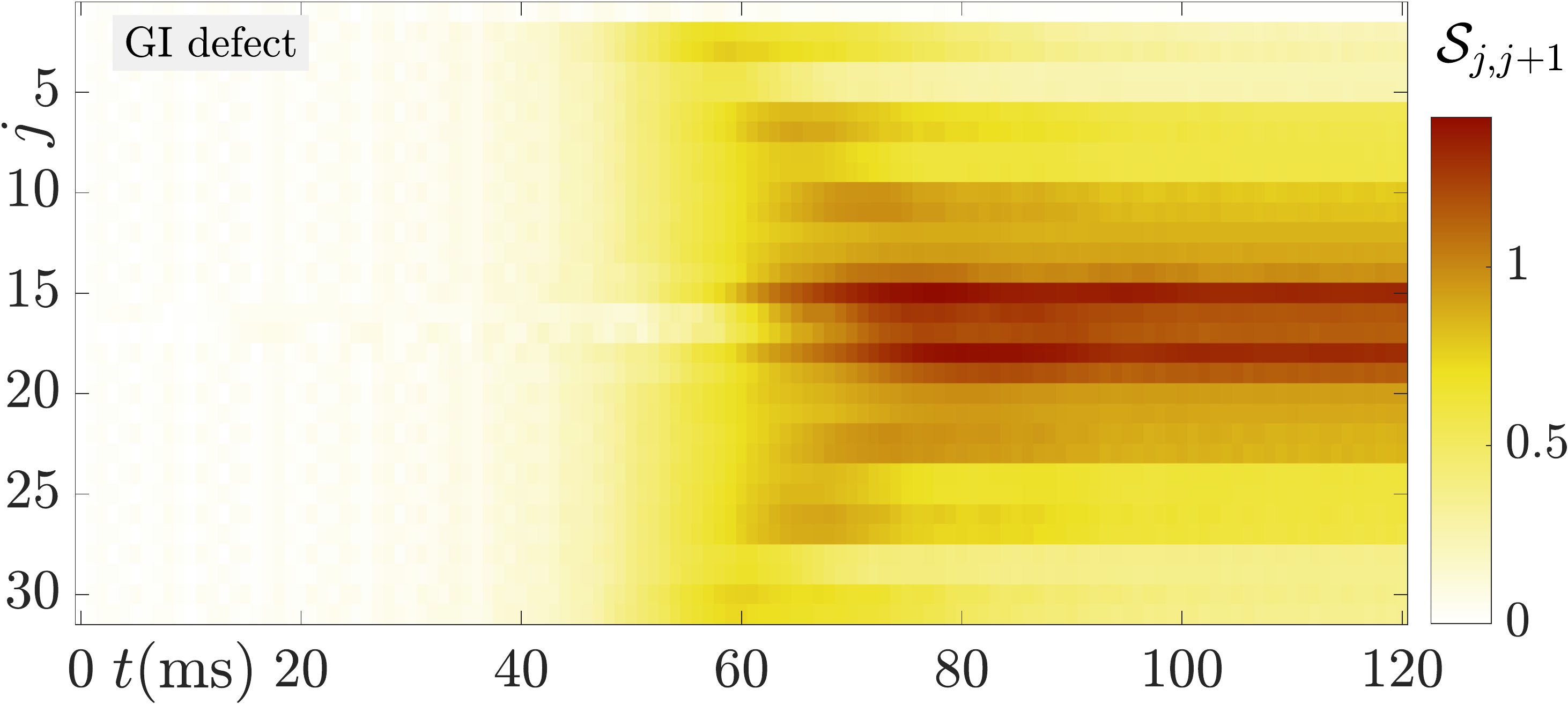}}\quad
	\includegraphics[width=.48\textwidth]{{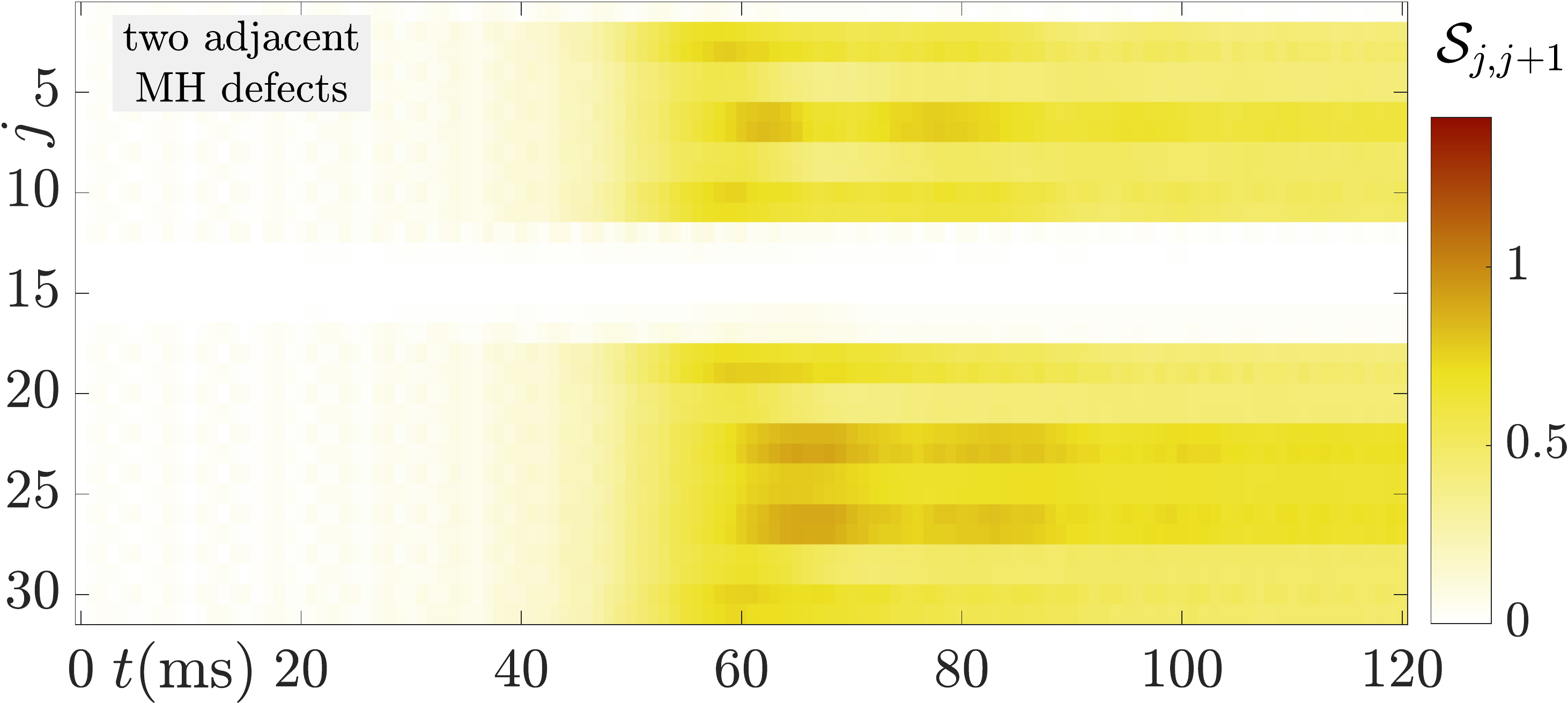}}\quad
	\caption{(Color online). Dynamics of the von Neumann entanglement entropy for various initial states. Whereas the GI defect acts only partially as an edge (see text and Fig.~\ref{fig:FigGI}), the MH defect exhibits a much stronger edge effect (right-column panels). Indeed, going from an initial state with a single MH defect at site $j=16$ to one with two adjacent MH defects at sites $j=14,16$, we find that the entanglement dynamics profile of the right subsystem is unchanged (while the left one is altered since there is one fewer atom compared to the case of a single MH defect).}
	\label{fig:entropy}
\end{figure*}

\begin{figure*}[!ht]
	\centering
	\hspace{-.01 cm}
	\includegraphics[width=.48\textwidth]{{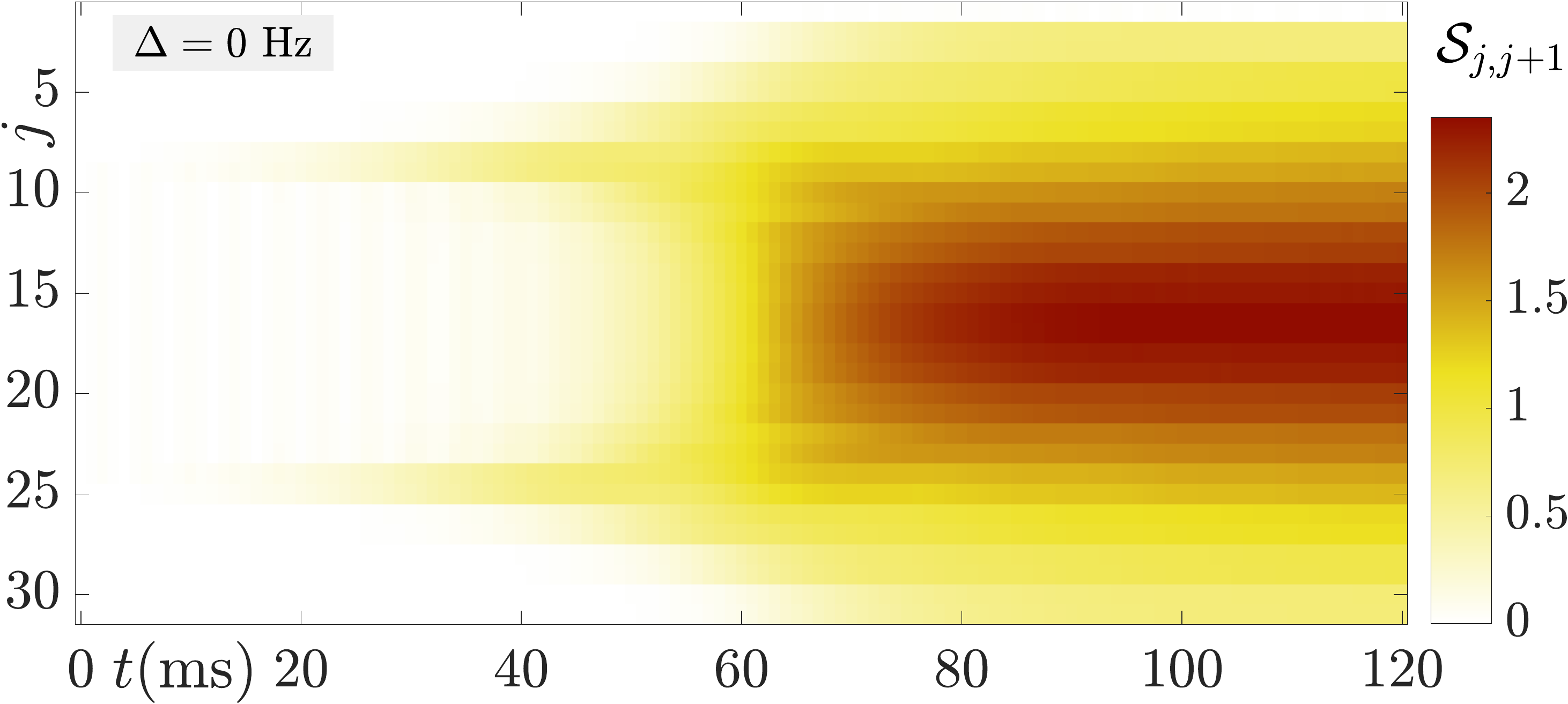}}\quad
	\includegraphics[width=.48\textwidth]{{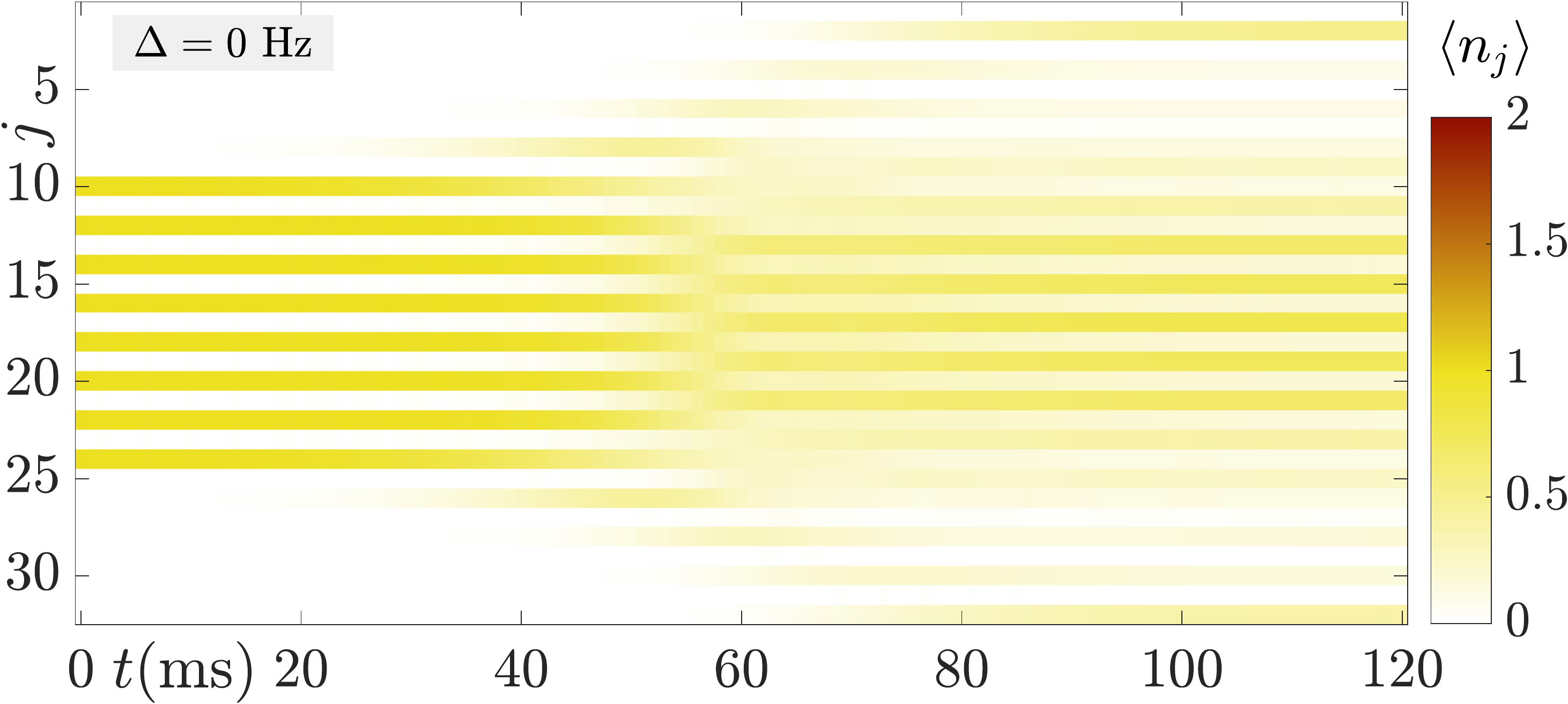}}\quad\\
	\hspace{-.01 cm}
	\includegraphics[width=.48\textwidth]{{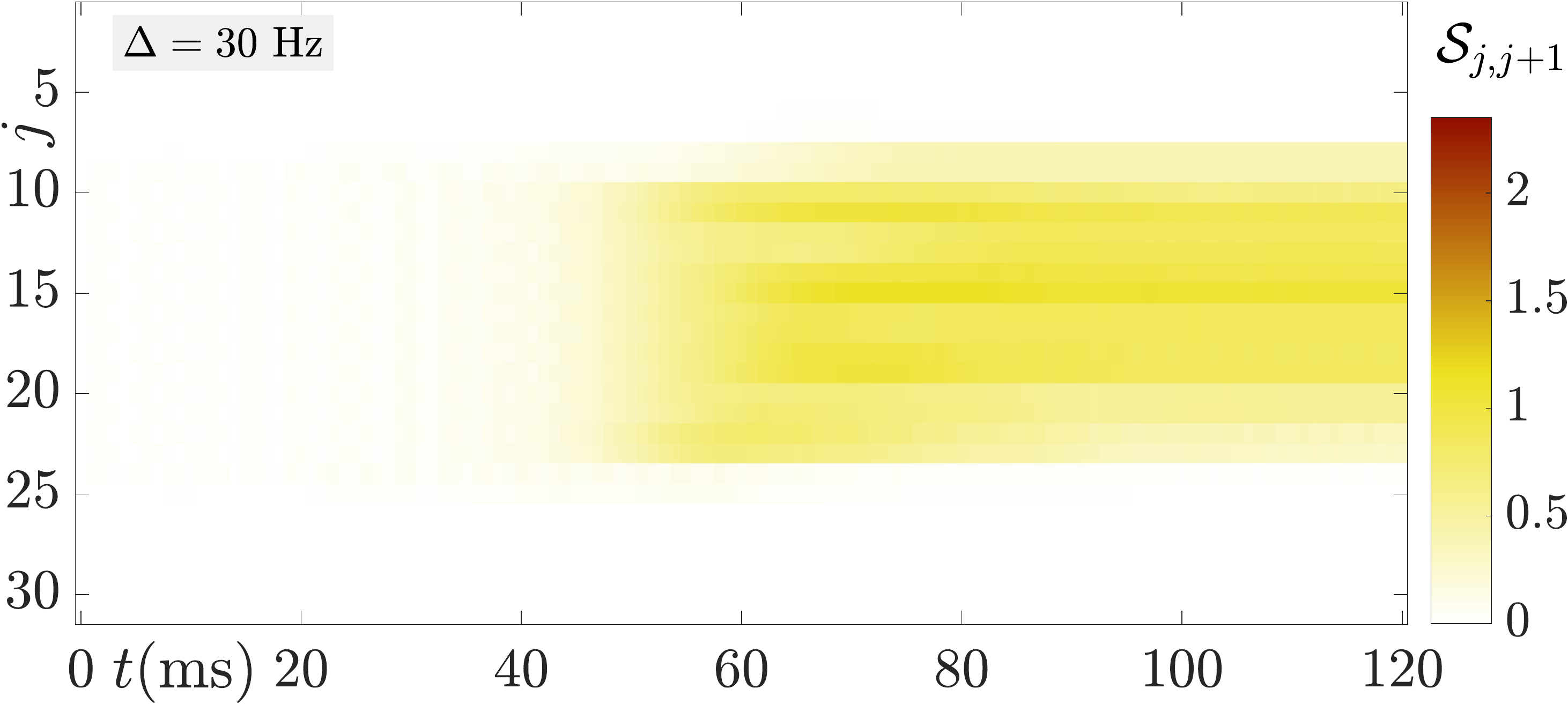}}\quad
	\includegraphics[width=.48\textwidth]{{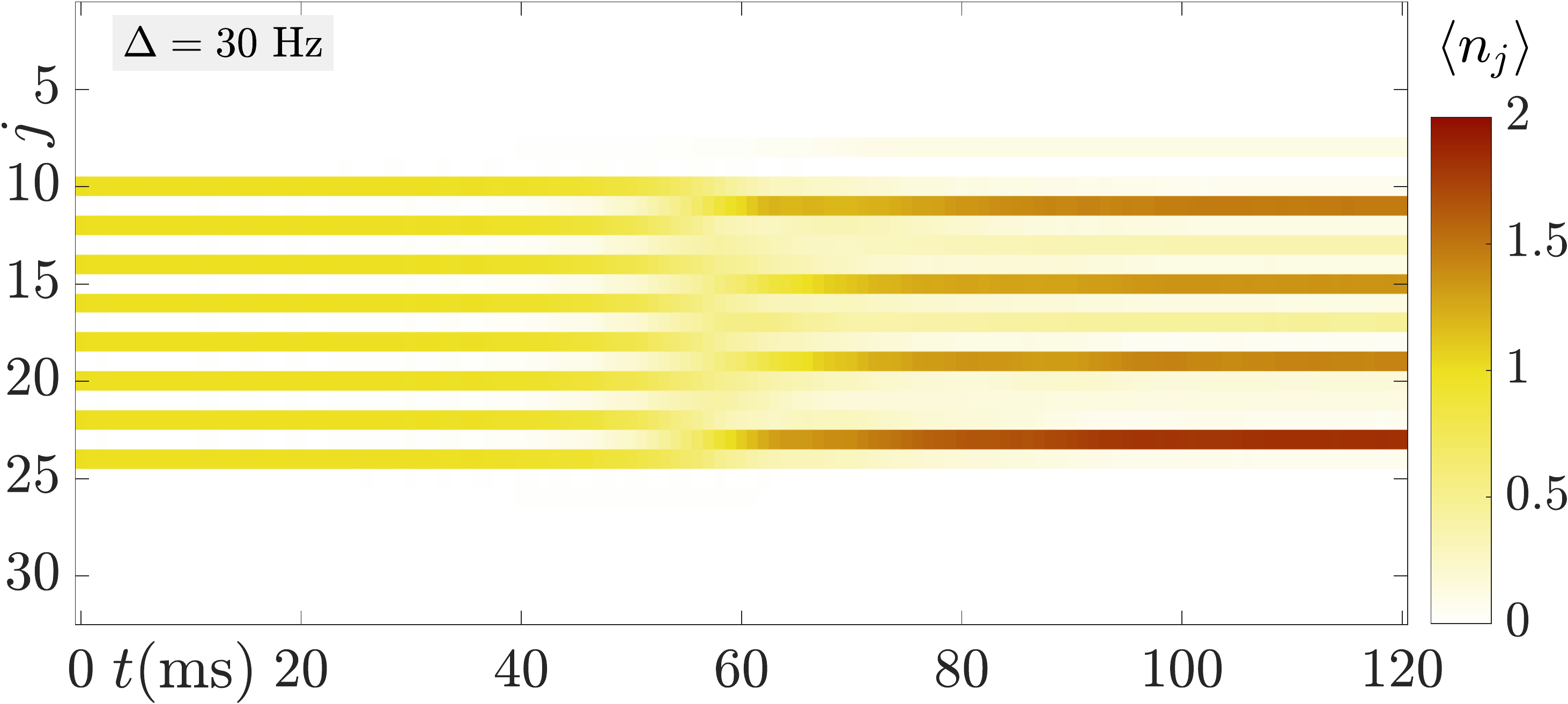}}\quad\\
	\hspace{-.01 cm}
	\includegraphics[width=.48\textwidth]{{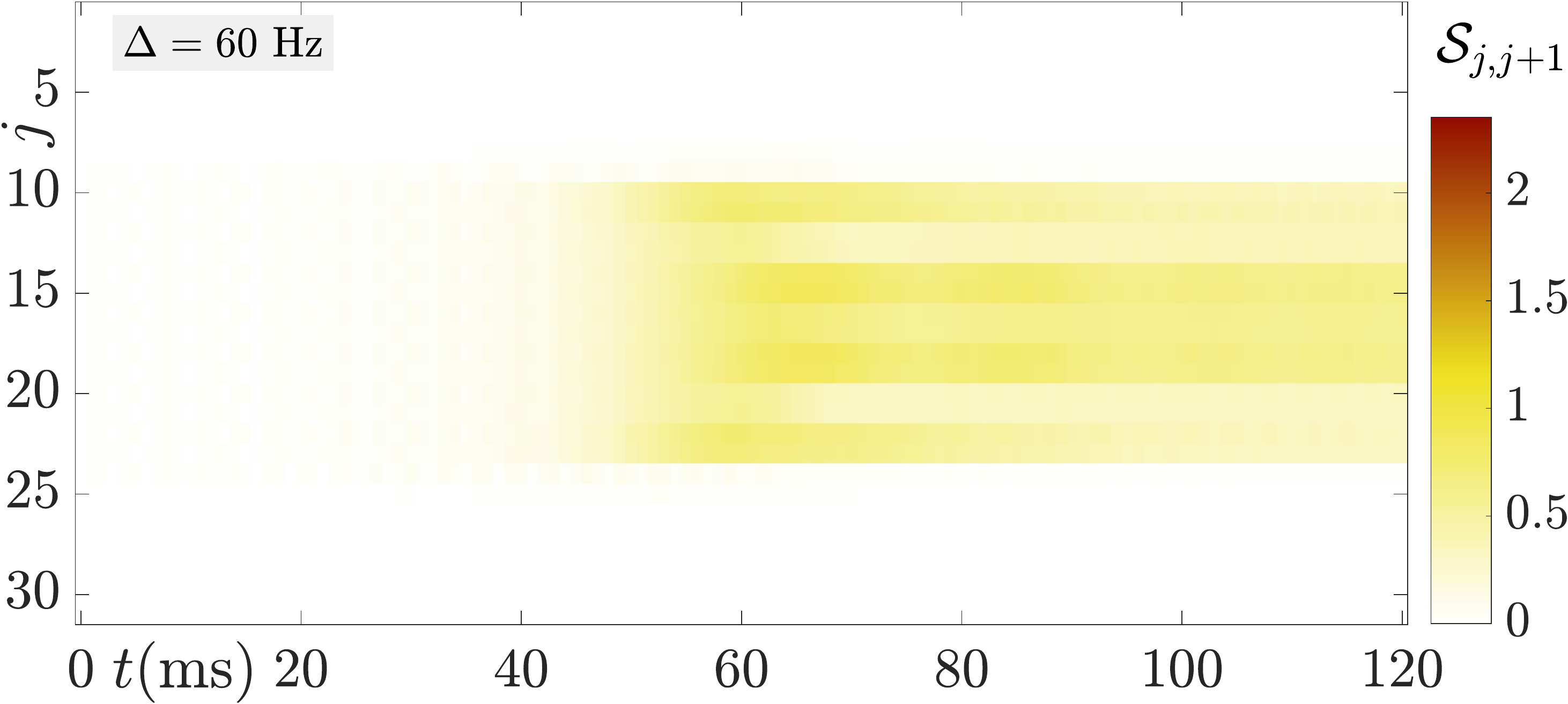}}\quad
	\includegraphics[width=.48\textwidth]{{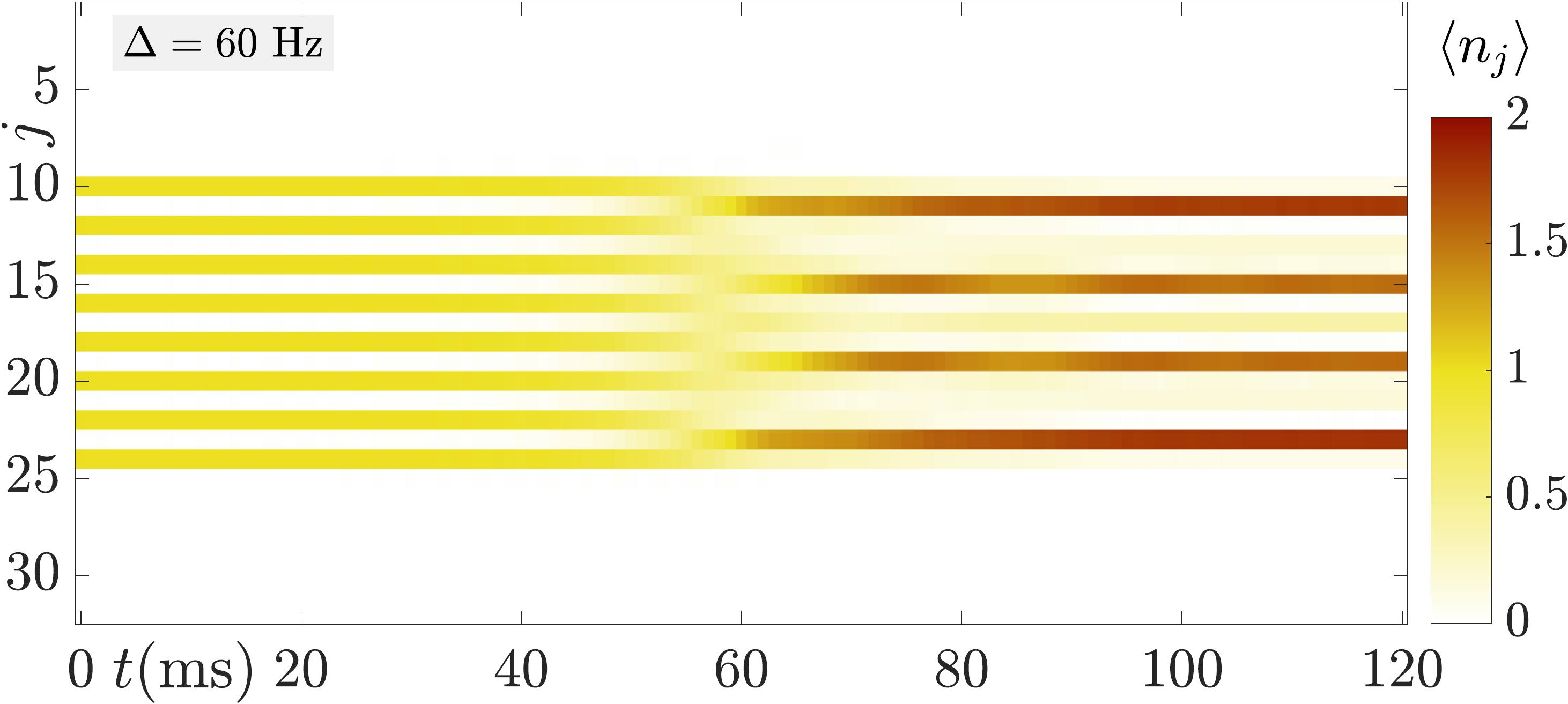}}\quad
	\caption{(Color online). Transport dynamics of the model given by Eq.~\eqref{eq:BHM} for different values of the tilt parameter $\Delta$ (see legends) starting in an initial state where only the middle eight matter sites are occupied by a single boson each, while all other sites are empty.
	The absence of transport outside of the center region at $\Delta=57$ Hz is a strong indication of having a gauge theory. While for the Bose--Hubbard model, only the global charge is conserved, which means atoms can move freely, for the gauge theory the charge is conserved locally, so the atoms cannot move into a region where the Gauss's law has a wrong eigenvalue. Said differently, there are fixed background charges everywhere in the outside region, which prevent the electrons and positrons to move there.}
	\label{fig:DeltaScan}
\end{figure*}

The main time-evolution calculations in this study \cite{HalimehChannel} are carried out using the time-dependent density-matrix renormalization group \cite{Vidal2004,White2004,Daley2004} ($t$-DMRG) based on the Krylov subspace method \cite{Schmitteckert2004,Feiguin2005,GarciaRipoll2006,McCulloch2007}. Upon repeated truncations of small Schmidt coefficients, one can represent the quantum many-body wave function in $t$-DMRG by matrix product states \cite{Uli2010}. The accumulated error of the simulation over evolution time is controlled through the so-called fidelity threshold \cite{White1992,White1993,Uli2005} (also known as truncation error).
We limit the total Hilbert space by setting the maximal on-site boson occupation to $N_\text{max}$. We find excellent convergence for a fidelity threshold of $10^{-6}$ for each time-step $\d t=10^{-4}$ ms and $N_\text{max}=3$ for the most demanding calculations, congruent with the relative insignificance of on-site occupation numbers larger than $2$ in the experiment \cite{Yang2020}.
Throughout the paper, we consider an optical superlattice of length $L=32$ sites ($16$ matter and $16$ gauge sites), though our conclusions on the proliferation of gauge-invariance violations due to the defects are independent of system size $L$, as we shall see in the following. In our code, we implement the bosonic Hamiltonian~\eqref{eq:BHM}, including a slight inhomogeneity in the on-site potential profile $U$. This mimics the experimental reality of Ref.~\cite{Yang2020}, where due to the large extent of the chain ($\sim71$ sites) there is a slight Gaussian inhomogeneity, such that the on-site potential at the edges is roughly $10$ Hz smaller than in the middle of the chain. Our numerical ramp involves tuning $U$ in the middle of the chain from $1.82$ to $1.34$ kHz, $J$ between $26$ and $93$ Hz, and slightly tuning $\delta$ between $725$ and $740$ Hz, while fixing $\Delta$ at $57$ Hz. Our conclusions do not depend on these exact quantitative specifics and also hold when accounting for Gaussian fluctuations in the parameter $\delta$ \cite{Yang2020}. In $t$-DMRG, the $32$-site lattice follows the site indexing convention specified in Fig.~\ref{fig:mapping} and Sec.~\ref{sec:model}. Odd (even) sites represent gauge (matter) sites. The first ($j=1$) site on this lattice is therefore a gauge site, and the last site ($j=32$) is a matter site. Directionality is chosen such that left to right is from first to last site.

Our main observable is the gauge-invariance violation.
For an individual constraint at matter site $j_\mathrm{m}$, it is defined as
\begin{align}\label{eq:error}
\varepsilon_{j_\mathrm{m}}(t)=1-\Big[\bra{\psi(t)} P_{j_\mathrm{m}}^{\ket{002}+\ket{200}}+P_{j_\mathrm{m}}^{\ket{010}}\ket{\psi(t)}\Big]^2.
\end{align}
Here, $j_\mathrm{m}$ is an \textit{even}---or, equivalently, \textit{matter}---site of the optical lattice, which, along with its two adjacent odd---or, equivalently, gauge---neighbors, forms the bosonic three-site unit cell on which the local constraint is defined. Using as basis the set of bosonic Fock states, $P_{j_\mathrm{m}}^{\ket{002}+\ket{200}}$ projects the wave function onto the subspace of the total Hilbert space where the unit cell $\{j_\mathrm{m}-1,j_\mathrm{m},j_\mathrm{m}+1\}$ hosts the particle-number configurations $\ket{002}$ or $\ket{200}$, whereas $P_{j_\mathrm{m}}^{\ket{010}}$ projects on the subspace where these occupations are $\ket{010}$.

For a valid comparison between the cases with different defect configurations, we define the \textit{dynamics-induced gauge violation} as
\begin{align}\label{eq:TotErr}
\varepsilon(t)=\sum_{j_\mathrm{m}}\frac{\varepsilon_{j_\mathrm{m}}(t)-\varepsilon_{j_\mathrm{m}}(0)}{L_\mathrm{m}},
\end{align}
where $L_\mathrm{m}=L/2$ is the number of matter sites---equivalently, local gauge constraints---in our system.
This quantity averages the violation over matter sites and is calibrated with respect to the initial cumulative violation at $t=0$.
Thus, $\varepsilon(t)$ measures the \emph{dynamics-induced gauge violation} accumulated on top of the presence of the initial defect(s).

In the following, we first study the dynamics-induced gauge violation $\varepsilon(t)$, followed by an in-depth analysis of the site-resolved dynamics of gauge-invariance violation $\varepsilon_{j_\mathrm{m}}(t)$ and other local observables such as the particle and doublon occupations. We provide ``time-slice'' videos \cite{HalimehChannel} of both these quantities. We conclude this section with an analysis of the von Neumann entanglement entropy, which provides a further diagnostic to probe the buildup of dynamics in the system over time, and how it is affected by the different defects.

\subsection{Total dynamics-induced gauge violation}\label{sec:TotalViolation}

In Fig.~\ref{fig:FigTotalError}, we show the time evolution of $\varepsilon$ for various defective initial states as compared to the ideal defect-free ``clean'' case. Figure~\ref{fig:FigTotalError}(a) displays results for an initial state with a single, two separate, and two adjacent MH defects. Intriguingly, the total dynamics-induced gauge-invariance violation due to the MH defects remains close to that of the clean case for all times. Moreover, the number of MH defects and the spacing between them does not seem to have any discernible effect on $\varepsilon$. We find the same qualitative picture for MI defects [see Fig.~\ref{fig:FigTotalError}(b)] as well as combinations of MI and MH defects [see Fig.~\ref{fig:FigTotalError}(c)]. Thus, the gauge-invariance error due to matter defects is very benign.

It is worth noting here that for $t\lesssim60$ ms the dynamics-induced violation due to MH defects seems to be slightly lower than the clean case. As we shall see, this also occurs for the other two defects, and for combinations of the considered defects in this work. The reason this happens is because a defect will reduce the number of gauge-invariant constraints by at least one. However, any error spreading due to this violation will be averaged over $L_\mathrm{m}$ sites according to Eq.~\eqref{eq:TotErr}, leading to this apparent discrepancy. We choose not to compensate for this trivial effect, because at times $t>60$ ms the violation might become delocalized, in which case the normalization in Eq.~\eqref{eq:TotErr} is the proper one.

In contrast to the case of matter-field defects, the effect of GI defects in the initial state on the gauge-invariant dynamics is not as innocuous, which can be seen in Fig.~\ref{fig:FigTotalError}(d). Already a single GI defect significantly increases the violation compared to the clean case after crossing Coleman's phase transition. Indeed, at the final ramp time of $t=120$ ms, the gauge violation in the presence of a GI defect is more than double that of the clean case. Adding a second GI defect at a few sites from the first one exacerbates the violation further, making it three times that of the clean case at the end of the ramp. Interestingly though, placing two GI defects adjacent to each other leads to a gauge violation that is qualitatively similar in its dynamics to the case of a single GI defect. This indicates that either GI defect blocks some of the violation proliferation coming from the other. We will come back to this point later.

Finally, Fig.~\ref{fig:FigTotalError}(e,f) display the dynamics of $\varepsilon$ when a GI defect is present in the initial state along with either an MH or MI defect, respectively. Interestingly, both scenarios are qualitatively identical whether the matter-field defect is MH or MI. When the matter defect is separated from its GI counterpart, the total dynamics-induced violation rises most above the clean case.
There is a small asymmetry with respect to the relative position of GI and matter-field defect, which we attribute to the tilt in the optical lattice setup [see Eq.~\eqref{eq:BHM} and associated discussion in Sec.~\ref{sec:model}].
When the matter-field defect is to the left of its GI counterpart, the dynamics-induced violation proliferates somewhat less than in the opposite arrangement, albeit it remains comparable in both cases so long as the GI and matter defects are separated.
This situation changes significantly when both gauge and matter defects are placed adjacent to one other. Whereas the dynamics-induced violation is still relatively large when the GI defect is to the left of its matter-field counterpart, the opposite arrangement leads to a violation that is very comparable to that of the clean case. Given that the matter-field defect does not itself generate a further gauge-invariance violation spread with respect to the clean case, one can conclude that in the right arrangement, the matter-field defect can act as a suppressor of gauge violations caused by its GI counterpart. In Sec.~\ref{sec:SiteResolvedDynamics}, we further investigate this intriguing possibility, where we consider the \textit{site-resolved} gauge-invariance violation $\varepsilon_{j_\mathrm{m}}(t)$.

Nevertheless, we can already conclude from the \textit{dynamics-induced} gauge violation of Eq.~\eqref{eq:TotErr} that the influence of the considered defects on the experiment of Ref.~\cite{Yang2020} is insignificant (matter-field defects) to mild (gauge impurity). Our numerical calculations indicate that the defects on the matter sites do not contribute to additional gauge violations. Only the defect on the gauge site can increase the violations on the states after $120$ ms evolution. By the end of the ramp in the $71$-site experimental chain \cite{Yang2020}, our numerical results indicate that one GI defect causes $18(12)\%$ increase in the gauge violation, and two identical GI defects lead to additional $3(2)\%$ increase in the gauge violation.

\subsection{Site-resolved gauge-theory dynamics}\label{sec:SiteResolvedDynamics}
We can go beyond the results of the averaged dynamics-induced gauge violation of Eq.~\eqref{eq:TotErr} shown in Fig.~\ref{fig:FigTotalError} by further investigating the dynamics of the site-resolved gauge-invariance violation $\varepsilon_{j_\mathrm{m}}(t)$ given in Eq.~\eqref{eq:error} as well as other local observables for the various considered initial states; cf.~Fig.~\ref{fig:violations}. In this spirit, we first show the dynamics of $\varepsilon_{j_\mathrm{m}}(t)$ for the case of a defect-free initial state in the top panel of Fig.~\ref{fig:FigClean}, along with the site-resolved dynamics of the particle occupation (middle panel) and the doublon number (bottom panel). The gauge-invariance violation $\varepsilon_{j_\mathrm{m}}$ at matter site $j_\mathrm{m}$ seems to increase over time on average, with considerable oscillations, up until the vicinity of Coleman's phase transition at $t\approx60$ ms, and afterwards it steadily decreases over time on average. This phase transition involves the spontaneous breaking of a $\mathrm{Z}_2$ symmetry, where the associated equilibrium phase is doubly degenerate. This is quite evident in the dynamics of the particle number $n_j$. At $t=0$ the initial state has a single boson on every matter site and is $\mathrm{Z}_2$ symmetric. This is equivalent to the QLM ground state at $m\to-\infty$ (see Fig.~\ref{fig:mapping}, top left). This symmetry is preserved during the ramp until the critical point is crossed at $t\approx60$ ms, where the $\mathrm{Z}_2$ symmetry is spontaneously broken, and now the gauge fields begin to order by the formation of doublons on odd sites of the optical lattice representing gauge fields.

After the phase transition, there are two degenerate ground states in equilibrium, representing a left-(right-)pointing electric-field configuration everywhere when the bosonic superlattice hosts doublons on gauge sites of index $j_\mathrm{g}$ satisfying $j_\mathrm{g}\bmod4=1$ ($3$) with all other sites empty; cf.~Fig.~\ref{fig:mapping}, bottom left. The system settles into the right-pointing electric-field configuration at the edges, as the way we initialize the system breaks the symmetry at the boundaries: the single bosons on sites $j=2,4$ are likely to form a doublon on site $j=3$, while the gauge field on site $j=1$ is unlikely to host a doublon. In the middle of the chain, however, there seems to be a superposition of both states, because away from the edges, it is approximately equally likely for doublons to occupy the gauge sites with index $j_\mathrm{g}$ satisfying $j_\mathrm{g}\bmod4=1$ or $3$. This picture is further confirmed when looking at the doublon number in the bottom panel of Fig.~\ref{fig:FigClean}. Indeed, at early times, there are no doublons in the system because the system is in the charge-proliferated state, but once the ramp sweeps through the critical point, doublons begin to proliferate. Nevertheless, the doublon occupations on gauge sites in the middle of the chain carry intermediate values (rather than $0$ or $1$ as at the edges), and this is because the system has roughly equal probability to be in the left- and right-pointing electric-field configurations there, indicative of a superposition state that spontaneously breaks the $\mathrm{Z}_2$ symmetry. Thus, the final state at $t=120$ ms in the middle of the chain represents a Greenberger--Horne--Zeilinger (GHZ) type entangled state.

We now consider the case where the initial state hosts a matter-field defect, as displayed in Fig.~\ref{fig:FigMatterDefects}, where the panels on the left (right) column show the dynamics in the presence of a single MH (MI) defect. The site-resolved gauge violation is qualitatively identical, where we see that $\epsilon_{j_\mathrm{m}}$ ($j_\mathrm{m}$ is even as it representes a matter site) is unity at site $j_\mathrm{m}=16$ where the defect is located, but very close to zero elsewhere. Indeed, this violation remains extremely localized throughout the entire time evolution. Moreover, the results for the particle and doublon occupations show that the matter-field defect acts as a quasi-hard edge. It leads to the state arranging itself into the right-pointing electric-field configuration around the middle of the chain, where in the clean case a GHZ state exists instead. In contrast, the upper half of the chain does not exhibit a clear $4$-site-ordering configuration due to the presence of an odd number (seven) of bosons there.

Whereas the presence of matter-field defects in the initial state does not compromise the fidelity of gauge invariance throughout the dynamics, the GI defect is not as innocuous, which can be seen in Fig.~\ref{fig:FigGI}. At times before crossing Coleman's phase transition, $t\lesssim60$ ms, the gauge violation is strongly localized at two adjacent constraints defined at the matter sites $j_\mathrm{m}=16,18$---the GI defect itself is a single boson at site $j=17$ in the initial state. However, after the ramp has passed the critical point at $t\gtrsim60$ ms, there is a clear yet contained proliferation of the associated gauge violation, which by the end of the ramp has spread to the neighboring two constraints on either side at $j_\mathrm{m}=12,14,20,22$; see also associated ``time-slice'' video for a different visualization of this dynamics \cite{HalimehChannel}.
In particular, we also find that the gauge-violation spread is more adverse on the left side of the GI defect due to the tilt of the lattice. Looking at the particle- and doublon-number expectation values, we find that the GI defect also partially behaves as an edge, albeit not a hard one as in the case of matter-field defects. This becomes clear by noticing how, in the middle of the chain, the system shows at long times a greater probability to settle into the right-pointing electric-field configuration (i.e., doublons emerging on gauge sites $j_\mathrm{g}$ such that $j_\mathrm{g}\bmod4=3$; see Fig.~\ref{fig:mapping}, bottom left) as compared to the clean case in Fig.~\ref{fig:FigClean}. Nevertheless, this edge effect is more prominent in the case of a matter-field defect.

The site-resolved dynamics in presence of multiple GI defects in the initial state is presented in Fig.~\ref{fig:FigMultipleGI}. Since the gauge violation generated by a single GI defect exhibits limited spreading into the neighboring constraints (see Fig.~\ref{fig:FigGI}), it is not surprising that when two GI defects are apart, the overall gauge violation will spread over a larger region of the lattice, which also explains the corresponding spatially averaged dynamics-induced violation shown in Fig.~\ref{fig:FigTotalError}(d). However, once the GI defects are placed closer to one another, each suppresses the spread of the other's gauge violation on its side, which explains why the spatially averaged dynamics-induced violation is qualitatively similar to that due to a single GI defect, as observed in Fig.~\ref{fig:FigTotalError}(d). This similarity extends to the particle and doublon occupations, where again we see that two adjacent GI defects, much the same way as in the case of a single GI defect, increase the probability of our system to order into the right-pointing electric-field configuration at late times. However, in the case of two separate GI defects, the picture is slightly changed because of the particular location of the leftmost GI defect (the superlattice site indexing is counted from left to right as $j=1,\ldots,32$). We can now think of the system as composed of three blocks, with the first being to the left of site $j=11$ (initial location of first GI defect), the second between sites $j=11$ and $17$ (initial location of second GI defect), and the third being right of site $j=17$. The first block shows a right-pointing electric-field configuration, albeit not at site $j=11$ where the leftmost GI defect is located, as is clear by looking at the corresponding doublon occupation at late times. Nevertheless, the probability of this configuration is not close to unity since this block contains an odd number (five) of bosons. The leftmost GI defect also acts partially as an edge, and leads to a relatively large probability of a left-pointing electric-field configuration (i.e., doublons emerging on gauge sites $j_\mathrm{g}$ such that $j_\mathrm{g}\bmod4=1$; see Fig.~\ref{fig:mapping}, bottom left) in the second block at late times. This configuration does not have a probability close to unity either, since that block also contains an odd number (three) of bosons. On the other hand, the third block shows a clear right-pointing electric-field configuration at large times, with the GI defect at site $j=17$ acting in part as an edge to give this configuration a higher probability than in the clean case.

Figure~\ref{fig:FigMHGI} shows the dynamics for both an MH and a GI defect placed in the vicinity of one another in the initial state. In the left-column panels, we show results for when the MH and GI defects are located at sites $j=14$ and $17$, respectively, thereby leading to the violation of Gauss's law on the three adjacent constraints denoted by matter sites $j_\mathrm{m}=14,16,18$. In the right-column panels, the GI and MH defects are present at sites $j=13$ and $16$, respectively, which breaks Gauss's law on the three adjacent constraints denoted by the matter sites $j_\mathrm{m}=12,14,16$. These two seemingly similar arrangements actually lead to significantly different dynamics, as can also be seen in the associated ``time-slice'' videos \cite{HalimehChannel}. Due to the lattice tilt, the GI defect-induced gauge violation spreads more to the left than to the right at late times, as shown in Fig.~\ref{fig:FigGI}. Consequently, when an MH defect is placed to its left, the gauge-violation spread is considerably suppressed, which further confirms our conclusion that matter-field defects act as quasi-hard edges. This also validates the corresponding spatially averaged \textit{dynamics-induced} gauge violation shown in Fig.~\ref{fig:FigTotalError}(e), where it is qualitatively similar to that of the clean case. When the order is changed, and now the MH defect lies immediately to the right of its GI counterpart, the gauge violation due to the latter proliferates leftwards at late times, which explains the large deviation from the clean case in Fig.~\ref{fig:FigTotalError}(e). Similarly to the case of a single MH defect, the two cases in Fig.~\ref{fig:FigMHGI} both show a higher probability of the subsystems, which are divided by the quasi-hard edge-like MH defect, to arrange into the right-pointing electric-field configuration at late times. We note here that the qualitative picture in Fig.~\ref{fig:FigMHGI} is unchanged if we replace the MH defect with its MI counterpart in the associated initial states (not shown).

In this Section we have shown how matter- and gauge-field defects lead to fundamentally different dynamics of the gauge-invariance violation in the experiment of Ref.~\cite{Yang2020}. We have demonstrated this through $t$-DMRG calculations of the site-resolved dynamics of gauge invariance. However, we can also understand this fundamental difference between these defects by determining the most dominant violating processes at the end of the ramp. This is shown in Fig.~\ref{fig:violations} for four initial states: the defect-free one, two states each with a single matter-field (MH or MI) defect, and a state with the GI defect. The dominant violations in the case of an MH or MI defect do not involve the site in which the defect resides. Moreover, the dominant violations are the same for the MH and MI defects. This explains why qualitatively, the gauge-invariant dynamics is identical for both matter-field defects, and it also consolidates the conclusion that they resemble quasi-hard edges. On the other hand, we see that for the GI defect the dominant violation processes all involve the defect site. Indeed, the GI defect facilitates gauge-violating processes that are energetically highly suppressed in the case of the matter-field defects. See Appendix~\ref{sec:hist} for a quantitative histogram analysis of the dominant processes in the final wave function at $t=120$ ms for these different initial states.

\subsection{Dynamics of von Neumann entanglement entropy}

To further confirm that matter-field defects act as quasi-hard edges while the GI defect not so much, we examine in Fig.~\ref{fig:entropy}, for the case of several initial states, the time-evolved bond-resolved von Neumann entanglement entropy
\begin{align}
\mathcal{S}_{j,j+1}=-\Tr\big\{\rho_{1\to j}\log\rho_{1\to j}\big\},
\end{align}
where $\rho_{1\to j}$ is the reduced density matrix of the subsystem formed on the lattice along sites $1,2,\ldots j$.

In the clean case, $\mathcal{S}_{j,j+1}$ grows extremely slowly until the ramp passes the critical point at $t\gtrsim 60$ ms, where the von Neumann entanglement entropy grows markedly on average. The only exception is site $j=1$ which, due to the lattice geometry, involves very little of the dynamics. [Within the gauge theory, a boson from a matter site can only hop if it combines with a particle from the next matter site into the gauge-field well that lies between the two, which corresponds to the annihilation of a charge--anti-charge pair with a concomitant flip of the electric field, see Eq.~\eqref{eq:effective-Hamiltonian}; cf.~Fig.~\ref{fig:mapping}. Since the first site is a gauge-field site, there is no allowed gauge-invariant process that leads the boson from the second site (a matter site) to hop there.]

Upon introducing a single GI defect at site $j=17$, the von Neumann entropy changes qualitatively nontrivially only in the middle due to the GI defect there, albeit no hard-edge effect is present to isolate the subsystems on either side of the defect from one another. The situation fundamentally changes when the initial state hosts an MH defect. Placed at site $j=16$, the latter leads to a strong suppression of entanglement at all evolution times between both subsystems on either side of the MH defect. Interestingly, placing a second MH defect on site $j=14$ does not change much in the entanglement-entropy pattern of the right subsystem, indicating almost complete isolation of the latter. The left subsystem shows a different structure as the number of bosons is reduced by one there due to the addition of the MH defect, thereby altering the dynamics. The smallness of the change in the entanglement entropy of the right subsystem despite changing its left counterpart strongly indicates a near-complete isolation of both subsystems due to the MH defects. Again, when replacing the MH defect with its MI counterpart, our conclusions remains unchanged.

\subsection{Transport dynamics}
Furthermore, we investigate in our lattice gauge model the transport property out of a well-localized initial domain. In the BHM Hamiltonian~\eqref{eq:BHM}, a linear potential with $\Delta$ per site is employed to suppress the long-range transport of the atoms; see Sec.~\ref{sec:model}. Such a tilt is provided by the projection of gravity, which gives $57$ Hz per lattice site to the atoms. This linear energy offset breaks the translational symmetry of the lattice potential, whose effects are investigated numerically here.
	
In Figure \ref{fig:DeltaScan}, we show the evolution of a state at three different $\Delta$. We choose an initial configuration with $8$ atoms located at the center of the $32$-site chain, in the region from site $10$ to site $24$ in alternating filling. The other lattice sites are initialized free of atoms. Without the linear gradient, meaning $\Delta=0$, the atoms have a large probability to escape from the local constrains, as shown in Fig.~\ref{fig:DeltaScan}(a). In this case, the on-site energy on different matter sites are degenerate (same for different gauge sites). The atoms can undergo a second-order tunneling to their next-nearest-neighbor sites, with similar coupling strength as our elementary gauge-invariant interaction, which is also generated in second-order perturbation theory; cf.~Eq.~\eqref{eq:effective-Hamiltonian}. With a weak linear potential, we can strongly suppress the undesired second-order tunneling by lifting the degeneracy of the on-site energies. Figure \ref{fig:DeltaScan}(c) shows that the dynamical evolution becomes localized to the region between site $10$ and site $24$, without discernable propagation to the outer parts. The gauge-symmetry breaking process is suppressed, further indicating the conservation of the local invariant of our lattice gauge theory.

\section{Proposal for future experiments}\label{sec:future}
The above results provide a clear way forward to improve the precision of experimental implementations along the lines of Ref.~\cite{Yang2020}.
Most importantly, the gauge-impurity defects can significantly increase the gauge violations in our quantum simulator. However, we can improve the fidelity of our initial state by modifying the procedure of the state preparation, thereby eliminate the effects induced by gauge-impurity errors. The preparation error origins from the imperfection of the site-selective addressing technique, which has an efficiency of 99.5(3)\%. After this spin-flipping process, the atoms on the other spin states are removed from the lattice confinement. To reduce the error, we can simply repeat this spin flipping and atom removal twice to fully clean the atoms on the gauge sites. Previous experiments \cite{Yang:2019} did not observe significant heating in this preparation stage, indicating the practicability of this scheme.

From the numerical calculations, we find that any kind of defects can break the 1D system into smaller subsystems. Above, we analyzed defects only due to the imperfection of the state preparation, while they could also be generated during the dynamics, such as the gauge violations in the clean case. To reduce the probability of gauge violations appearing during the dynamics, we should enhance the coupling strength $\tilde{t}$ while reducing the bare tunneling by adding a stronger energy penalty. Following this, the Hubbard parameters can be further optimized to suppress the gauge violations.

Our results also open an interesting perspective to experimentally certify the emergence of a gauge theory, namely by purposefully imprinting well-localized modifications in the occupation number of the initial state, such as the MH defects considered in this work. If these represent defects of an otherwise fully intact gauge symmetry, they correspond to a locally conserved quantity and thus cannot propagate, as we have seen in our above numerical results. In contrast, in absence of a gauge symmetry only the global atom number is conserved, and an initially localized MH defect will spread through the entire system. This strongly different behavior may thus serve as an indicator for an emergent local conservation law.

\section{Conclusions}\label{sec:conclusion}
In this work, we have numerically investigated the robustness of gauge-invariant dynamics in a bosonic mapping of the one-dimensional $\mathrm{U}(1)$ quantum link model, which undergoes a controlled ramp through Coleman's phase transition, as has recently been experimentally realized in Ref.~\cite{Yang2020}. Our results, obtained through the time-dependent density-matrix renormalization group, demonstrate a high level of fidelity in the gauge-invariant dynamics of this experimental setup even in the presence of defects in the initial state due to imperfect preparation. Our numerical calculations have included both matter- and gauge-field defects, and various combinations thereof.
	
Even though matter-hole and matter-impurity defects act as quasi-hard edges that separate the system into smaller subsystems, the resulting gauge violation is extremely localized such that the total spatially averaged \textit{dynamics-induced} gauge violation is qualitatively unaltered with respect to the clean case. Indeed, whether the initial state is defect-free or contains matter-field defects, dynamics-induced violations are of roughly the same number and weight. Results of the site-resolved von Neumann entanglement entropy confirm this picture.

In the presence of a gauge-impurity defect, the dynamics gives rise to violating processes that are much larger than in the case of clean or matter-defect initial states, thereby increasing the overall dynamics-induced gauge violation. Findings such as these help to focus experimental efforts onto the most pertinent sources of gauge-violating errors.

Finally, we have discussed how our results inspire future experiments that can utilize well-localized defects in the initial state in order to probe the presence of a local conservation law. As shown in this work, gauge-theory dynamics involves local constraints, which inherently prohibit the spread of such defects in case the gauge theory is faithfully realized. This can usher in a practical diagnostic for detecting local gauge symmetries in an experiment.

\section*{Acknowledgments}
This work is part of and supported by the DFG Collaborative Research Centre SFB 1225 (ISOQUANT), the Provincia Autonoma di Trento, and the ERC Starting Grant StrEnQTh (Project-ID 804305). I.P.M.~acknowledges support from the ARC Future Fellowships scheme, FT140100625.

\appendix

\begin{figure*}[!ht]
	\centering
	\includegraphics[width=0.75\textwidth]{{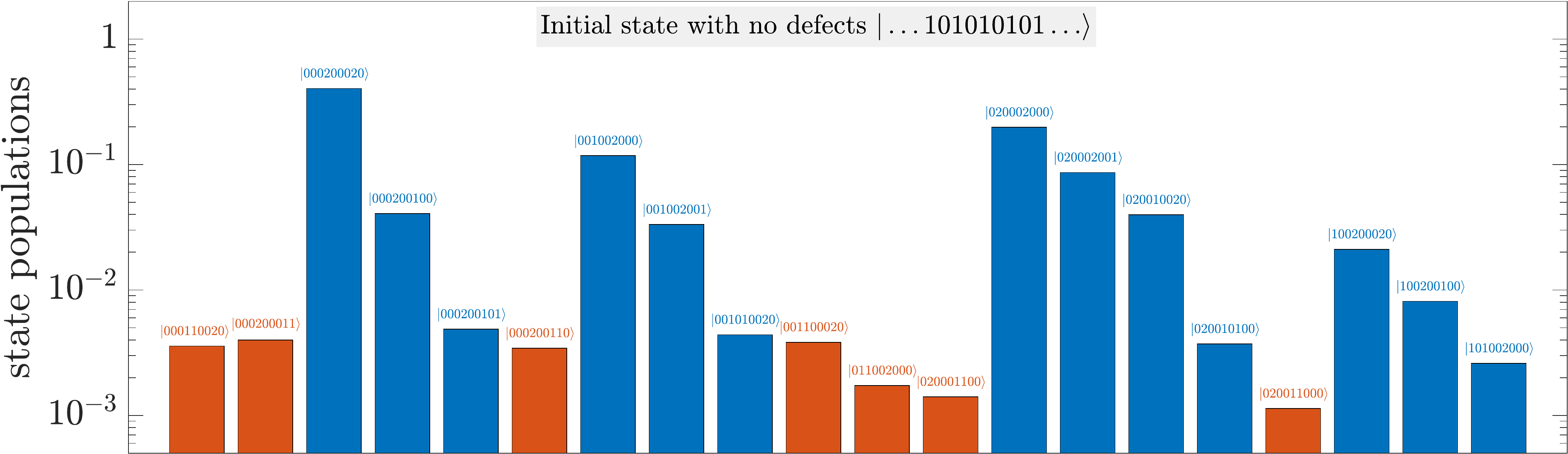}}\\
	\includegraphics[width=0.75\textwidth]{{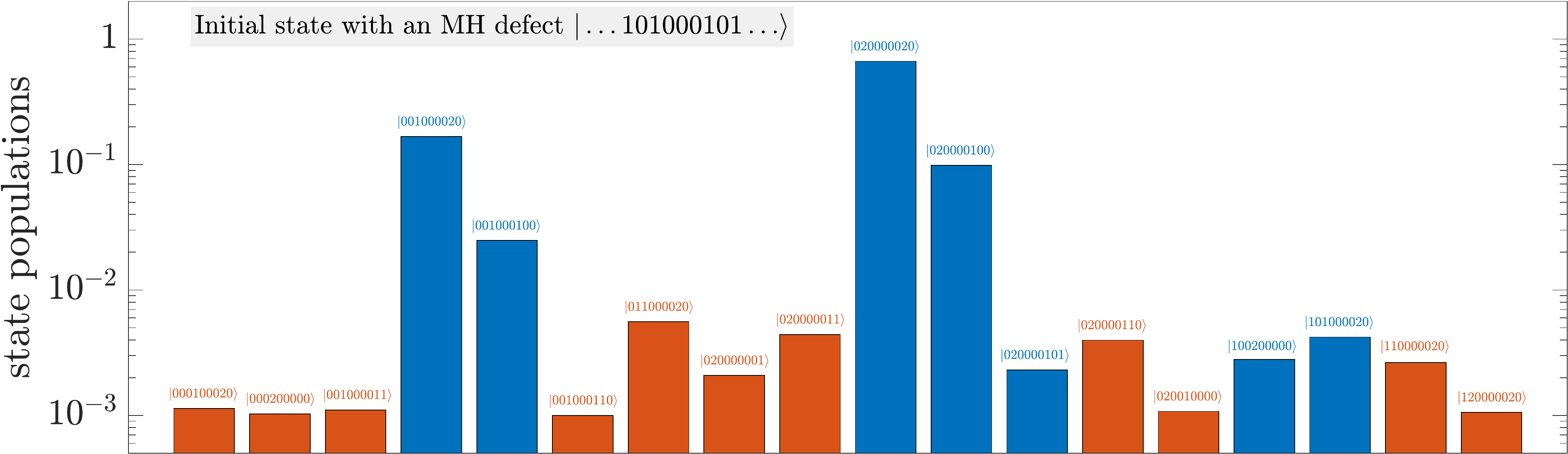}}\\
	\includegraphics[width=0.75\textwidth]{{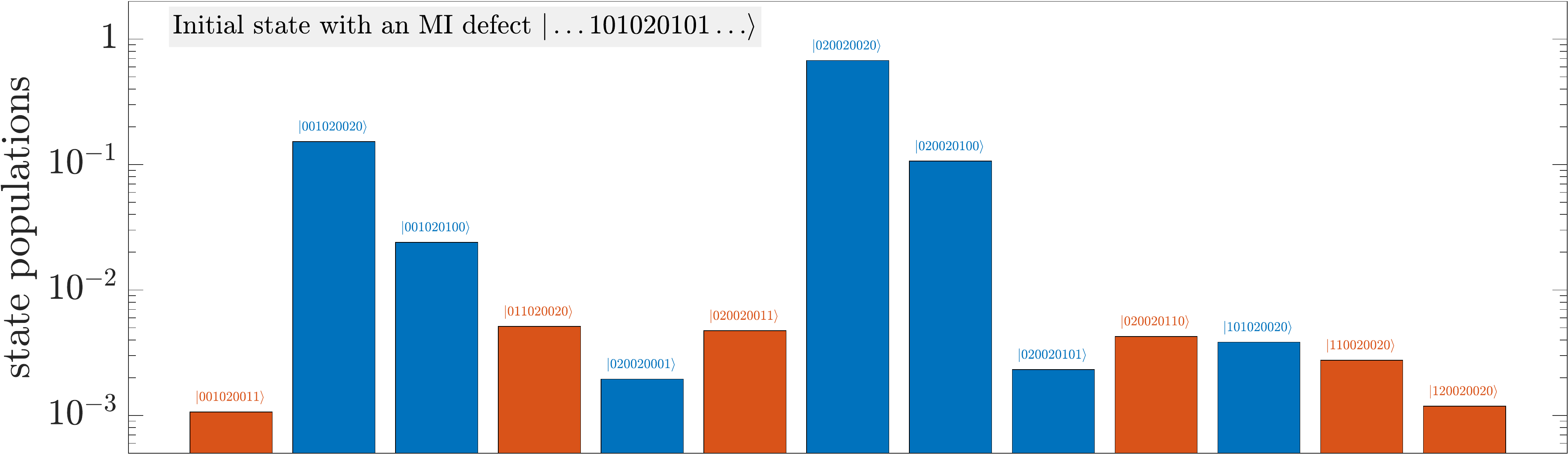}}\\
	\includegraphics[width=0.75\textwidth]{{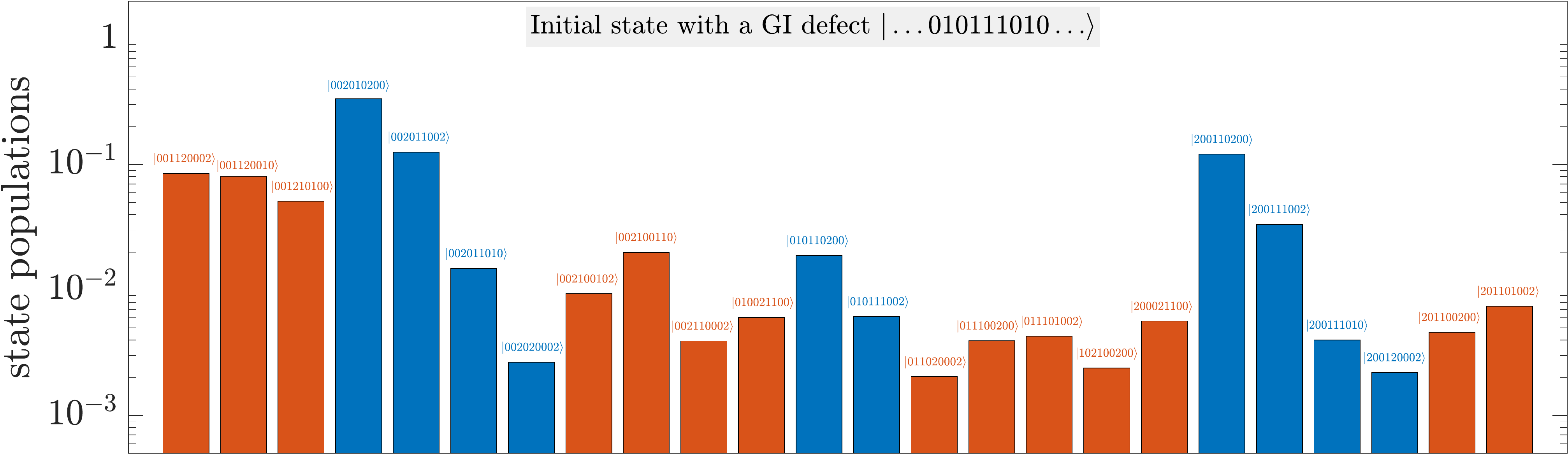}}
	\caption{(Color online). Histogram analysis of the most dominant processes in the final wave function at $t=120$ ms for several initial states considered in this work. Calculated in $t$-DMRG, these are the populations that the final wave function occupies in each dominant product Fock state of the $9$-site configuration $\{j_\mathrm{c}-4,\ldots,j_\mathrm{c},\ldots,j_\mathrm{c}+4\}$ at the middle of the chain, where the cutoff population is $\lambda_\text{thresh}$. From top to bottom: initial state with no defects ($j_\mathrm{c}=16$, $\lambda_\text{thresh}=10^{-3}$), a single MH defect ($j_\mathrm{c}=16$, $\lambda_\text{thresh}=10^{-3}$), a single MI defect ($j_\mathrm{c}=16$, $\lambda_\text{thresh}=10^{-3}$), and a single GI state ($j_\mathrm{c}=17$, $\lambda_\text{thresh}=2\times10^{-3}$). Note that for the defective initial states, the defect is also located at $j_\mathrm{c}$. Configurations in blue color are gauge-allowed states, while configurations colored orange are caused by gauge-violating dynamics.
	}
	\label{fig:hist}
\end{figure*}

\section{Derivational details of mapping}\label{sec:mapping}
In this section, we show the effective emergence of the quantum link model from the ideal/defect-free bosonic system with energy constraints on local occupations.
We start with a QLM Hamiltonian describing a U(1) lattice gauge theory in one spatial dimension \cite{Wiese_review,Chandrasekharan1997},
\begin{align}
H 
&= \sum_{\ell } \left[-\frac{\tilde{t}}{2}\left( \psi^{\dagger}_{\ell}S^+_{\ell,\ell+1}\psi_{\ell+1} +\text{H.c.}\right) +  (-1)^{\ell} m \psi_{\ell}^{\dagger} \psi_{\ell} \right]. 
\end{align}
Now, we perform a particle-hole transformation on every second site, i.e.,
\begin{equation}
\psi_{\ell} \rightarrow \psi^{\dagger}_{\ell} \quad \& \quad \psi^{\dagger}_{\ell} \rightarrow \psi_{\ell} \quad \forall \ell \text{ odd} \ \text{($\ell \in \mathrm{o}$)}.
\end{equation}
This mapping preserves the fundamental anti-commutation relations of fermions, i.e.,
\begin{equation}
\left\{\psi_{\ell},\psi^{\dagger}_{m}\right\} = \delta_{\ell m} \quad, \quad  \left\{\psi_{\ell}^{(\dagger)},\psi^{(\dagger)}_{m}\right\} = 0.
\end{equation}
As a next step, we exchange $\ket{\uparrow}\leftrightarrow\ket{\downarrow}$ on even sites ($\ell \in \mathrm{e}$), i.e. \begin{equation}
S^{\pm}_{\ell,\ell+1} \rightarrow -S^{\mp}_{\ell,\ell+1} \quad \& \quad S^{z}_{\ell,\ell+1} \rightarrow -S^{z}_{\ell,\ell+1}, \quad \forall \ell \in \mathrm{e},
\end{equation}
which preserves the spin algebra $[S^+,S^-] = 2S^z$ and $[S^z,S^\pm] = \pm S^\pm$.
Discarding an inconsequential constant, these transformations lead to the Hamiltonian 
\begin{equation}
H = \sum_{\ell} \left[-\frac{\tilde{t}}{2}\left(\psi_{\ell}S^+_{\ell,\ell+1}\psi_{\ell+1} + \text{H.c.}\right) + m \psi_{\ell}^{\dagger} \psi_{\ell} \right],
\end{equation}
whose first terms differs by an inconsequential factor $-\mathrm{i}$ from that of $H_{\text{QLM}}$ given in Eq.~\eqref{eq:H-QLM} of the main text.
The above analysis assumes $\tilde{t}$ to be a real number. Some derivations starting from the Hamiltonian of quantum electrodynamics, as in Ref.~\cite{Kasper2016}, use a notation with an imaginary $\tilde{t}$. These different notations have no physical consequence, as we can map $S^{\pm}\rightarrow \pm (-\mathrm{i}) S^{\pm}$, which preserves the spin algebra and makes the prefactor real.

To arrive at the Bose--Hubbard model, we further map the fermions to spins via a Jordan--Wigner transformation, 
\begin{equation}
\psi_{\ell}^{\dagger}\psi_{\ell} \rightarrow s^{z}_{\ell} + \frac{1}{2} \quad \& \quad \psi_{\ell}^{\dagger}\psi_{\ell+1} \rightarrow \pm s^{+}_{\ell} s^{-}_{\ell+1} \; ,
\end{equation}
where we defined a second set of spin operators $s^{\pm}, s^{z}$, that now describe the matter fields. The sign in the second mapping can be chosen at will. Using this transformation, the Hamiltonian (up to an irrelevant constant) takes the form \cite{Hauke2013,Yang2016}
\begin{align}
H= \sum_{\ell } \left(\mp\frac{\tilde{t}}{2}\left[ s^{-}_{\ell}S^+_{\ell,\ell+1}s^-_{\ell+1} + \text{H.c.}\right] +   m s_{\ell}^{z}  \right)\; .
\end{align}

In turn, the matter-spins can be mapped to hard-core bosons on sites $\ell$, via $\ket{\uparrow/\downarrow} \rightarrow \ket{1/0}$, and
\begin{equation}
s^+ \rightarrow b^{\dagger}, \quad s^- \rightarrow b, \quad s^z \rightarrow b^{\dagger}b-\frac{1}{2}.
\end{equation}
Up to an irrelevant constant, this leads to
\begin{align}
H = \sum_{\ell } \left[\mp\frac{\tilde{t}}{2}\left(b_{\ell}S^+_{\ell,\ell+1}b_{\ell+1} + \text{H.c.}\right) +   m b_{\ell}^{\dagger}b_{\ell}  \right].
\end{align}
Finally, interpreting the spin on the links as the boson occupation states $\ket{2/0}$, one arrives at
\begin{equation}
H_{\text{eff}} = \sum_{j_\text{m}} \left\{m b_{j_\text{m}}^{\dagger}b_{j_\text{m}} + \frac{\tilde{t}}{\sqrt{8}} \Big[b_{j_\text{m}-1} \big(b^{\dagger}_{j_\text{m}}\big)^2 b_{j_\text{m}+1} +\text{H.c.} \Big]\right\},
\end{equation}
where we have relabeled the index and have made a choice of sign.
This is the effective Hamiltonian quoted in Eq.~\eqref{eq:effective-Hamiltonian} of the main text. 

As described in detail in Ref.~\cite{Yang2020} this Hamiltonian can be obtained from the Bose--Hubbard model in Eq.~\eqref{eq:BHM} through second-order degenerate perturbation theory. The factor $\tilde{t}/2$ describes the strength of the operation $\ket{101} \rightarrow \ket{020}$, derived as  
\begin{align}\nonumber
\tilde{t} &=\sqrt{2} J^2 \left(\frac{1}{\delta+\Delta} + \frac{1}{U-\delta +\Delta} + \frac{1}{\delta-\Delta} + \frac{1}{U-\delta-\Delta}\right)\\ &\approx\frac{8\sqrt{2} J^2}{U}\,,
\end{align}
where the approximation is valid close to the resonance condition.

\section{Histogram of most dominant processes in final wave function}\label{sec:hist}

In this section, we provide information about the gauge-violation processes beyond what is presented in Fig.~\ref{fig:violations} and the associated text. For this purpose, we consider the Fock basis of particle-number product states $\ket{N_1,N_2,\dots,N_L}$, and see how much these states are occupied by the final wave function after the ramp has concluded at $t=120$ ms. For numerical feasibility, we only consider the $9$ sites centered around the defect at site $j_\mathrm{c}$ for the imperfect initial states, or around the middle of the chain for the clean state. For the clean case and those of matter-field defects $j_\mathrm{c}=16$, while for the gauge impurity $j_\mathrm{c}=17$. As such, we calculate the probability that we will find the final wave function in product states with a given configuration $|N_{j_\mathrm{c}-4},\ldots,N_{j_\mathrm{c}},\ldots,N_{j_\mathrm{c}+4}\rangle$ around the site $j_\mathrm{c}$, for the cases of a defect-free initial state, and an initial state with one of the three main defects (MH, MI, GI) that we have considered in this work.
	
The corresponding results are presented in Fig.~\ref{fig:hist}. We see that in the cases of a clean initial state, or an initial state with a matter-field defect, the dominant product states are gauge-allowed ones (blue color). This consolidates our conclusions that matter-field defects are benign, and have a negligible effect on the gauge-invariant dynamics. Indeed, the ratio, be it in terms of number or weight, of gauge-allowed processes to their counterparts due to gauge-violation dynamics is qualitatively the same between the clean case and its matter-field defect counterparts. For these three initial states we have only shown the dominant processes with a population greater than $10^{-3}$.
	
On the other hand, we see that in the case of a GI defect in the initial state, the number and weight of dynamics-induced gauge-violating processes and their gauge-allowed counterparts are much closer. This provides further understanding as to why the GI defect contributes to the total dynamics-induced gauge violation significantly more than its matter-field counterparts. Here, we have chosen a population cutoff of $2\times10^{-3}$.

\bibliography{DefectsBiblio}
\end{document}